\def\ARXIVVERSION{1}
\documentclass[conference]{IEEEtran}
\usepackage{cite}
\usepackage{amsmath,amssymb,amsfonts}
\usepackage{graphicx}
\usepackage{textcomp}
\usepackage{xcolor}
\usepackage[utf8]{inputenc}

\DeclareUnicodeCharacter{2013}{--}
\DeclareUnicodeCharacter{2014}{---}
\DeclareUnicodeCharacter{2019}{'}
\DeclareUnicodeCharacter{2212}{-}

\DeclareMathOperator*{\argmax}{argmax}
\usepackage{xspace}
\usepackage{amsthm}
\usepackage{float}
\usepackage{multirow}
\usepackage{makecell}
\usepackage{colortbl}
\graphicspath{{figs/}}
\DeclareGraphicsExtensions{.pdf}

\usepackage{algorithm}
\usepackage{algpseudocode}
\usepackage{multicol}
\usepackage{cuted}
\usepackage{lipsum}

\usepackage{caption}
\usepackage{subfigure}
\usepackage{wrapfig}
\usepackage{gensymb}
\usepackage{booktabs}
\usepackage{comment}

\PassOptionsToPackage{hyphens}{url}
\usepackage{hyperref}

\newcommand{\diff}[1]{\textcolor{black}{#1}}
\newcommand{\shepherd}[1]{\textcolor{black}{#1}}
\newcommand{\todo}[1]{\textcolor{purple}{TODO: #1}}
\newcommand{\hl}[1]{\textcolor{black}{#1}}

\newcommand{\update}[1]{\textcolor{orange}{Update: #1}}

\newcommand{\ie}{\textit{i.e.,}\xspace}
\newcommand{\eg}{\textit{e.g.,}\xspace}

\newcommand{\fig}{Fig.~}

\newcommand{\att}{OpZ}
\newcommand{\vzw}{OpY}
\newcommand{\tmb}{OpX}

\providecommand{\Description}[1]{}

\def\BibTeX{{\rm B\kern-.05em{\sc i\kern-.025em b}\kern-.08em
    T\kern-.1667em\lower.7ex\hbox{E}\kern-.125emX}}

\IEEEoverridecommandlockouts
\IEEEpubid{\makebox[\columnwidth]{979-8-3195-0662-7/26/\$31.00 $\copyright$2026 IEEE \hfill}\hspace{\columnsep}\makebox[\columnwidth]{ }}

\begin{document}

\title{Multipath Adaptive Video Streaming with Multiple Description Neural Video Codec over 5G Networks}

% Double-blind: authors omitted for review
%\author{\IEEEauthorblockN{Paper ID: 368}}
\author{
\IEEEauthorblockN{Xinyue Hu, Ziyan Wu, Jiaxiang Tang, Wei Ye, Qixin Zhang, Eman Ramadan, Ali Anwar, Zhi-Li Zhang}
\IEEEauthorblockA{University of Minnesota Twin Cities, Minneapolis, USA \\
 %   Minneapolis, USA, 55455 \\
    \{hu000007, wu000598, tang0836, ye000094, zhan8548, eman, aanwar, zhzhang\}@umn.edu}
}

\maketitle

\begin{abstract}

\diff{5G networks employ multiple radio channels to meet growing demands for bandwidth and high-resolution video streaming for emerging applications. However, existing multipath video systems are largely designed around monolithic codecs, which require sufficiently complete chunk delivery, or layered codecs, which depend on timely base-layer delivery. Under fast-varying 5G conditions with blockage, handovers, and heterogeneous path capacities, 
%such codec-level decoding dependencies make path mapping fragile:
\shepherd{we observe that decoding dependencies in existing codecs make multipath delivery fragile:} 
transient under-delivery of critical video data can directly trigger stalls and degrade QoE.}
%While multipath video streaming has been extensively studied, existing studies primarily focus on adaptive bit rate (ABR) streaming using videos encoded using \hl{monolithic coding (e.g., Advanced Video Coding) or layered coding (e.g., Scalable Video Coding)}. 5G poses unique challenges that require re-thinking multipath adaptive video streaming. We argue that neither \hl{monolithic coding or layered coding} is best suited for streaming high-resolution videos using diverse 5G channels with rapidly varying conditions and temporary ``outages'' due to blockage or handovers. 

%Through extensive trace-driven evaluations and experiments over operational 5G networks, we demonstrate that our system enhances overall QoE by 26\%-44\%, improves video quality by 41.8\%, and maintains video stall rates below 0.13\%.

\diff{This paper proposes NeuralMDC, a neural multiple-description video codec co-designed with multipath streaming for dynamic 5G networks. NeuralMDC encodes each video chunk into \emph{independently-decodable} and \emph{mutually-refinable} description streams, each spanning the full chunk. This design changes the multipath delivery unit from dependent packets or layers to independent chunk-level streams, so missing streams primarily reduce quality rather than making the chunk undecodable. Built on NeuralMDC, we develop a user-space multipath streaming system that maps description streams to heterogeneous 5G paths with simple yet effective scheduling logic. Across trace-driven emulation and operational 5G experiments, NeuralMDC improves QoE by 26\%–44\% over existing monolithic, layered, and neural streaming systems, improves video quality by up to 41.8\%, and keeps stall ratios below 0.32\%.}

%In this paper, we tackle the challenges posed by 5G networks by advocating \emph{ multi-description coding} (MDC), which produces \emph{independent} video streams. Leveraging recent advances in deep learning, we design a novel \emph{neural} MDC video codec, named NeuralMDC, that uses masked transformers to encode each video chunk into multiple \emph{independently-decodable} and \emph{mutually-refinable} streams. Built on NeuralMDC, we develop a simple yet effective multipath MDC video streaming system over 5G networks.  Through extensive trace-driven evaluations and experiments over operational 5G networks,  we demonstrate how our multipath MDC streaming system can better cope with rapidly varying 5G channels and deliver superior performance over multipath  \hl{monolithic-coding and layered-coding} streaming systems. Our system enhances overall QoE by 26\%-44\%, improves video quality by 41.8\%, and maintains video stall rates below 0.32\%. 
%representing a 97\% reduction in 99th-percentile stall time. }

\end{abstract}

\section{Introduction}\label{sec:intro}

\diff{Video streaming is central to mobile applications ranging from conventional video services to} cloud gaming, augmented/virtual/extended reality (A/V/XR), tele-operated robots, autonomous driving, and Digital Twins, where streaming of either pre-recoded or real-time video is central. %With peak throughput reaching multiple Gbps~\cite{wong20205g,xu2020understanding,narayanan2020first}, 
5G networks, \diff{with Gbps peak throughput~\cite{wong20205g,xu2020understanding,narayanan2020first},} offer the potential to support ultra-high-resolution video streaming. To increase coverage and support growing bandwidth requirements, 5G networks employ multiple radio channels from diverse radio bands and multiple operators~\cite{liu20205g,ahokangas2019business},
%, and multiple operators often coexist in a given (urban) location~\cite{liu20205g,ahokangas2019business}, 
suggesting the possibility of utilizing multiple 5G channels from either the same or multiple operators to improve throughput and robustness as required by emerging applications (\eg CellFusion system\cite{cellfusion-sigcomm23}  for tele-operated driving).

%  and more sophisticated systems such as Chorus introduce coupled QoE optimization and fine-grained packet scheduling. However, prior studies typically consider moderate-throughput settings such as one WiFi path plus one 4G/5G path, whereas high-resolution 5G streaming must cope with aggregate bitrates of hundreds of Mbps.

Multipath video streaming has been extensively studied, %(see,~\cite{lv2024chorus} and the references therein)
\diff{but existing studies} largely assume that each \diff{video chunk is} encoded \hl{into multiple \emph{single} bitstreams at different quality levels using monolithic codecs such as Advanced Video Coding (AVC/H.264) or High Efficiency Video Coding (HEVC/H.265).}
\diff{Once a quality level is selected, the key} challenge lies in \diff{\emph{path mapping}: deciding how to distribute each encoded chunk across paths with heterogeneous and time-varying bandwidths while meeting playback deadlines.} %\shepherd{For monolithic codecs, this usually means packetizing and striping one interdependent chunk bitstream across paths; in segment-based ABR systems, late packets can leave the chunk’s HTTP response incomplete and trigger stalls.}
Previous studies have shown that standard multipath transport protocols such as MPTCP~\cite{mptcp} and MP-QUIC~\cite{mpquic:1,zheng2021xlink} are ineffective for multipath video streaming over wireless networks~\cite{saha2019musher,lv2024chorus, de2019multipathtester}. More sophisticated systems such as Chorus~\cite{lv2024chorus} couple QoE optimization with path mapping through coarse-grained decisions and fine-grained packet scheduling. However, Chorus was conducted under moderate-throughput settings with one WiFi path and one 4G/5G path, where aggregate chunk throughput remains below 70~Mbps.
\diff{5G multipath streaming faces more challenges}, especially when mobility is involved~\cite{narayanan2020lumos5g,5g-mobility}: i) diverse radio characteristics across 5G channels, with bandwidth ranging from 20 MHz to 100 MHz; ii) highly dynamic channel conditions that cause rapid and significant throughput fluctuations; iii) temporary link failures due to blockage or handovers %, where throughput may drop to nearly zero for 10's or 100s of milliseconds (ms) up to a few seconds
-- the \emph{``bad'' path} problem; and iv) the absence of a consistently superior 5G channel %, as no single channel always outperforms others 
(see~\S\ref{sec:5G-challenges}). 
Such channel heterogeneity and dynamics make multipath streaming with \emph{monolithic coding} particularly problematic: over-estimating available bandwidth \diff{and ``bad'' paths  lead to  incomplete chunk delivery that directly causes video stalls and degrades overall QoE. 
They also limit \emph{layered coding} such as SVC/SHVC: although late enhancement layers can be discarded, playback still depends on timely base-layer delivery. \emph{Striping} layers across paths leaves the base layer vulnerable to ``bad'' paths, while \emph{pinning} layers requires reliably selecting the best path and accurately matching uneven layer sizes to fluctuating path bandwidth.} Therefore, %We therefore argue that 
neither \hl{monolithic coding} nor \hl{layered coding} is best suited for multipath streaming over 5G networks. %This discussion illustrates that there are interesting interplays among video codec designs, multipath streaming systems, and network dynamics. 

In this paper, we tackle the challenges in multipath video streaming over 5G networks \diff{ through a \emph{video codec--streaming system co-design.}} %by \emph{joint} video coding and streaming system design.  
We revisit Multiple Description Coding (MDC)\diff{, a long-standing coding principle that generates independently decodable \shepherd{descriptions}, and show why this abstraction} is best suited for streaming high-resolution videos utilizing multiple 5G channels (\S\ref{sec:overview}). 
%We revisit Multi-Description Coding (MDC) which produces \emph{independent} video streams and argue that MDC is best suited for streaming high-resolution videos utilizing multiple 5G channels (\S\ref{sec:overview}). 
\diff{However, traditional MDC designs~\cite{franchi2005multiple, le2023multiple} are difficult to use in our setting: they often incur high coding overhead, support only a small number of descriptions, require complex multi-decoder architectures, or provide limited control over description sizes. To overcome these limitations,} we %leverage} recent advances in neural video coding and
\diff{design \emph{NeuralMDC},} a novel \emph{neural} MDC video codec using masked transformers, which encodes each video \diff{chunk} into multiple \emph{independently-decodable} and  \emph{mutually-refinable} \shepherd{descriptions}. \diff{Each description spans the full chunk \shepherd{and is transmitted as one stream\footnote{\shepherd{We use description to refer to the codec-level independently decodable unit, and stream to refer to its transmitted bitstream in the streaming system; unless otherwise noted, the two terms correspond one-to-one.}}};
%and can be decoded independently; 
additional descriptions progressively improve reconstruction quality.} Our NeuralMDC is efficient, offering fine-grained control over stream size distribution, enabling flexible combinations of streams \diff{and robust recovery from missing descriptions with modest compression overhead} (see~\S\ref{sec:MDC}). 
%-- even \emph{partial} streams are decodable and thus \emph{loss-resilient},  and achieving superior rate-distortion performance without incurring significant compression overheads (see~\S\ref{sec:MDC}). 

Built on NeuralMDC codec, we develop a \diff{user-space} multipath MDC video streaming system for 5G networks (\S\ref{sec:MDC-streaming}). 
\diff{The key insight is that NeuralMDC changes the unit of multipath delivery\shepherd{: instead of striping packets from a monolithic encoded chunk or assigning dependent codec layers to paths, the system schedules independently decodable chunk-level description streams}. %from dependent packets or dependent layers to independent chunk-level streams. 
As a result, transient under-delivery on a path primarily reduces video quality rather than making the chunk undecodable. Unlike generic packet-loss recovery~\cite{cheng2024grace}, NeuralMDC targets deadline-driven partial delivery in 5G multipath streaming: fast-varying and heterogeneous paths may deliver only a subset of scheduled descriptions before playback, and any received subset remains useful for chunk-level reconstruction. 
This property simplifies path mapping: instead of protecting a monolithic whole chunk or a critical base layer, the scheduler maps independently useful streams to heterogeneous paths. While NeuralMDC} works with both striping and pinning, our evaluation results show that pinning offers the best QoE performance with nearly zero stalls: each independent stream is delivered over a single path, so a ``bad'' path only affects its assigned streams without \diff{triggering dependency-driven \shepherd{retransmission} over other channels, or} cascading stalls.  Thanks to the fine-grained control over stream size distribution using Pyramid Source Coding (\S\ref{s:spliting}), independent streams of varying sizes can be flexibly combined, greatly simplifying dynamic stream-to-path mapping. The independent decodability and loss-resiliency of NeuralMDC reduce the need for accurate bandwidth estimation and path quality monitoring, making the system robust to estimation errors and wrong decisions.

Through extensive trace-driven evaluations using real-world 5G traces and experiments over operational 5G networks, 
we demonstrate how our multipath \diff{NeuralMDC} streaming system can better cope with rapidly varying 5G channels 
and deliver superior performance over existing multipath \hl{monolithic coding and layered coding} streaming systems.  
In emulated 5G networks, \diff{NeuralMDC} streaming system enhances overall QoE by 26\%-44\%, achieving up to 15.4\% improvement in video bitrate, a 41.8\% increase in video quality, and keeping video stalls below 0.32\%. Real-world experiments further validate its effectiveness, demonstrating a 25.0\% QoE improvement and near-zero average stall time.

%a 97\% reduction in 99th-percentile stall time. 
%compared to streaming with SVC. %\xinyue{update number}

\diff{In summary, this paper makes the following contributions:}
\begin{itemize}
\item \diff{We identify why monolithic and layered coding are poorly matched to dynamic 5G multipath delivery.}

\item \diff{We design \emph{NeuralMDC}, a neural MDC video codec that realizes MDC abstraction as fine-grained, configurable, independently decodable, and mutually refinable chunk-level descriptions, overcoming the limited scalability and inflexible stream sizing of traditional MDC designs. }%a neural multiple-description video codec that encodes each video chunk into multiple independently decodable and mutually refinable descriptions. Unlike traditional MDC designs, NeuralMDC supports many fine-grained descriptions with configurable sizes, enabling flexible stream combinations with modest compression overhead.}

\item \diff{We develop a user-space multipath streaming system that exploits NeuralMDC’s chunk-level independent decodability to simplify path mapping over heterogeneous 5G channels. Missing streams reduce reconstruction quality rather than making a chunk undecodable, improving robustness to bandwidth variation, blockage, handovers, and path-estimation errors.}

\item \diff{We evaluate NeuralMDC with trace-driven emulation and operational 5G experiments, showing 26\%--44\% QoE gains and stall ratios below 0.32\%.}
\end{itemize}

\section{Background and Motivation}\label{sec:background}
%1. Background of 5G and CA.
%2. Video Codecs. How AVC is generally tied to streaming nowadays. Why we need MDC.

\begin{figure}[t]
    \centering
    \includegraphics[width=\columnwidth]{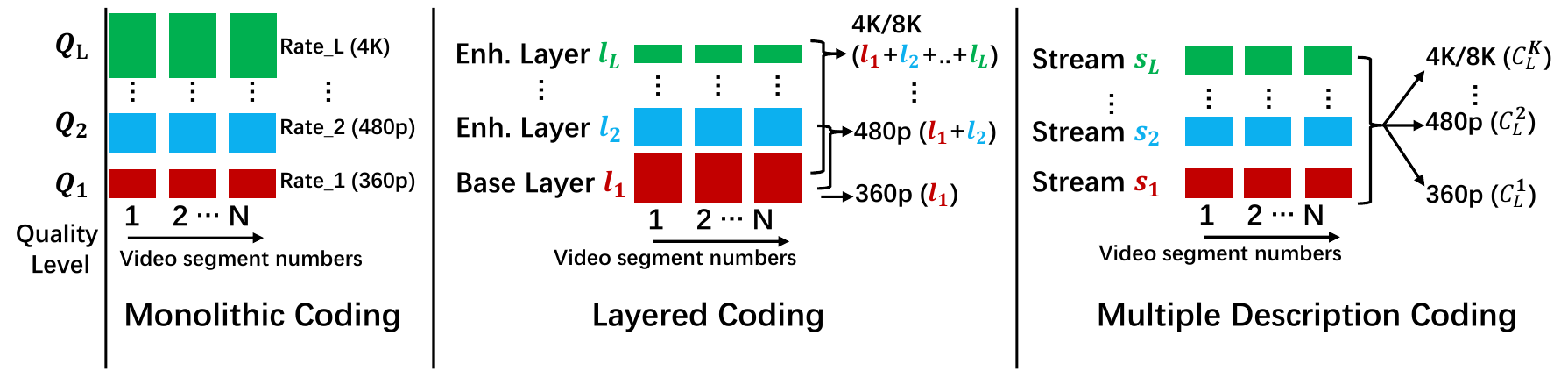}
    \vspace{-6mm}
    \caption{Three types of video codecs to encode a video at different quality levels.} % In \hl{monolithic coding}, one quality level is achieved by selecting a single encoded bitstream. In \hl{layered coding }, a quality is achieved by combining multiple \hl{dependent} layers in a sequential order. In multiple description coding, a quality is achieved by a combination of multiple streams.}
    \vspace{-4mm}
    \label{f:avc_svc_mdc}
\end{figure}
 
\subsection{Video Codecs and Video Streaming}\label{sec:codecs} 
Video codecs fall broadly into three categories \hl{(see Fig.~\ref{f:avc_svc_mdc}): (i) monolithic coding (\eg AVC/H.264 and HEVC/H.265), which encodes video into a single bitstream that must be received in full for correct decoding; (ii) layered coding (\eg SVC~\cite{schwarz2007overview} and SHVC~\cite{shvc}), which encodes video into a base layer and enhancement layers, allowing decoding from the base layer alone while progressively improving quality as additional layers are received; and (iii) multiple description coding (MDC)~\cite{goyal2001multiple, kazemi2014review}, which encodes the video into multiple independently decodable and mutually refinable streams, so that any subset of streams can be used to reconstruct the video. 
Both traditional layered codecs (e.g., SVC/SHVC) and multiple description coding schemes are largely extensions of AVC/HEVC and have not been widely used due to their high complexity and significant bitrate overhead. SVC/SHVC introduces inter-layer dependencies and signaling overhead, and suffers from reduced compression efficiency because inter-layer frame prediction is largely avoided to prevent reconstruction drift. 
Existing MDC codecs~\cite{franchi2005multiple, le2023multiple} rely on cumbersome multi-decoder architectures, scale poorly beyond two streams, and offer limited loss resilience due to de-correlated nature of DCT transforms and complex state synchronization.} %Improving robustness typically requires oversampling or duplicating source data, further reducing compression efficiency.
%(a) AVC and its variants; (b) Scalable (Layered) Video Coding (SVC)~\cite{schwarz2007overview} such as SHVC~\cite{shvc}; and iii) Multiple Description Coding (MDC)~\cite{goyal2001multiple, kazemi2014review}, see Fig.~\ref{f:avc_svc_mdc}.  
%Due to their complexity and high overheads, SVC and MDC codecs based on conventional compression methods have not been widely used. 
With rapid advances in deep learning, many \emph{neural} video codecs have been proposed to \hl{improve compression efficiency}: for example,  those in~\cite{lu2019dvc, hu2021fvc, rippel2021elf} are neural \hl{monolithic} codecs;  SWIFT~\cite{swift} is a neural \hl{layered} codec. 
In contrast, neural MDC codecs remain largely unexplored, with the only prior work~\cite{hu2021multiple} proposing a GNN-based super-resolution method to enhance the reconstruction quality of a traditional MDC codec.

Today's video streaming systems employ \emph{adaptive bit rate} (ABR) algorithms to dynamically select the quality level of the next (\hl{monolithic-}encoded) video chunk or the number of layers/streams of the next (\hl{layered}/MDC encoded) video chunk. The ABR decision is typically based on client buffer occupancy and estimated network throughput, aiming to optimize overall QoE~\cite{robust-mpc, yan2020learning}. Differences in various video streaming systems lie primarily in their ABR algorithms. ABR algorithms have been extensively studied for both \hl{monolithic}~\cite{robust-mpc} and \hl{layered} codecs~\cite{liu2020grad,swift}, although the latter have not been deployed yet in practice. %Due to the vast literature on ABR algorithm designs, we will not delve into these topics further. 

%Today's video streaming systems employ \emph{adaptive bit rate} (ABR) algorithms. An ABR algorithm typically comprises two basic components: \emph{Network Estimation}, which estimates network throughput, measures client buffer occupancy, and possibly monitors additional network performance metrics (e.g., round-trip latency); \emph{QoE Decision} which uses the information from Network Estimation to select the quality level of the next (\hl{monolithic-}encoded) video chunk or the number of layers/streams of the next (\hl{layered}/MDC encoded) video chunk to prefetch; the goal is to optimize the overall QoE metric.  Differences in various video streaming systems lie primarily in the ABR algorithms used for dynamic video quality or layer selection. ABR algorithms have been extensively studied in the literature. Various deep learning-based video streaming algorithms have also been developed in recent years.  Although not yet deployed in practical systems, ABR algorithm designs for \hl{layered coding} have also been studied in the literature (see, e.g.,~\cite{liu2020grad,swift}). Due to the vast literature on ABR algorithm designs, we will not delve into these topics further. %\todo{cite some ABR papers}
%In~\S\ref{sec:related}, we will highlight a few studies that are most relevant to our study.

\vspace{-4pt} 

\subsection{5G: Opportunities and Challenges}\label{sec:5G-challenges} 
5G networks employ a diverse set of radio channels from low/mid-band frequency ranges to high-band (mmWave) frequency ranges to support various applications. 
With wide deployments of 5G networks, in a typical urban location, there are often multiple channels offered by different operators~\cite{Ahmad-Hotmobile23,ye2024dissecting,Dimitrios-Uplink-IMC23}. For example, according to the analysis of~\cite{Ahmad-Hotmobile23} based on measurement studies conducted in the US and Europe~\cite{sigcomm-ross}, there are at least four radio channels available in each location under study.  Furthermore, multiple 5G operators often coexist, offering competing 5G services~\cite{sigcomm-ross}.  By establishing multiple connections (``paths'') using multiple channels (either from the same operator or different operators), it is now feasible to support applications that require streaming of ultra-high-resolution videos over 5G networks. 
Combining multiple 5G channels not only increases the overall system throughput, but also helps mitigate challenges posed by individual channels. \hl{This path diversity is particularly valuable for high-bandwidth and mission-critical applications such as immersive XR services, UAV video feeds, and autonomous vehicle sensor sharing. In practice, emerging Dual-SIM Dual-Active (DSDA) technologies~\cite{car_DSDA,qualcomm_DSDA} enable simultaneous use of two independent 5G data connections on a single platform.} %for latency- and bandwidth-intensive applications, such as automotive telematics

\begin{figure}[t]
    \centering
    \begin{minipage}{0.48\columnwidth}
        \centering
        \includegraphics[width=\linewidth]{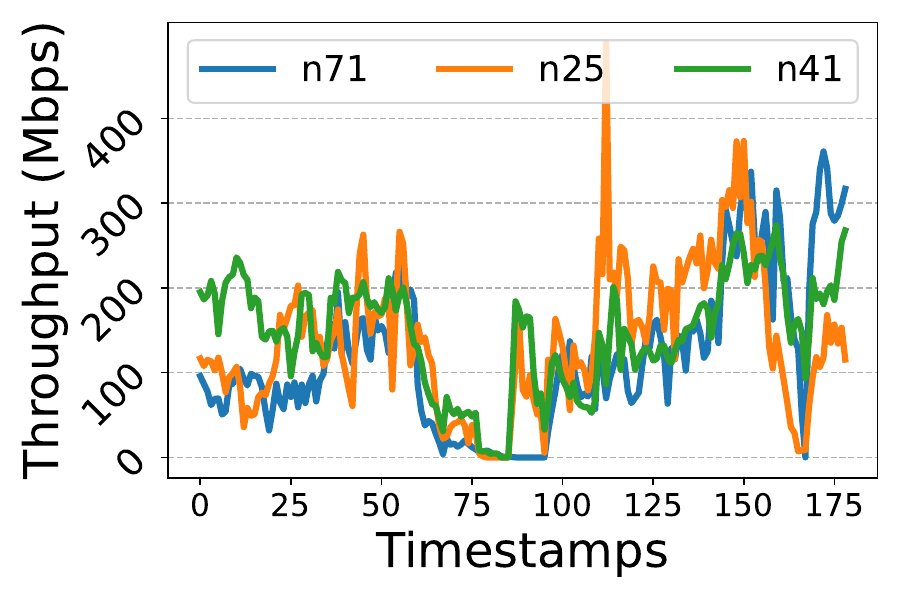}
        \vspace{-6mm}
        \caption{Traces of different bands.}
        \vspace{-6mm}
        \label{f:multi-band-traces}
    \end{minipage}
    \hfill
    \begin{minipage}{0.48\columnwidth}
        \centering
        \includegraphics[width=\linewidth]{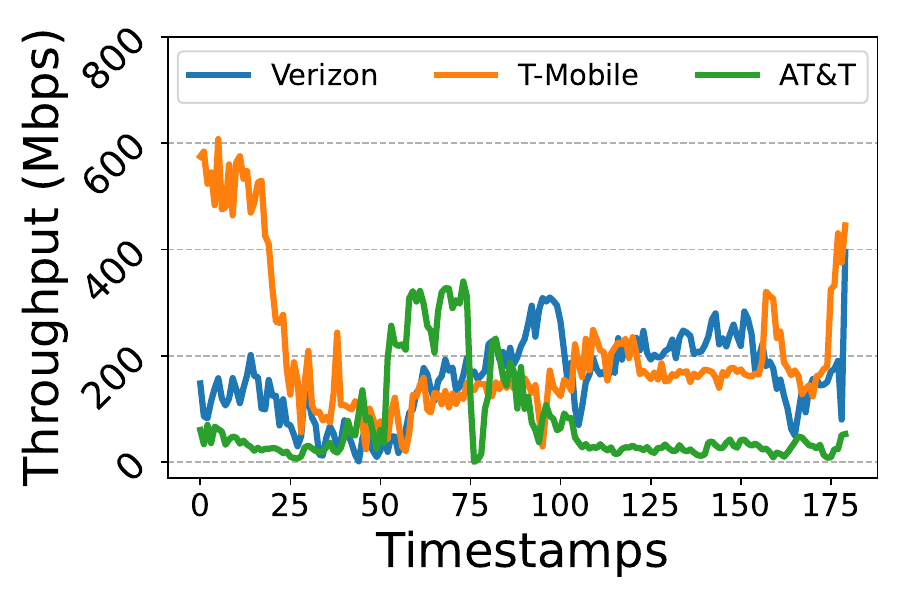}
        \vspace{-6mm}
        \caption{Traces of different operators.}
        \vspace{-6mm}
        \label{f:multi-operator-traces}
    \end{minipage}
\end{figure}

While offering great opportunities, 5G networks also pose significant challenges for multipath video streaming, as outlined in~\S\ref{sec:intro}. 
\begin{comment}
First, 5G radio channels can be very diverse -- depending on the bands used, their bandwidth can range from 20 MHz to 100s MHz, and their radio propagation characteristics may also differ drastically. Second, 5G channel bandwidth can fluctuate rapidly and wildly due to interference, radio resource competition, and other factors; worse, 
a 5G channel may temporarily ``fail'' (i.e., throughput tentatively drops to nearly zero), e.g., due to blockage or handovers (when the user is mobile). Third, availability of radio channels may also vary over time or locations. Fourth, as 5G channel conditions can vary rapidly and significantly over time, there is often no single ``best'' channel (``best path'')~\cite{Ahmad-Hotmobile23}.
\end{comment}
Using  publicly released 5G measurement data from~\cite{ye2024dissecting}, Figs.~\ref{f:multi-band-traces} and~\ref{f:multi-operator-traces} illustrate throughput traces from three radio channels of the same operator (n25, n41, n71 bands) and from three different operators, respectively. The traces show that throughput of each channel fluctuates significantly, can drop to near zero, and no single channel consistently outperforms others. These challenges make multipath video streaming over 5G networks using \hl{monolithic} and \hl{layered} codecs ineffective, as we elaborate next. \emph{Effectively tackling these 5G challenges motivates us to design a novel neural MDC codec and advocate for multipath neural MDC video streaming over 5G networks}.

%we present some real-world examples to help illustrate these challenges. Fig.~\ref{f:multi-band-traces} plots the throughput traces collected over a duration of 3 minutes from three radio channels used by the same operator. These channels are from three different 5G mid-bands: n25 (1.9 GHz), n41 (2.5 GHz), and n71 (0.6 GHz) bands. Fig.~\ref{f:multi-operator-traces} plots the throughput traces of 5G radio channels collected from three different operators. In both cases, the traces from the radio channels are collected simultaneously using several smartphones. These sample traces show that the throughput of each channel can fluctuate significantly; and can sometimes drop to near zero. None of the channels consistently outperforms other channels.
%These challenges make multipath video streaming over 5G networks using \hl{monolithic} and \hl{layered} codecs ineffective, as will be elaborated in~\S\ref{sec:multi-path-design-challenges}. \emph{Effectively tackling these 5G challenges motivates us to design a novel neural MDC codec and advocate for multipath neural MDC video streaming over 5G networks}. 

%%%%%%%%%%%%%%%%%%%%

\subsection{Limitations of Monolithic and Layered Codecs for Multipath Streaming}\label{sec:multi-path-design-challenges}
Multipath streaming introduces an additional design component beyond ABR: \emph{Path Mapping}, that decides what video data must be transmitted over which path. This creates interesting interplay among multipath characteristics, video codecs, and ABR algorithms  that have not been explored before.

%When streaming \hl{monolithic}-encoded videos, \emph{striping} the single video bitstream across paths is the only option. HHowever, this leads to significant QoE degradation over 5G networks due to two primarily reasons: 1) Fast varying 5G channels make it difficult to accurately estimate the channel bandwidth; over-estimating the aggregate bandwidth causes stalls with cascading effects on future chunks. 2) Temporary path failures (``bad'' paths) further exacerbate the problem, leading to prolonged video stalls. 
When streaming \hl{monolithic}-encoded videos, \emph{striping} the single video bitstream across paths is the only option.
%Striping is used by all existing multipath video streaming studies (see, e.g.,~\cite{zheng2021xlink,lv2024chorus} and references therein). 
As already stated in the introduction, multipath \hl{monolithic} video streaming (with striping)  over 5G networks suffers significant QoE degradation due to two primarily reasons: 1)~Fast varying 5G channels make it difficult to accurately estimate the channel bandwidth; over-estimating the aggregate throughput not only causes video stalls, but may also have a cascading impact on the timely delivery of future chunks. 2)~Temporary path failures (``bad'' paths) further exacerbate the problem, leading to prolonged video stalls.

%With \hl{layered}-encoded videos, we can employ either \emph{striping} or \emph{pinning}, but neither is well-suited for 5G multipath streaming. 
%\textbf{Layered streaming with striping} can mitigate path dynamics and ``bad'' path effects when disruptions affect only enhancement layers, allowing playback to fall back to use lower-quality layers without stalling. However, failures during base-layer delivery require rescheduling remaining packets across paths and still cause stalls if the entire base layer cannot be received before the playback deadline. Thus, striping inherits the limitations of \hl{monolithic} streaming, albeit to a lesser extent. 

When streaming \hl{layered} encoded videos,  we can employ either \emph{striping} or \emph{pinning}, but neither is well-suited for 5G multipath streaming.

\noindent
%\textbf{Problem of Multipath \hl{Layered-Coding} Streaming with \emph{Striping}.}
\textbf{\hl{Layered-Codec} w. \emph{Striping}.}
Multipath \hl{layered} streaming with \emph{striping} can mitigate path dynamics and ``bad'' paths when disruptions affect only enhancement layers, allowing playback to fall back to lower-quality layers without stalling. However, failures during base-layer delivery require rescheduling remaining packets across paths and may still cause stalls if the entire base layer cannot
be received before the playback deadline. Thus, striping inherits the limitations of \hl{monolithic} streaming, albeit to a lesser extent, while offering improved resilience to bandwidth overestimation by discarding late enhancement layers.

\noindent
\textbf{\hl{Layered-Codec} Streaming w. \emph{Pinning}.}
\hl{Layered} streaming with pinning can avoid the ``bad'' path problem \emph{if} the ``best'' path is selected for the base layer, the next ``best'' path is selected for the first enhancement layer, and so forth. However, this requires: 1) best path selection \& ordering, and 2) careful video-to-path mapping based on layer bitrates and path qualities. 
In 5G networks, there may not always be a consistently ``best'' path. 
%It is also unclear what metrics should be used to quantify the ``best'' path: is it highest throughput, minimal roundtrip time (minRTT~\cite{mptcp}), throughput variability~\cite{sigcomm-ross} or some other measures of path reliability? 
Moreover, due to layered codec designs and path diversity, the \emph{mismatch} between layer bitrate requirements and available path bandwidths is often inevitable\footnote{For example,  due to the limitation of its neural codec design, SWIFT's~\cite{swift} base layer is significantly larger than its enhancement layers. Pinning it to the ``best'' path may not be feasible if the path's bandwidth is insufficient, requiring striping across multiple ``best/next-best'' paths.}. The diverse and fast-varying 5G channel characteristics and \emph{dependency} among video layers make designing accurate path quality monitoring mechanisms for
best path selection and effective bandwidth estimation for
dynamic path mapping all the more daunting.

\section{System Overview}\label{sec:overview}
The problems plaguing multipath \hl{monolithic-encoded} and \hl{layered-encoded} video streaming over 5G networks lead us to advocate jointly co-designing video codec and multipath streaming using MDC. To overcome the limitations of conventional MDC codecs, we propose a neural MDC codec (NeuralMDC).  We also develop a novel multipath neural MDC video streaming system, specifically designed to overcome the challenges posed by 5G networks. 
%Both the neural MDC and multipath streaming system are specifically designed to overcome the challenges posed by 5G networks.
%Below we provide an overview of NeuralMDC and the streaming system. %and illustrate how NeuralMDC simplifies the overall system design while enabling effective utilization of the high (aggregate) bandwidth offered by multiple 5G channels.
%In the following, we provide an overview of our proposed neural codec and multipath streaming system. In particular, we will illustrate how our neural MDC codec makes the overall system design far simpler while enabling us to effectively take advantage of the high (aggregate) bandwidth offered by multiple 5G channels.

\begin{table}[t]
\centering
\vspace{-2mm}
\caption{NeuralMDC features and mechanisms}
\vspace{-2mm}
\resizebox{\columnwidth}{!}{
\begin{tabular}{|c|c|}
\hline
\rowcolor[gray]{0.9}
\textbf{Features} & \textbf{Mechanisms}   \\
\hline
        Dependency-free decoding & \makecell[l]{Utilizes conditional-coding compression\\ No motion vectors, residuals, or I-frames} \\
        \hline
        Adjustable stream sizing & \makecell[l]{Pyramid Source Information Splitting} \\
        \hline
        Compression efficiency & \makecell[l]{Entropy encoding w/ more accurate \\ distribution estimation by Transformer}
        \\
        \hline
        \makecell[c]{Any (partial) \\ stream combination} 
        & \makecell[l]{Masked Transformer to reconstruct \\ frames w/ any amount of tokens}
        \\
        \hline
        Runtime & \makecell[l]{Model quantization, batch processing} \\
        \bottomrule
    \end{tabular}
}
\vspace{-2mm}
\label{tab:MDC}
\end{table}

\noindent
\textbf{NeuralMDC Codec Overview.}
Our NeuralMDC is designed with four goals:  1)~\textit{Dependency free decoding}: ensure video streams are independent, allowing any combination to be decoded for enhanced video quality. 
2)~\textit{Adjustable stream size}: split video information into multiple streams of varying sizes to adapt to 5G throughput dynamics. 
3)~\textit{Compression efficiency}: maintain good compression efficiency, and 
4)~\textit{Runtime}: minimize compression latency overhead.
%as splitting video reduces spatial-temporal redundancy,
To this end, we leverage the conditional-coding compression technique, an emerging paradigm in neural compression~\cite{li2021deep,li2023neural,vct}.

NeuralMDC employs an AutoEncoder~\cite{he2022elic} to transform video frames into correlated latent tokens, which are split into multiple descriptions and independently entropy-encoded into bitstreams using a Masked Transformer~\cite{yu2023magvit, m2t}. Each stream is independently decodable; when any streams are missing, the transformer infers the missing tokens from the received ones. Any combination of received streams enhances token accuracy and decoded frame quality.  NeuralMDC distributes tokens across descriptions in a pyramidal structure, generating streams of varying sizes. Model quantization and batch processing are applied to accelerate encoding/decoding. Table~\ref{tab:MDC} lists the main features and key mechanisms (details in \S\ref{sec:MDC}).

\iffalse
NeuralMDC first employs an AutoEncoder~\cite{he2022elic} to transform video frames into correlated latent tokens, which are subsequently split into multiple descriptions. A Masked Transformer~\cite{yu2023magvit, m2t} is used to capture the spatial-temporal correlations among tokens and independently entropy-encodes each description into a bitstream. As a result, each stream is independently decodable. When any streams are missing, NeuralMDC uses the transformer to infer the missing tokens based on the received ones. Any combination of received streams enhances token accuracy and decoded frame quality. The size of an MDC stream is determined by the number of tokens it contains, with more tokens yielding a larger stream. NeuralMDC distributes tokens across multiple descriptions in a pyramidal structure, generating streams of varying sizes. For efficient video compression, NeuralMDC employs the Masked Transformer to %exploit temporal correlations between frames and spatial redundancy within a description. It 
estimate the token distribution of a description conditioned on tokens from the previous frame, enabling entropy coding to allocate fewer bits to a more predictable description~\cite{minnen2018joint, minnen2020channel}.  Finally, model quantization and batch processing are applied to accelerate NeuralMDC. Table~\ref{tab:MDC} lists its main features and key mechanisms used.
\fi

\noindent
\textbf{Multipath NeuralMDC Streaming System Overview.}
Our multipath MDC video streaming system leverages three key features of our NeuralMDC codec: i)~independently decodable video streams eliminate the need for best path selection and minimize the impact of ``bad'' paths on the overall video chunk delivery; ii)~flexible, fine-grained control over stream number and their sizes (bitrates) simplifies dynamic stream-to-path mapping; and iii)~resilience to data losses -- even \emph{partial} streams are decodable (see \S\ref{sec:codec-evaluation}) -- reduces the need for accurate bandwidth estimation and precise path quality monitoring, making the system more robust to estimation errors and wrong decisions. Together, these features enable more ``aggressive'' streaming decisions while minimizing their negative impacts, allowing the system to fully utilize the high aggregate bandwidth offered by multiple 5G channels while more effectively accommodating and coping with their diverse characteristics and fast-varying dynamics. 
Table~\ref{tab:system-comparison-summary} lists the major design challenges and advantages of multipath MDC streaming over multipath \hl{monolithic} \& \hl{layered} streaming.

\begin{figure*}[t]
% \vspace{-2mm}
    \centering
    \includegraphics[width=0.8\textwidth]{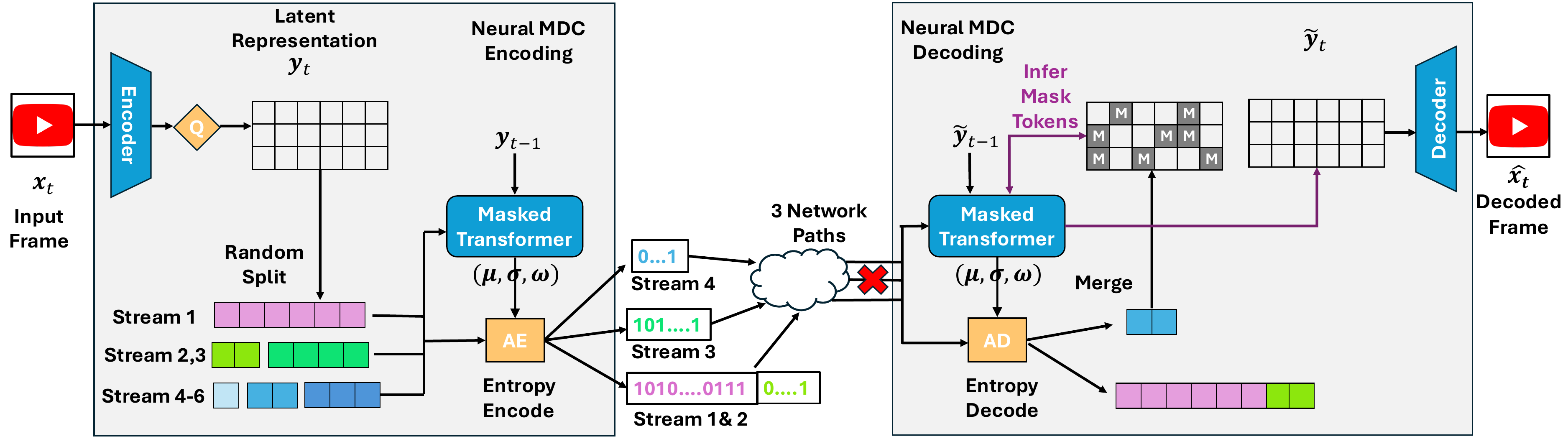}
    \vspace{-2mm}
    \Description{}
    \caption{NeuralMDC Workflow: an example that 1) encodes a video frame to 6 streams and transmits 4 of them over 3 network paths to match path bandwidth; 2) decodes 3 received streams, infers missing streams, and decodes the frame.}%\footnote{AE and AD refer to entropy encoding and decoding using Arithmetic Coding.}
   \vspace{-6mm}
    \label{f:framework}
\end{figure*}

\begin{table}[t]
% \scriptsize
  \centering
  \vspace{-2mm}
  \caption{Multipath Streaming: \hl{monolithic coding vs. layered coding} vs. MDC}
  \vspace{-2mm}
  \resizebox{\columnwidth}{!}{
  \begin{tabular}{|c|c|c|c|}
    \hline
       \rowcolor[gray]{0.9} &&&\\
\rowcolor[gray]{0.9}
\multirow{-2}{*}{\textbf{Codec}} & \multirow{-2}{*}{\makecell{\textbf{Path}\\\textbf{Mapping}}} 
& \multirow{-2}{*}{\textbf{Path Dynamics}} & \multirow{-2}{*}{\textbf{Path Failures}} \\
\hline
% \rowcolor[gray]{0.9}     \multirow{2}{*}{\textbf{Codec}} & \multirow{2}{*}{\makecell{\textbf{Path}\\\textbf{Mapping}}} & \multirow{2}{*}{\textbf{Path Dynamics} }& \multirow{2}{*}{\textbf{Path Failures}} \\
     \textbf{\hl{Monolithic codec}} & Striping & 
    \makecell[l]{Require accurate estimation\\ Overestimation degrades QoE}
    & Complete stalls \\
    \hline
    \multirow{2}{*}{\textbf{\hl{Layered codec}}} & Striping & \multirow{2}{*}{\makecell[l]{Require best path selection \\
    Complex path mapping}} & \multirow{2}{*}{Stall w/o base layer} \\
    & Pinning & & \\
    \hline
    \multirow[b]{2}{*}{\textbf{MDC codec}} & Striping & 
    \multirow[b]{2}{*}{\centering\makecell[l]{No need for accurate estimation\\ Robust to wrong decisions}} 
    & Minimal Stalls \\
    \cline{2-2}\cline{4-4}
    & Pinning & & \makecell[l]{Solve \textbf{Bad} Path\\
    (Nearly) No stalls} \\
    \hline
  \end{tabular}}
  \label{tab:system-comparison-summary}
  \vspace{-2mm}
\end{table}

\section{Neural MDC}\label{sec:MDC}
This section presents the design details of NeuralMDC. %which generates \emph{fine-grained} \emph{independently-decodable} and \emph{mutually-refinable} video streams for rapid adaptation to fluctuating 5G throughput. 
%It also covers runtime optimizations for fast encoding and decoding.
\fig \ref{f:framework}  depicts the architecture of NeuralMDC. %It consists of three main components: AutoEncoder, Source information splitting \& merging, and  Masked Transformer. 
We first describe the overall encoding and decoding processes before detailing each component.

\subsection{Encoding \& Decoding Process}

\noindent\textbf{Encoding process:} NeuralMDC Encoder uses an  AutoEncoder~\cite{he2022elic} to independently  transform each input frame $x_t$ into a quantized latent representation $y_t=\lfloor E(x_t) \rfloor $. 
Unlike DCT transforms in H.26x, which decorrelate  coefficients, the AutoEncoder transforms retain spatial-temporal correlations in the latent domain~\cite{li2023mage, yu2023magvit}. 
The tokens of latent representation  $y_t$ are then split into multiple descriptions (\S\ref{s:spliting}). 
Masked transformers (\S\ref{s:transformer}) are then used to estimate the distribution of the current latent representation based on the prior one, $p(y_t | y_{t-1})$. The distribution, combined with Arithmetic Coding~\cite{witten1987arithmetic}, is used to independently entropy-encode each description into a bitstream. %The more accurately the transformer predicts the distribution, the fewer bits are needed to transmit $y_t$.

\noindent\textbf{Decoding process:} 
NeuralMDC Decoder uses the same Masked Transformer to estimate the distribution $p(y_t | \widetilde{y}_{t-1})$, Arithmetic Coding then uses the estimated distribution to entropy-decode each bitstream into latent tokens. Since each stream is independently entropy encoded, each is independently decodable. 
Next, the reconstructed latent tokens are merged together (\S\ref{s:spliting}). 
In case of missing streams, their tokens are assigned with a special masked token and predicted by the Masked Transformer using the estimated distribution (\S\ref{s:transformer}). The AutoEncoder~\cite{he2022elic} then decodes the reconstructed latent representation  $\widetilde{y}_t$ to generate the frame $\widetilde{x}_t=D(\widetilde{y}_t)$.
Any combination of received streams improves the latent representation accuracy and thus the decoded frame quality.

\subsection{Pyramid Source Information Splitting} \label{s:spliting}
%We split the latent representation by randomly masking parts of it with a special learnable mask token to form multiple masked latent representations that, when combined, equal the original representation. \fig \ref{f:seperation_example } \todo{illustrates an example of forming 4 descriptions}. The mask token is assigned a high probability, leading to negligible bitrate overhead as arithmetic coding encodes the mask token with only a few bits. Streams with more masked tokens have smaller bitrates after entropy encoding.

%Given the wide fluctuations in 5G throughput, ranging from a few Mbps to 1 Gbps, and the varying characteristics of different 5G channels, MDC streams should have differing bitrates. % to adapt effectively to these conditions. 
To create asymmetric MDC streams and enable fine-grained bitrate adaptation for varying 5G channels, we distribute different amounts of latent tokens across multiple streams in a pyramidal manner~\cite{adelson1984pyramid}. %with token ratios following a pyramid structure
The splitting  process begins by randomly and evenly distributing tokens across 
$S$  node descriptions. Each node description is then further divided into $m$ streams with token ratios following a pyramid structure. The value of $m$ follows a pyramid structure as well. The first stream contains the most amount of tokens (\ie source information) and thus has the largest bitrate, with progressively fewer tokens added in subsequent streams.
%the token ratio for the $m$-th stream out of $M$ streams is defined as $\frac{2m-1}{M^2}$. 
\fig \ref{f:framework} illustrates an example of forming six MDC streams and merging three of them. 
\shepherd{We implement token-to-description assignment using a seed-controlled pseudo-random permutation over latent-token indices. The seed is included in stream metadata, allowing the decoder to regenerate and invert the permutation to recover received tokens to their original latent positions.} When a description stream is missing, NeuralMDC assigns a special mask token to the corresponding token positions.
%To enable the receiver to recover the original token positions in the latent representation, we use a reversible pseudo-random function for mapping, with both sender and receiver using the same random seed. When a stream is missing, NeuralMDC Decoder assigns a special masked token to its corresponding token positions. 

Note that NeuralMDC exclusively uses the latent representation as source information, because  other types of source information, such as motion, optical flow, and residuals~\cite{lu2019dvc,xiang2022mimt,li2023neural, cheng2024grace}, carry different important source information and lack strong correlations with each other. Consequently, the loss of one type (\eg motion) cannot be efficiently estimated from the received other types (\eg residual).

%Note that NeuralMDC exclusively uses the latent representation as source information and does not utilize other types of source information, such as motion, optical flow, or residuals~\cite{lu2019dvc,xiang2022mimt,li2023neural, cheng2024grace}, as they carry different important information and lack strong correlations with each other. Consequently, the loss of one type (\eg motion) cannot be efficiently estimated from the received other types (\eg residual). 
%The distinct impacts of loss on motion vectors and residuals on reconstructed video quality are shown in \fig \ref{f:source_type_loss_perf}. 
%It is evident that motion information is more critical than residuals, and the loss of motion cannot be effectively compensated for, even if the residuals are fully received.  
%Therefore, NeuralMDC exclusively uses the latent representation as source information, letting the transformer extract diverse contexts from representations for compression.

%Note that when partitioning the source information into multiple descriptions, our approach avoids introducing any additional redundant information, as commonly seen in traditional MDC methods (e.g., through oversampling or duplication). the DCT transform de-correlates the coefficients, thereby preventing the estimation of one coefficient from others within the same DCT block.

\subsection{Masked Transformer for Entropy Coding and Inference}\label{s:transformer}

By avoiding motion-residual wrapping, NeuralMDC uses masked transformers to \diff{leverage} spatial-temporal correlations between frames to improve compression efficiency. It uses masked transformers to estimate the conditional distribution of the latent representations to allow entropy coding to allocate fewer bits to more likely symbols. %NeuralMDC uses masked transformers to leverage spatial-temporal correlations between frames to improve compression efficiency. 
Given a chunk of frames $\{x_t\}_{t=1}^T$ and their latent representations $\{y_t\}_{t=1}^T$, where each is split into $M$ descriptions $y_t = \{y_{t,m}\}_{m=1}^M$, the masked transformer predicts the conditional distribution of each description $p(y_{t,m} | y_{t-1})$, which is used to entropy code the description $y_{t,m}$ into a bitstream.  The average bitrate of an MDC stream is $\approx  E_{y_{t,m} \sim Q}[ -\log_2 p(y_{t,m} | y_{t-1})]$, where Q is the true symbol distribution of the description~\cite{minnen2020channel}. We model the conditional distribution as a mixture of Gaussians with $3$ mixtures, each parameterized by a mean $\mu$, scale $\sigma$, and weight $\omega$. 
The transformer runs independently on each description, %enabling parallel execution.
trading reduced spatial context for parallel execution. 
%To compress the full video, this process is applied iteratively to each frame, with the transformer predicting the conditional distributions for each latent representation. 
For the first frame, distributions are predicted by padding with zeros.

\noindent\textbf{Training:} We train the Autoencoder and Masked Transformer to  minimize the rate-distortion loss function

\vspace{-6mm}
\begin{equation}
    E_{(x_1, x_2)\sim p_{X_{1:2}}}[ -\log_2 p(y_{2,m}  | y_{1,m}) + \lambda MSE(x_2, \widetilde{x}_2)]
    \label{eq:bitrate}
\end{equation}
where $(x_1, x_2)\sim p_{X_{1:2}}$ are two adjacent video frames drawn from the training dataset, the parameter $\lambda$ with a larger value results in a higher bitrate but lower pixel distortion %(\ie better quality) 
of the reconstructed frame. To simulate arbitrary \diff{missing} descriptions during training, we randomly sample a subset ($0\%$ to $100\%$) of tokens in $y$ and replace them with a special learnable mask token. The masked transformer is trained to predict the removed tokens.% by minimizing the cross entropy between the predicted distributions of masked tokens and the true distributions of the removed tokens~\cite{yu2023magvit}. %\diff{The random masking used in training does not model random network packet loss; instead, it trains the transformer to reconstruct latent-token subsets corresponding to missing streams under deadline-driven partial delivery.}

\noindent\textbf{Inferring Missing Streams:}
%The inference of missing streams involves distribution prediction and sampling. 
%After entropy decoding the received streams and merging the available tokens, 
The masked transformer predicts the distribution of missing tokens  $p(y_{t, M} |\widetilde y_{t,\overline M}, \widetilde y_{t-1})$, where $\widetilde y_{t,\overline M}$ represents the received tokens and, along with the previous representation, provide temporal and spatial context for the transformer. %to predict the distributions of the missing tokens. 
Each missing token $\widetilde y_{t}^j$ is inferred by sampling from its most probable values.

\vspace{-4mm}
\begin{equation}
    \widetilde y_{t}^j = \argmax_{y_{t}^j} p(y_{t, M} |\widetilde y_{t,\overline M}, \widetilde y_{t-1})
\end{equation}

\subsection{Implementation and Optimization} \label{s:runtime_opt}
We implement NeuralMDC in Pytorch by extending M2T~\cite{m2t}, a masked transformer-based image codec. \diff{Unlike M2T, NeuralMDC uses masked modeling to realize video MDC.} %, and VCT~\cite{vct}, a  transformer-based video codec. 
%two recent work utilizing masked and unmasked transformers for image and video compression. 
We train NeuralMDC on an Nvidia A6000 GPU using the Vimeo-90K dataset \cite{xue2019video}.%, the standard training dataset used by many neural codecs~\cite{lu2019dvc, li2023neural,cheng2024grace}.  
To achieve various bitrate control, we optimize the training loss for six $\lambda$ values (0.0001 to 1). %The training is conducted using Nvidia A6000 GPU. %To further optimize NeuralMDC for mobile devices, we export our NeuralMDC model to ONNX~\cite{onnx} and use ONNX Runtime's quantization (\texttt{FP16}) to reduce the model size and improve inference speed.

The encoding and decoding latency of NeuralMDC primarily stems from the Autoencoder and Masked Transformer runtimes. 
%By eliminating motion-residual wrapping operations, NeuralMDC's streamlined architecture enables faster encoding and decoding than other neural video codecs. 
Two runtime optimizations are applied: (1) ONNX export with \texttt{FP16} quantization~\cite{onnx} to reduce the precision of neural network weights from 32-bit to 16-bit, nearly halving computation time; and (2) batch processing~\cite{chen2024lifter}, where the Autoencoder processes a batch of frames in parallel to reduce the Transformer’s idle time, since each frame is converted to its latent representation independently. 
For edge deployment on mobile devices and unmanned ground vehicles, we follow Qualcomm AI Hub’s edge deployment pipeline~\cite{qai_mae}. We first convert the PyTorch model into a TorchScript-traced representation and then export it into ONNX format to achieve platform-agnostic \shepherd{deployment}. We apply post-training \texttt{FP16} quantization to reduce computation and memory overhead\shepherd{. We further apply  ONNX graph optimizations~\cite{onnx-transformer}, including constant folding, redundant-node elimination, operator fusion where supported, and hardware-specific kernel selection, before compiling} hardware-specific executables for efficient NPU inference~\cite{Qualcomm_NPU}. 
%, apply graph-level optimizations to eliminate redundant operators, and compile hardware-specific executables for efficient inference on NPUs~\cite{Qualcomm_NPU}. 
\diff{We discuss the compute, memory, and deployment implications of NeuralMDC in \S\ref{sec:conclusion}.}

\section{Multipath MDC Streaming}\label{sec:MDC-streaming}
%We present the \diff{design} of our multipath MDC video streaming system architecture and its core components.  %We also outline the testbed and prototype implementations.

% \input{tables/streaming-diff-codecs}

\begin{comment}
    Our multipath MDC video streaming system designs leverage the key features afforded by our neural MDC codec: i) video streams can be independently decoded -- this eliminates the need for best path selection and minimizes the effect of ``bad'' paths on the overall video chunk delivery; ii) the number of video streams and their sizes (bit rates) can be flexibly encoded and made fine-grained -- this greatly simplifies the problem of (dynamic) mapping of video streams to paths; and iii) our neural MDC codec is resilient to data losses, as \emph{partial}  streams can also be decoded (\todo{see~\S\ref{sec:codec-evaluation}}) -- this reduces the requirements of accurate bandwidth estimation and precise path quality monitoring, and makes the system designs more robust to estimation errors and wrong decisions. In particular, these features together allow the system to make more ``aggressive'' decisions while minimizing their negative impacts -- this is particularly important when streaming videos over 5G networks, as it enables us to more fully utilize the high (aggregate) bandwidth offered by multiple radio channels while more effectively accommodating and coping with the diverse and fast varying radio channel dynamics. 

\end{comment}

\subsection{Multipath NeuralMDC Video Streaming}  
\begin{figure}[t]
    \centering
    \includegraphics[width=0.9\columnwidth]{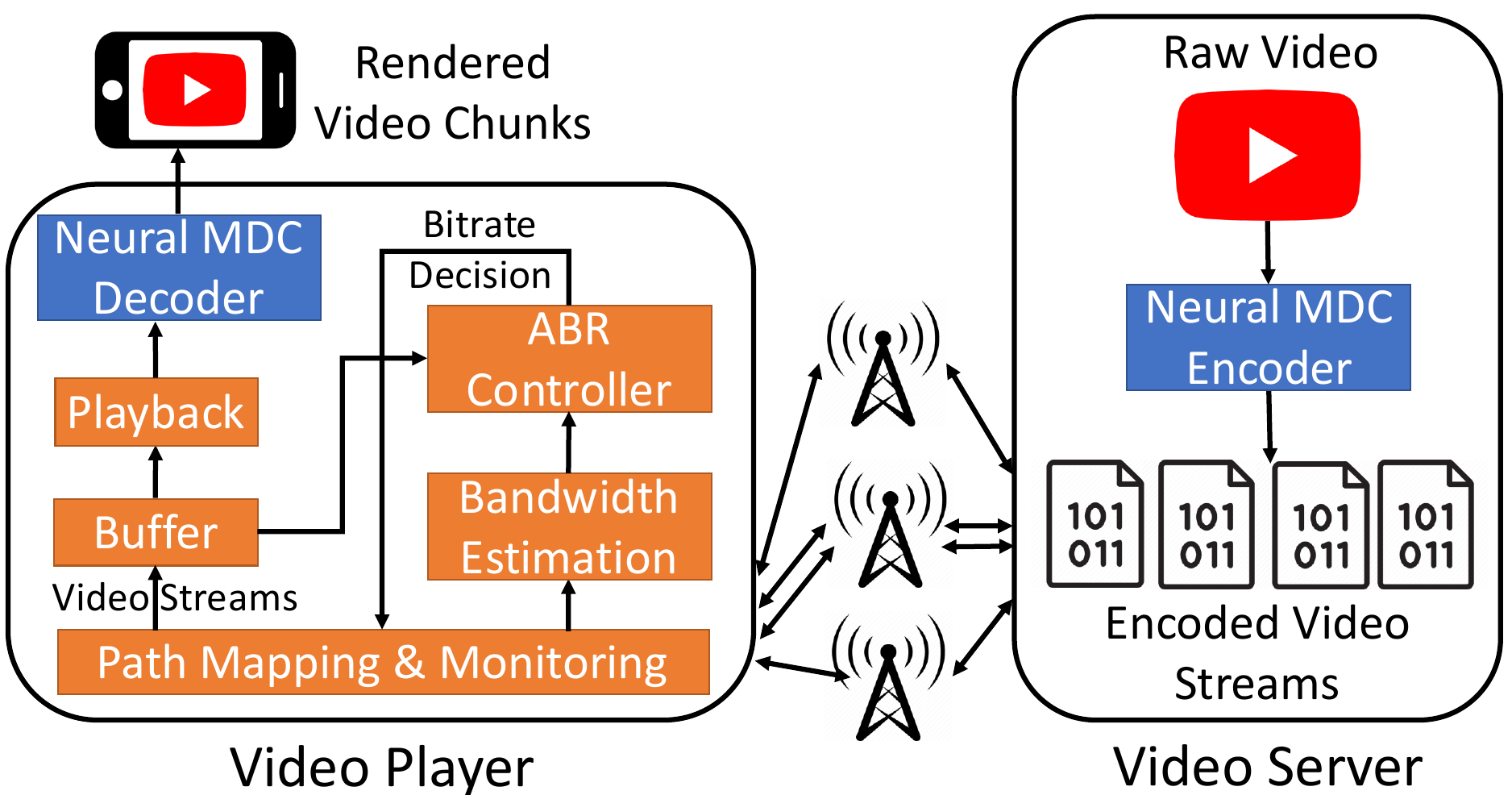}
    \vspace{-2mm}
    \caption{NeuralMDC Streaming System Architecture}
    \label{fig:video-streaming-arch}
    \vspace{-4mm}
\end{figure}

Our multipath neural MDC video stream system architecture is schematically depicted in Fig.~\ref{fig:video-streaming-arch}. It comprises three main components: 1) \emph{Bandwidth Estimation}; 2) \emph{ABR Controller}; and 3) \emph{Path Monitoring and Mapping}. Our system also includes an \emph{(optional) playback logic} to further optimize QoE at (near-) playback time. We will briefly describe the design of each component below. Both path striping or pinning can be used, although our extensive evaluations show multipath MDC streaming with \emph{pinning} provides the best performance results. %Hence we will emphasize the design of multipath MDC video streaming with pinning. 

\noindent\textbf{Bandwidth Estimation.} We use the harmonic mean method (as in~\cite{lv2024chorus}) to estimate per-path bandwidth. Although more accurate algorithms such as deep learning-based 5G throughput prediction~\cite{Eman-IMC21,ye2024dissecting,narayanan2020lumos5g,PREDICT} exist, these algorithms are sophisticated and computationally expensive, and we find the robustness afforded by NeuralMDC renders them unnecessary for meaningful QoE gains, so we opt for simpler designs.

%require additional feature collection and computation resources, which add significant complexity in the system design. Thanks to the robustness and flexibility afforded by our NeuralMDC codec, we do not find these sophisticated algorithms provide considerable improvements in streaming QoE. We thus opt for simpler designs.

\noindent\textbf{ABR Controller.} This module decides the overall bitrate of video streams/chunks to prefetch on a chunk-by-chunk basis to optimize the overall QoE performance by balancing video quality, stall ratio, and smoothness. 
%This decision is often made based on estimated throughput, client buffer occupancy, or a combination of both (and possibly other factors). There is a vast research literature devoted to the design of ABR algorithms to optimize video QoE performance. 
\hl{NeuralMDC is compatible with any ABR algorithms. Following Chorus~\cite{lv2024chorus}, a recent mobile multipath monolithic video streaming system, 
we also adopt (robust) Model Predictive Control (MPC)~\cite{robust-mpc}.}  Based on the estimated aggregate path bandwidth, MPC decides the \emph{overall} bitrate of the next chunk; \emph{what} and \emph{how many} streams to be fetched will then be left to the path mapping and dynamic scheduling module. %We use a simple threshold-based method to decide \emph{when} to prefetch the next chunk: when a sufficient amount of data (from one or multiple streams) has been received, the controller decides on the bit rate of the next chunk to fetch. 

\noindent\textbf{Path Monitoring and Mapping.} 
We monitor individual paths using standard methods such as minRTT~\cite{mptcp}. Given the target bitrate from the ABR controller and the manifest file specifying stream bitrate distribution, this module selects which streams to fetch based on the estimated per-path bandwidth. The pyramid source coding scheme in NeuralMDC (\S\ref{s:spliting}), together with the flexibility to combine independent streams of varying sizes, makes this process straightforward. 

In the case of multipath streaming with \emph{pinning}, the module selects and assigns video streams to paths such that each stream's bitrate does not exceed the path's estimated bandwidth; if no suitable stream exists, the closest larger stream is chosen. The aggregate bitrate of the selected streams matches the target bitrate set by the ABR controller. To enhance robustness, streams are chosen from different sub-branches of the pyramid to maximize diversity. For example, if the base stream size is 1 unit and a path has an estimated bandwidth of 7 units, the module may assign streams of 4, 2, and 1 units from different sub-branches, ensuring both efficient bandwidth utilization and diversity. For \emph{striping}, a similar approach is applied using the aggregate bandwidth of all paths, with the additional constraint that streams of different bitrates are selected to maintain diversity. Each stream is packetized and fetched sequentially (starting with the smallest). When path conditions change drastically, the current stream is aborted if its playback deadlines cannot be met; otherwise, its remaining packets are dynamically rescheduled to a higher-capacity path.

\subsection{Implementation}  
We implement two user-space versions of our multipath MDC streaming system for evaluations in \emph{controlled} and \emph{real-world} 5G environments.
%To facilitate evaluations in \emph{controlled} environments as well as \emph{real-world} 5G networks, we have implemented two versions of our multipath MDC streaming system. Both are implemented in the user space. 

\noindent\textbf{Emulation Testbed.} For performance evaluation in a controlled environment, we have built a testbed comprising a video server and a client video player in Python. The client and server are connected via the Mahimahi network simulator~\cite{netravali2015mahimahi}, where Mahimahi Linkshell replays commercial 5G channel bandwidth traces (\S \ref{s:setup}).  Our implementation is adapted from the state-of-the-art video streaming simulation platforms~\cite{pensieve,swift}. We employ a \emph{path-aware} MPC algorithm~\cite{robust-mpc} for the basic ABR decision logic, as in Chorus~\cite{lv2024chorus}. 
%We implement customized path mapping and dynamic scheduling algorithms, adopting a simplified multipath scheduling algorithm proposed in Chorus~\cite{lv2024chorus}. 

To ensure fair comparison, we have implemented the CD and FC algorithms\footnote{Implementation code for Chorus is not publicly available.} from \diff{Chorus} \cite{lv2024chorus} for multipath streaming \emph{with striping}. Hence, apart from different codecs, the mechanisms used are the same. For multipath \hl{layered video streaming} with pinning, algorithms for the best path selection and dynamic layer-to-path mapping (and base layer re-injection) based on changing path dynamics are implemented. For multipath MDC with pinning, we implement the path mapping and dynamic stream scheduling algorithms as outlined earlier. %For faster evaluations, the trace-driven emulations omit the actual video decoding and playback.

\noindent\textbf{Real-World Multipath Video Streaming.}
To evaluate operational 5G networks, we have implemented an end-to-end prototype using Go and Python. The server runs on an AWS EC2 instance hosting pre-encoded video chunks via HTTP. The video client is implemented in the user space and runs on a Linux laptop tethered (via USB) to multiple 5G Android smartphones, with one persistent HTTP connection per phone.  For single-operator experiments, each phone is locked to a specific 5G band; for multi-operator experiments, each phone connects to a different operator\footnote{Devices with Dual-SIM Dual-Active capability can also enable dual simultaneous data streams~\cite{qualcomm_DSDA, car_DSDA}, but such devices were not available during our experiments.}. \hl{To support seamless multipath communication, we implement the dataplane in Go and expose it to the Python client via a gopy-compiled module. The Go module spawns per-path goroutines with dedicated HTTP client instances bound to different source IPs (one per phone), dispatching download requests and collecting feedback via Go channels. 
%and uses channels to dispatch download requests to workers and collect status/throughput feedback via a shared feedback channel during and after each transfer. 
We export a single function to Python that takes multipath download requests, handles request scheduling, concurrent downloads, feedback aggregation, and timeouts internally, and returns the downloaded data along with per-path throughput to the ABR layer. 
This design avoids Python’s GIL contention and the overhead of multi-process IPC (Inter-Process Communication) or external queuing services, while providing efficient parallelism and asynchronous coordination.} %without requiring changes to the application logic or the HTTP server.}

\section{Multipath Streaming Evaluation}\label{sec:streaming-evaluation}
%\todo{ \textbf{quantitative results} of \textbf{How will previous ABR or system fail in 5G?}}
% \todo{Eurosys reviewer: \\
% 1. you only reported bitrate, which is weird. -> \textbf{ADD PSNR/MS-SSIM metric!}\\
% 2. I understand that 5G bandwidth is fluctuating for each channel. But could you show more \textbf{quantitative results}? \textbf{How will previous ABR or system fail?}\\
% 3. why do you separate into 10 stream?\\ 
% 4. In Section 6.1 MDC pinning is compared with the other relevant protocols. It will be good to see how neural MDC pinning works. Or it is the same thing? Authors should address that.\\
% 5. It is unclear how the system would perform when the number of paths / channels continuously goes up (beyond the case of dual-operators). It would be great if the authors can offer some intuitions on how the system would scale beyond 2 streams, and what the network performance would look like in that case.\\
% 6. The paper does not explain how the independent decodability of MDC streams impacts the overall video quality, especially when decoding with missing streams in \textbf{scenarios with path failures}. While the authors show good performance in terms of bitrate and stall metrics, \textbf{perceptual quality analysis} for partially received streams is missing.\\
% 7. throughput traces only. Important 5G characteristics like signal quality variations, handovers between cells, and carrier aggregation dynamics are not fully captured in the experiments.}

This section evaluates multipath video streaming with NeuralMDC in emulated and real-world 5G networks.

\subsection{Experiment Setup}
\label{s:setup}

We conducted experiments using a Linux laptop with a 14-core Intel Ultra CPU and three Samsung S21 Ultra phones. The video player was configured with a 30-second playback buffer and a path-aware MPC algorithm~\cite{lv2024chorus}. 
% a 14-core Intel Ultra 5 125U

\noindent\textbf{Baselines:} We compare NeuralMDC multipath streaming with the following codec and streaming combinations: \label{s:baselines}
\begin{itemize}
\item \textbf{H.265-Stripe:} \hl{Monolithic} codec H.265 via FFmpeg~\cite{ffmpeg} with path-striping, representing  Chorus~\cite{lv2024chorus}. %, a recent multipath streaming approach.
\item \textbf{SHVC-Stripe/Pin:} \hl{Layered} codec SHVC~\cite{shvc_implement}, a layered extension of H.265, with both path-striping and path-pinning scheduling algorithms.
\item \textbf{Swift-Stripe/Pin:}  Neural \hl{layered} codec Swift~\cite{swift}, constructed using a chain of AutoEncoders, with both scheduling algorithms.
\item \textbf{GRACE-Stripe:} Single-stream neural codec GRACE~\cite{cheng2024grace} with packet-level loss recovery, using a deadline-aware multipath striping that prioritizes sending a small prefix of packets for all frames in a chunk before transmitting additional packets\footnote{This reduces the probability that later frames receive no packets.}.
\end{itemize}

\noindent\textbf{Test Videos:} We evaluate on two raw video datasets: UVG~\cite{mercat2020uvg} and MCL-JCV~\cite{wang2016mcl}, totaling 37 videos with resolutions from 1080p to 4K. %\hl{The test videos have an average spatial index (SI) of 50.1 and temporal index (TI) of 25.2~\cite{itu1999subjective}, from entirely different sources than the training set.} 
Videos are segmented into 4-second  chunks and compressed offline using the above codecs. To account for 5G throughput characteristics, we configure H.265 and GRACE to encode each chunk into 8 bitrates\footnote{H.265 and GRACE bitrate: [4.1, 14.5, 24.5, 44, 94, 148, 250, 336] Mbps}. 
We configure SHVC to encode each chunk into eight layers\footnote{The SHVC implementation we use~\cite{shvc_implement} supports up to 8 layers.} at similar bitrates. 
The layer size of Swift is not configurable and Swift encodes each chunk \diff{into} 10 layers\footnote{Swift layer sizes are [66, 60, 52, 44, 36.8, 29.2, 22, 17, 11, 4] Mbps. The total bitrate of 10 layers is 342 Mbps.
%Swift bitrates are [66, 126, 178, 222, 258.8, 288, 310, 327, 338, 342] Mbps.
}.  
For NeuralMDC, we encode each chunk into 15 independent streams with varying sizes\footnote{MDC stream sizes are [4, 8, 8, 12, 12, 16, 19, 19, 23, 23, 28, 28, 38, 38, 56] Mbps. The total bitrate of 15 streams is 332 Mbps.}. \diff{Note, we do not use arbitrarily many streams because adaptation gains eventually saturate, while metadata, scheduling, and coding overheads increase.}
%to achieve more flexible adaptation
%For GRACE, we implement a deadline-aware multipath striping baseline that prioritizes sending a small prefix of packets for all frames in a chunk before transmitting additional packets, to reduce the probability that later frames receive no packets.
%\footnote{SHVC supports a maximum of 8 layers.}

% We rely on Accuver XCAL~\cite{xcal} -- a commercial grade tool which collects detailed 5G Radio Access Network protocol stack information.

\noindent\textbf{Network Traces:} 
We use XCAL~\cite{xcal} to collect network traces from the three major U.S. 5G operators concurrently. The recorded bandwidth fluctuates between 0 and 1.9 Gbps. Each trace also includes carrier aggregation (CA) components and round-trip time (RTT). \shepherd{In addition to 5G traces, our dataset contains periods where a device falls back to LTE/4G; we use these LTE/4G portions as the 4G traces.}
We filter out traces to ensure non-trivial bitrate selection. The final dataset consists of 63 traces, each with a duration of 500 seconds. 

\noindent
\textbf{Metrics:} We measure video quality using \diff{Multi-Scale Structural Similarity Index Measure  (MS-SSIM)~\cite{wang2003multiscale}} and overall streaming QoE as 
   $ QoE=\sum Q(k) - \mu\sum T(k) - \sum |Q(k+1) - Q(k)|$, where $Q(k)$ is the quality of k-th chunk, $T(k)$ is the stall time, and $|Q(k+1) - Q(k)|$ is the quality variation. %the value of $\mu$ is set to be 10. 
Following~\cite{yan2020learning}, $Q()$ uses MS-SSIM and $\mu=100$. 
%For a regular codec, $\mu$ is 336, and the maximum achievable QoE for a video streaming session is 20,160, representing the transmission of all 60 chunks at 336 Mbps with zero rebuffering.

%\input{04_MDC_Compression} %  codec performance

 % exp intro and setups
\begin{figure}[t]
    \centering
    \includegraphics[width=0.8 \columnwidth]{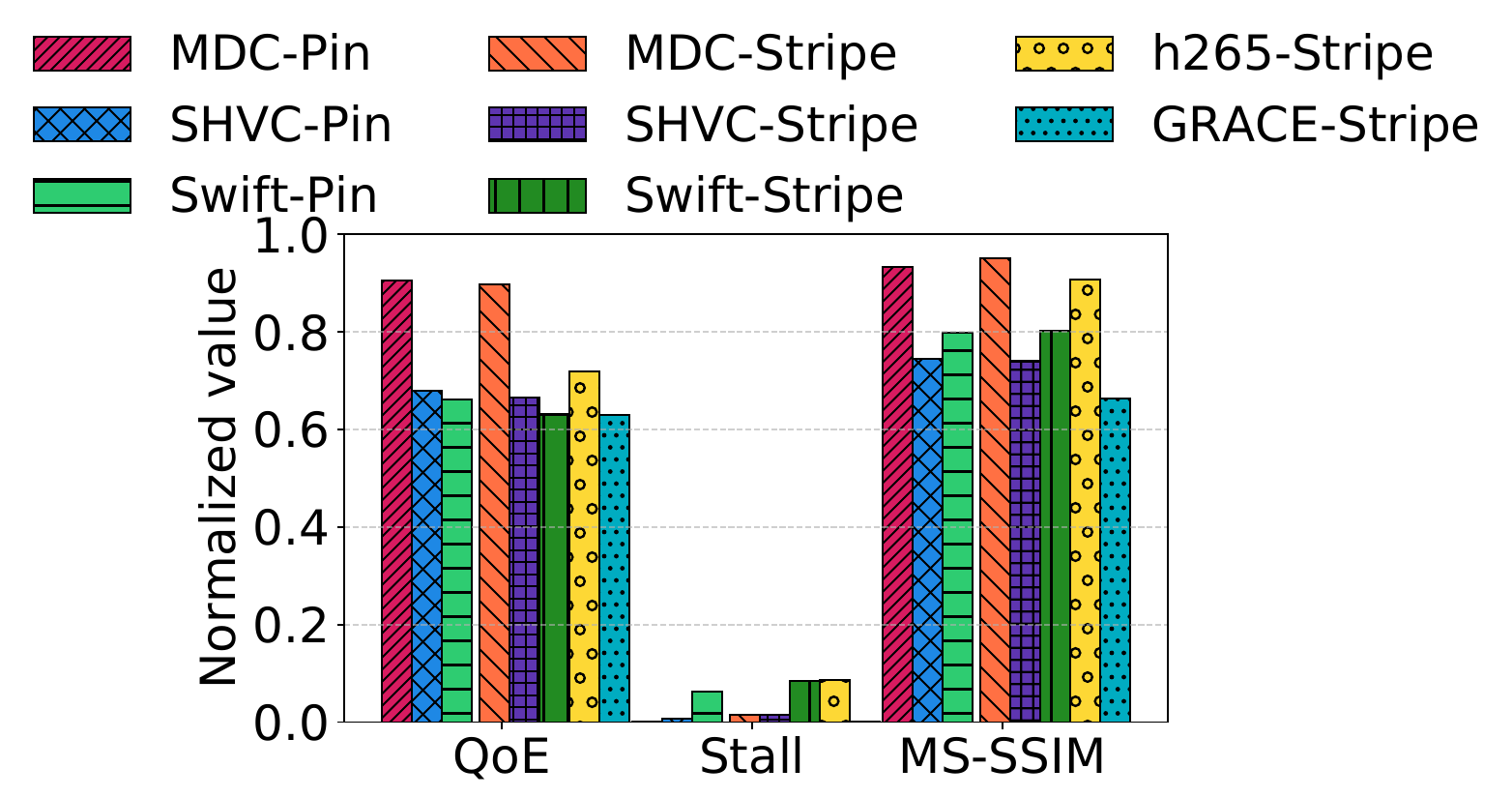}
    \vspace{-3mm}
    \caption{Overall emulated streaming performance.}
    \vspace{-4mm}
    \label{f:emulation_overal}
\end{figure}

\begin{figure}[t]
    \centering
    \subfigure[5G+5G]{\includegraphics[width=.39\columnwidth]{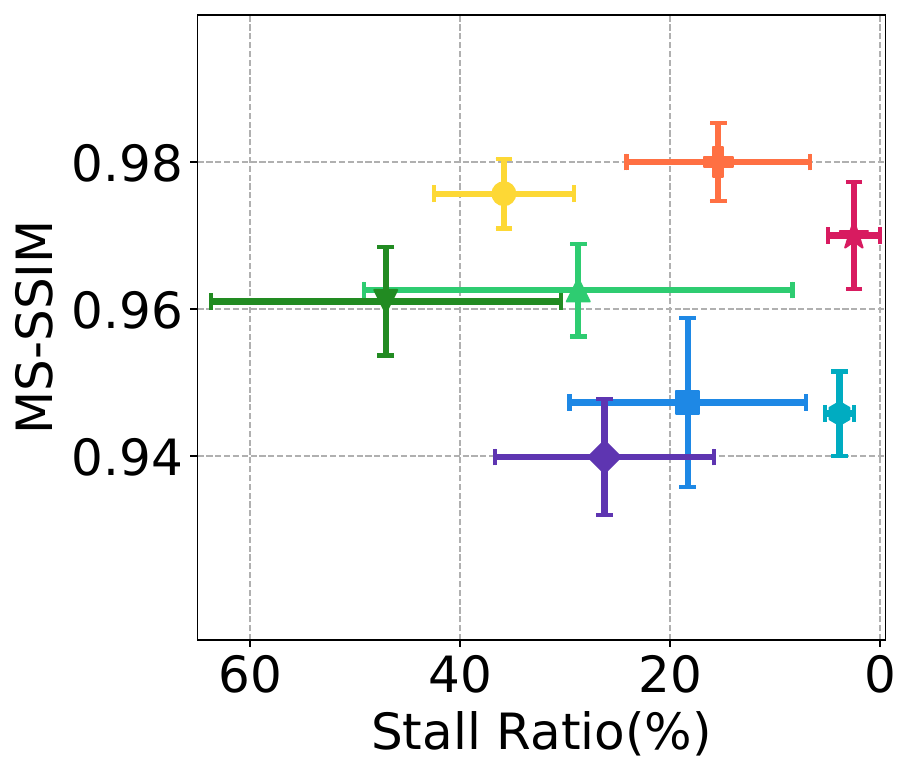}\label{f:emulation_5g_5g}}
    \subfigure[5G+4G]{\includegraphics[width=.54\columnwidth]{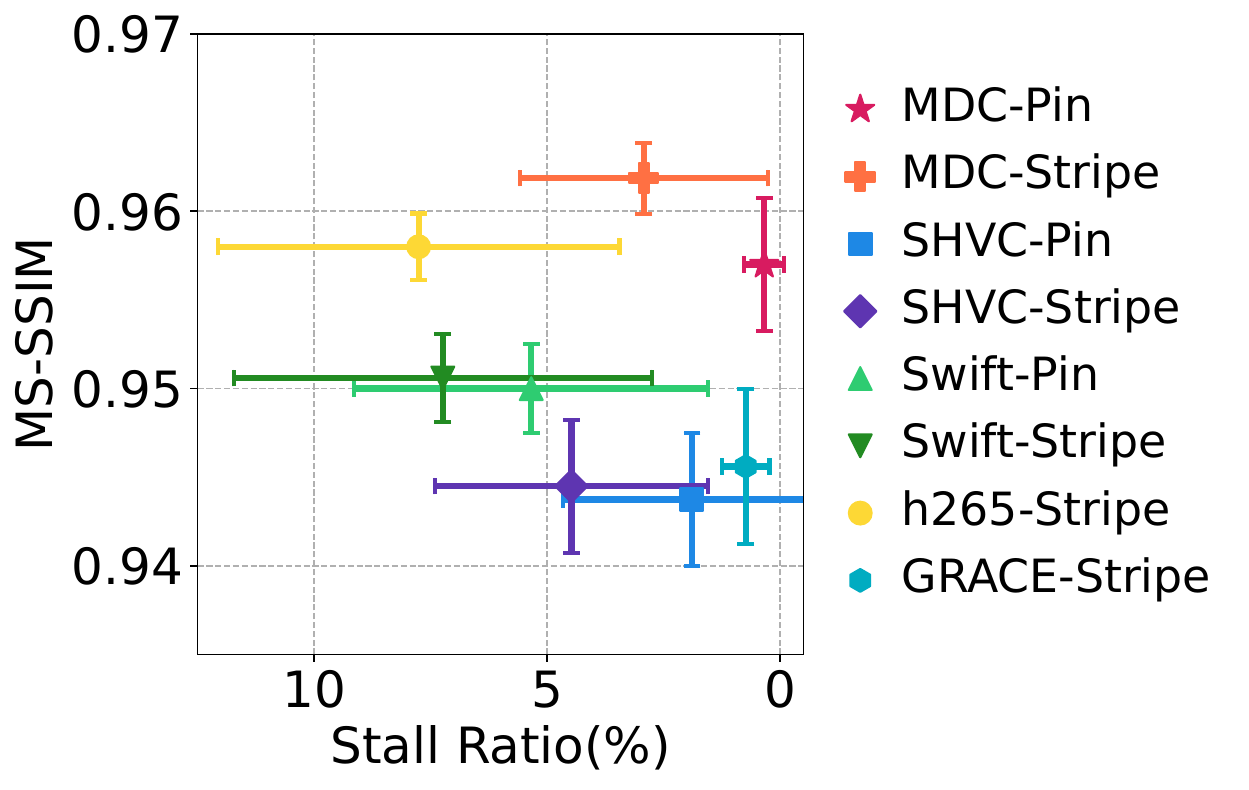}\label{f:emulation_5g_4g}}
    \vspace{-4mm}
    \caption{Performance using: 1) two unstable 5G channels and 2) one unstable 5G channel and one stable 4G.}
    \vspace{-4mm}
    \label{f:emulation_path_stability}
\end{figure}

%\subsection{Multi-path Video Streaming Emulation}
% \vspace{-0.2in}
\subsection{Trace-Driven Evaluation}
\label{s:emulation}
% \textbf{\hl{1.5 pages for simulation results }}
We first evaluate video streaming performance with NeuralMDC in emulated 5G networks using the collected traces. %We also provide detailed case studies that explain the rationale behind performance gains and demonstrate how NeuralMDC simplifies multi-path video streaming over 5G.

%%%%%%%%% emulation overall performance %%%%%%%%% 
% \begin{figure}[!ht]
%     \centering
%     \vspace{-2mm}    
%     \includegraphics[width=.8\linewidth]{figures/emulation_overall_barplot.pdf}
%     \vspace{-12pt}
%     \Description{}
%     \caption{Overall streaming performance in emulation.} %Error bars span ± one standard deviation from the average.
%     \vspace{-6mm}
%     \label{f:emulation_overal}
% \end{figure}
%%%%%%%%%%%%%%%%%%%%%%%%%%%%%%%%%%%% 

\noindent\textbf{Overall Performance:} 
Fig. \ref{f:emulation_overal} shows the streaming performance across all emulated 5G networks. 
Results show that NeuralMDC effectively improves streaming QoE regardless of the path mapping algorithm used. 
Specifically, 
NeuralMDC achieves 26\% and 44\% higher average QoE compared to H.265 and GRACE, respectively, delivering higher (1.0\%-41.8\%) visual quality while dramatically reducing rebuffering (cutting stall time by 99.4\% relative to H.265, while maintaining near-zero stalls comparable to GRACE). \diff{These gains come from both improved compression efficiency and NeuralMDC’s chunk-level description design: each description spans the full chunk, whereas GRACE remains sensitive to frame-level packet scheduling under bursty path under-delivery.}
Compared to streaming using SHVC and Swift, NeuralMDC improves average QoE by 34\% and 39.5\%, with 14.1\%-28.1\% higher \diff{video} quality and 34\%-92\% lower stall time.

%Compared to streaming with H.265 and GRACE, NeuralMDC achieves 26\% and 44\% higher average QoE, respectively. In terms of underlying QoE components, NeuralMDC delivers similar or higher (1.0\%-41.8\%) visual quality and dramatically reduces rebuffering (cutting stall time by 99.4\% relative to H.265, while maintaining near-zero stalls comparable to GRACE). 
%Compared to streaming using SHVC and Swift, NeuralMDC improves the average QoE by 34\% and 39.5\% respectively, and streaming using NeuralMDC delivers 14.1\%-28.1\% higher quality videos as well as reduces the average stall time by 34\%-92\%. 

\noindent\textbf{Impact of Channel Stability:} Fig \ref{f:emulation_path_stability} shows streaming performance across different 5G multipath scenarios.
Under two 5G channels %with high throughput but wild fluctuations,
 (Fig.~\ref{f:emulation_5g_5g}), streaming with NeuralMDC provides the lowest stall time, reducing stalls by 90.56\% compared to SHVC. While no single channel consistently outperforms the other, rapid 5G fluctuations make it difficult to accurately map SHVC layers to proper channels; incorrectly dispatching the base layer to a deteriorating channel leads to high stalls for SHVC, with worse impacts on Swift, due to its larger base layer, and H.265, which requires full reception for decoding. %GRACE has higher stalls than MDC-Pin as burst losses leave some frames with zero received packets. 
%Replacing one 5G channel with a more stable 4G channel (Fig.~\ref{f:emulation_5g_4g}) reduces SHVC stalls but lowers overall quality due to 4G's lower throughput.
%In Fig. \ref{f:emulation_5g_5g}, when using two 5G channels with high throughput but wild fluctuations, NeuralMDC provides the best streaming performance with the lowest stall time, reducing stalls by 90.56\% compared to SHVC. 
%While no single channel consistently outperforms the other, the rapid fluctuations of 5G throughput make it difficult to accurately predict future channel conditions and map SHVC layers to proper channels. Incorrectly dispatching the base layer to a deteriorating channel leads to high stall times for SHVC, with even more severe impacts on Swift, due to its larger base layer, and H.265, which requires full reception for decoding. 
Replacing one 5G channel with a 4G channel (Fig.~\ref{f:emulation_5g_4g}) %helps reduce the stall time of streaming with SHVC by mapping the base layer to 4G for more reliable delivery. However, the lower throughput of 4G decreases the overall aggregated bandwidth, leading to lower video quality.
 reduces stalls across all schemes, as the
  reliable 4G path anchors delivery; however, 4G's lower throughput reduces the overall aggregated bandwidth, leading to lower video quality.      
  MDC-Pin achieves near-zero stalls while maintaining the highest quality, and GRACE similarly achieves low stalls though at lower quality
  due to its weaker compression efficiency.

%5G + 5G (unstable + unsable, large stall, higher quality)
%5G + 4G(unstable + stable, small stall, lower quality)

%%%%%%%%%%%%%%%%%pin-strip  %%%%%%%%%%%%%%%%%%% 
% \begin{figure}[!htp]
%     \vspace{-4mm}
%     \centering
%     \captionsetup[subfigure]{skip=1pt} % Minimize space between subfigure and caption
%     \setlength{\abovecaptionskip}{1pt} % Reduce space above caption
%     \setlength{\belowcaptionskip}{1pt} % Reduce space below caption
%     \subfigure{
%         \includegraphics[width=0.22\textwidth]{plots/sequential_policy.pdf}
%         \label{fig:sequential_policy}
%     }
%     \hfill
%     \subfigure{
%         \includegraphics[width=0.22\textwidth]{plots/parallel_policy.pdf}
%         \label{fig:parallel_policy}
%     }
%     \Description{}
%     \caption{Streaming performance using path-striping and path-pinning.}
%     \label{fig:overall_comparison}
% \end{figure}

%%%%%%%%%%%%%%%%%%%%%%%%%%%%%%%%%%%% 
\begin{figure}[t]
    \centering
     \includegraphics[width=.44\columnwidth]{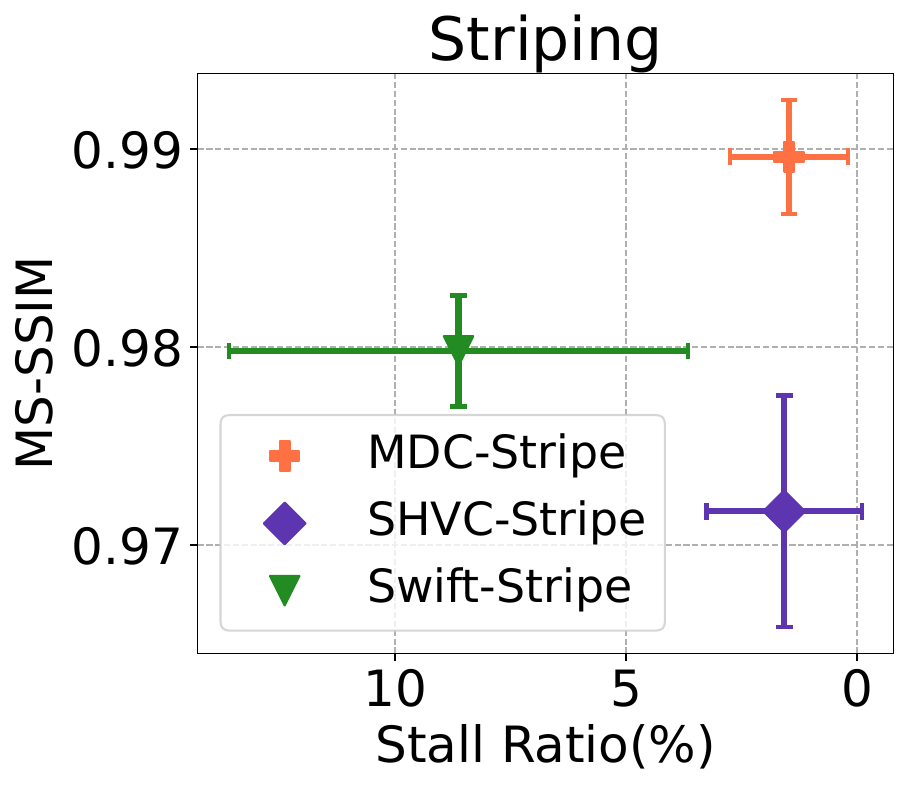}\label{fig:sequential_policy}
    \includegraphics[width=.44\columnwidth]{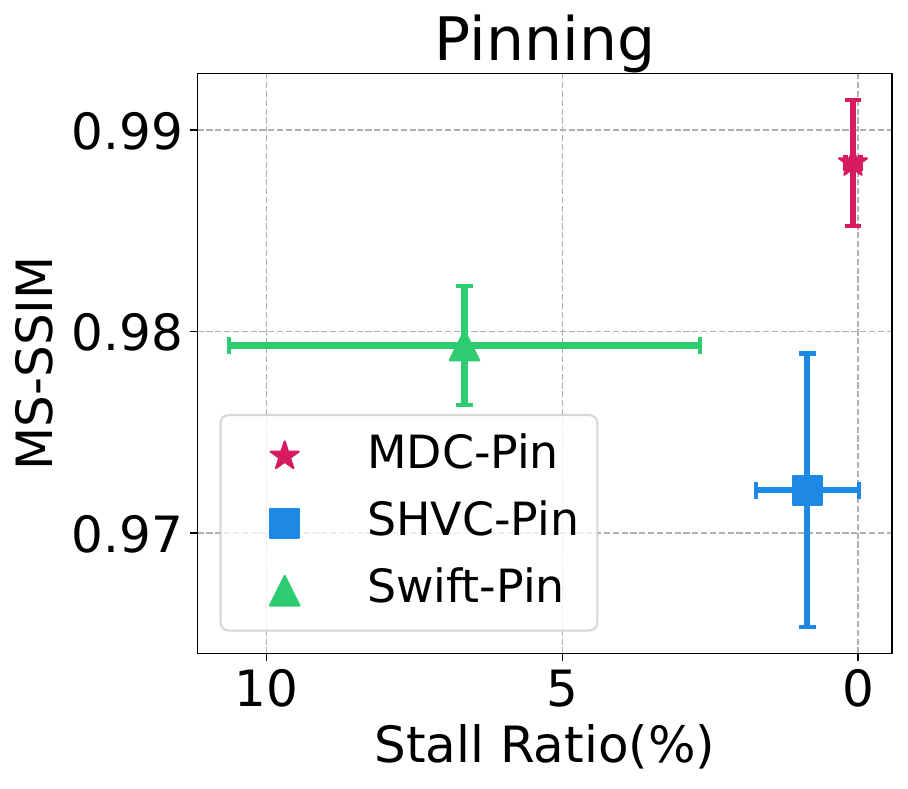}\label{fig:parallel_policy}
    \vspace{-3mm}
    \caption{Streaming performance: striping vs. pinning.}
    \vspace{-4mm}
    \label{fig:overall_comparison}
\end{figure}
\noindent\textbf{Path Pinning vs Path Striping:}
Fig. \ref{fig:overall_comparison} shows that path-pinning improves stall performance for all codecs, but gains depend on codec design. 
%Under striping, NeuralMDC  and SHVC achieve similarly low stall ratios, while Swift suffers from noticeably higher stalls due to its large base layer. Switching to pinning,  
%NeuralMDC  benefits most because its independent, fine-grained streams can be flexibly matched to heterogeneous paths, while SHVC and Swift remain limited by layer dependencies, size mismatch, and Swift’s large base layer.
NeuralMDC benefits the most, as its independent, fine-grained streams can be flexibly matched to paths of varying bandwidth without dependency constraints. SHVC improves more modestly, since layer dependencies and size mismatches still cause occasional stalls. Swift improves the least, as its disproportionately large base layer remains a bottleneck.  
Pinning, however, slightly reduces quality due to imperfect stream–path capacity matching, leaving some bandwidth underutilized.  Overall, the benefit of pinning depends on stream independence and flexible stream-to-path matching. NeuralMDC also scales well with more paths (see Fig. \ref{fig:increased_number_of_path}).  

\subsection{Real-World Evaluation}
\label{s:real_world_evaluation}

We evaluate video streaming with NeuralMDC over real-world 5G networks, namely AT\&T, T-Mobile, and Verizon, randomly shuffled and denoted as \tmb, \vzw, and \att.

\noindent
\textbf{Experimental Setup.} \diff{We evaluate streaming schemes under stationary, walking, and driving mobility patterns, and conducted tests using: 1)~different bands from the same operator and 2)~different operators. In the multiple-operator scenario, we evaluate streaming schemes exclusively under driving mobility, reflecting autonomous vehicles that rely on multiple operators.} Despite the non-repeatability of real-world experiments, we test seven streaming schemes 12 times per scenario under identical conditions.  Our evaluation includes 336  mobile streaming sessions.

\begin{figure}[t]
    \centering
    \subfigure[Average Performance]{\includegraphics[width=.63\columnwidth]{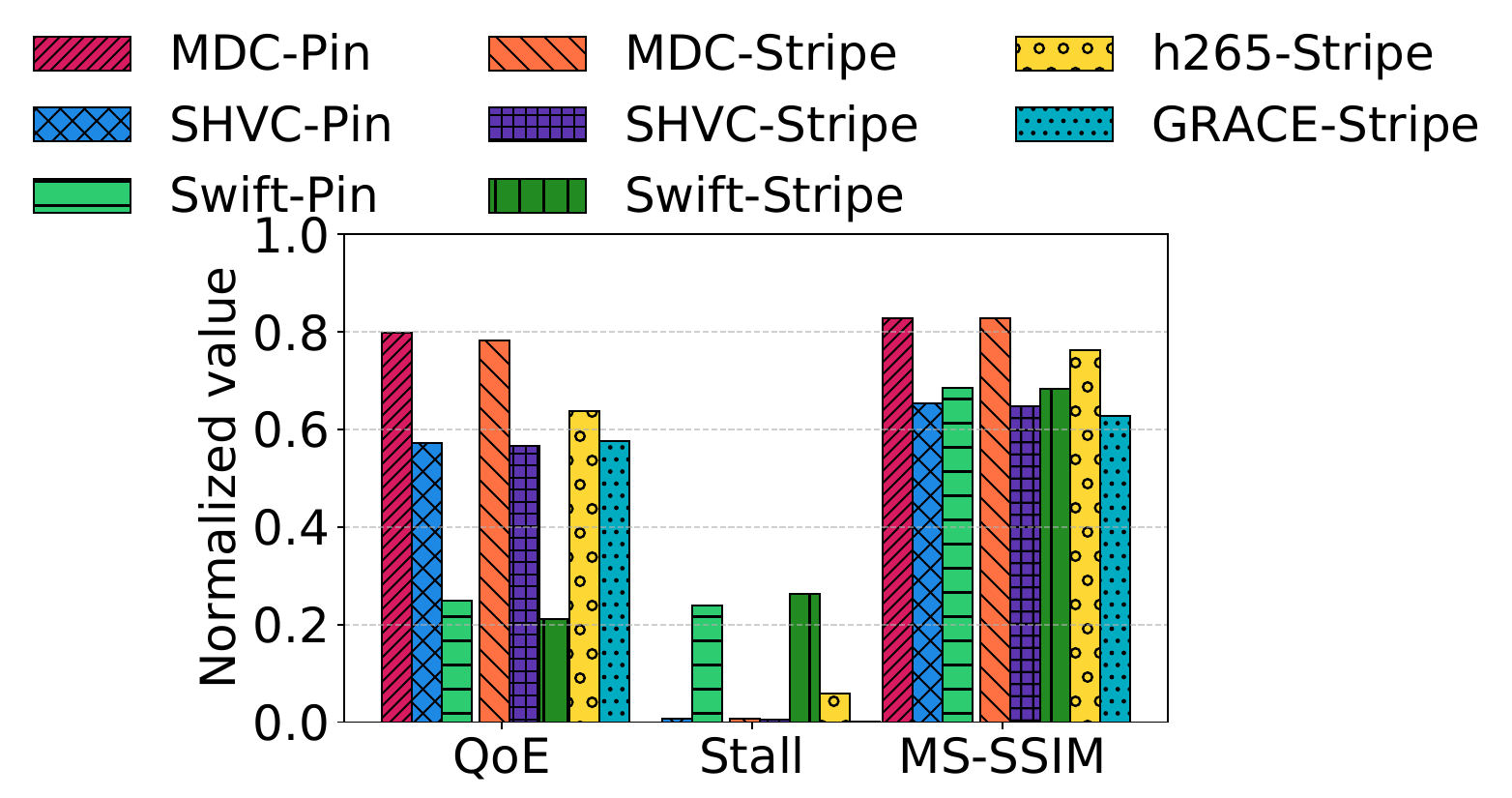}\label{f:real_world_average}}
    \hspace{-3.5mm}
    \subfigure[Tail Performance]{\includegraphics[width=.36\columnwidth]{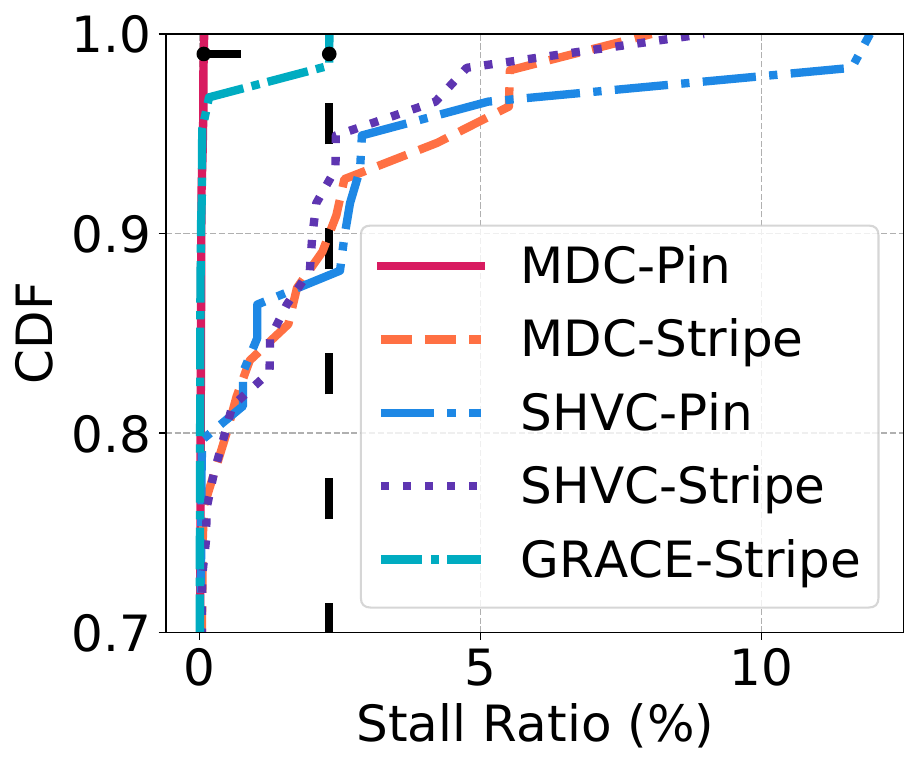}\label{f:real_world_tail}}
    \vspace{-4mm}
    \caption{Real-world streaming performance.}
    \vspace{-6mm}
    \label{f:real_world_overal}
\end{figure}

\noindent
\textbf{Average \& Tail Performance}
% db: −10 log(1 − SSIM)
Fig. \ref{f:real_world_overal} presents the overall real-world streaming performance. The performance is consistent with the emulation results.  
Fig.~\ref{f:real_world_average} shows the average performance. Video streaming with NeuralMDC achieves lower stall time, higher visual quality and a 25.0\% QoE improvement over the best existing scheme. 
In comparison, H.265 attains similar visual quality but incurs significantly more \diff{stalls}. GRACE exhibits minimal stalls, yet delivers much lower visual quality. 
SHVC, configured with a smaller base layer,  significantly reduces stalls compared to Swift, which features a large base layer. 
%Finally, compared to Swift, which features a large base layer, SHVC, configured with a smaller base layer, significantly reduces the average stall ratio. %from 18.2\% to 0.5\%. 
Fig.~\ref{f:real_world_tail} shows the stall ratio distributions of SHVC, GRACE, and NeuralMDC. 
%Streaming with NeuralMDC and GRACE exhibits near-zero average stall time. However, 
NeuralMDC achieves better tail performance  (96.6\% reduction of stall ratio at 99-percentile) than GRACE. This is because streaming with NeuralMDC preserves chunk-level decodability from any subset of streams, whereas GRACE requires receiving packets for each frame within a chunk and is therefore more sensitive to packet scheduling and burst losses. \shepherd{Note that real-world performance is generally lower than emulation because operational deployments introduce additional variability not captured by trace replay, including radio scheduling changes, device thermal behavior, background traffic, and carrier-side policies.}

\begin{figure*}[t]
    \centering
    \subfigure{\includegraphics[width=.24\textwidth]{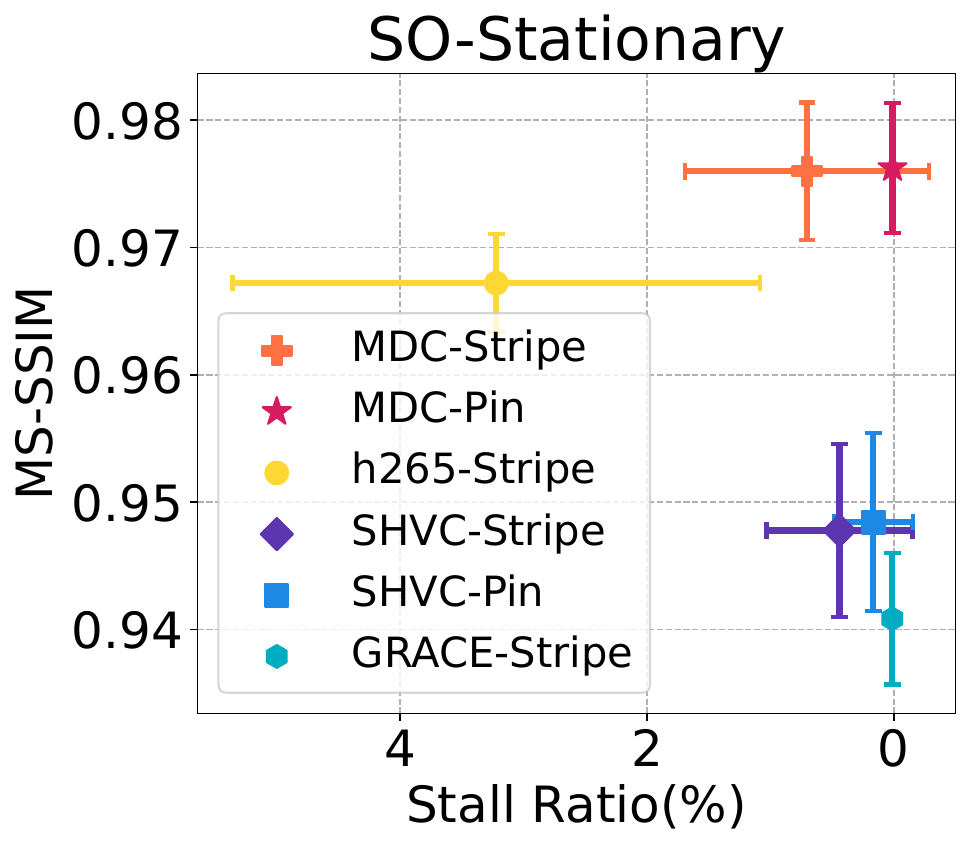}}
    \subfigure{\includegraphics[width=.24\textwidth]{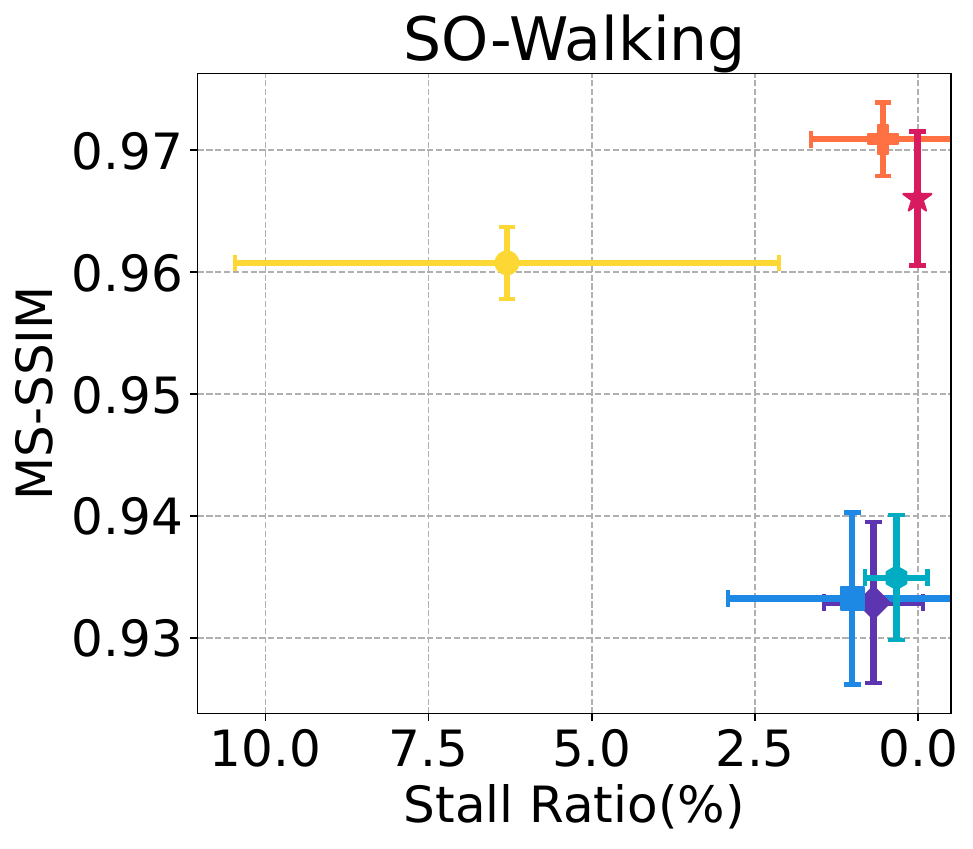}}
    \subfigure{\includegraphics[width=.24\textwidth]{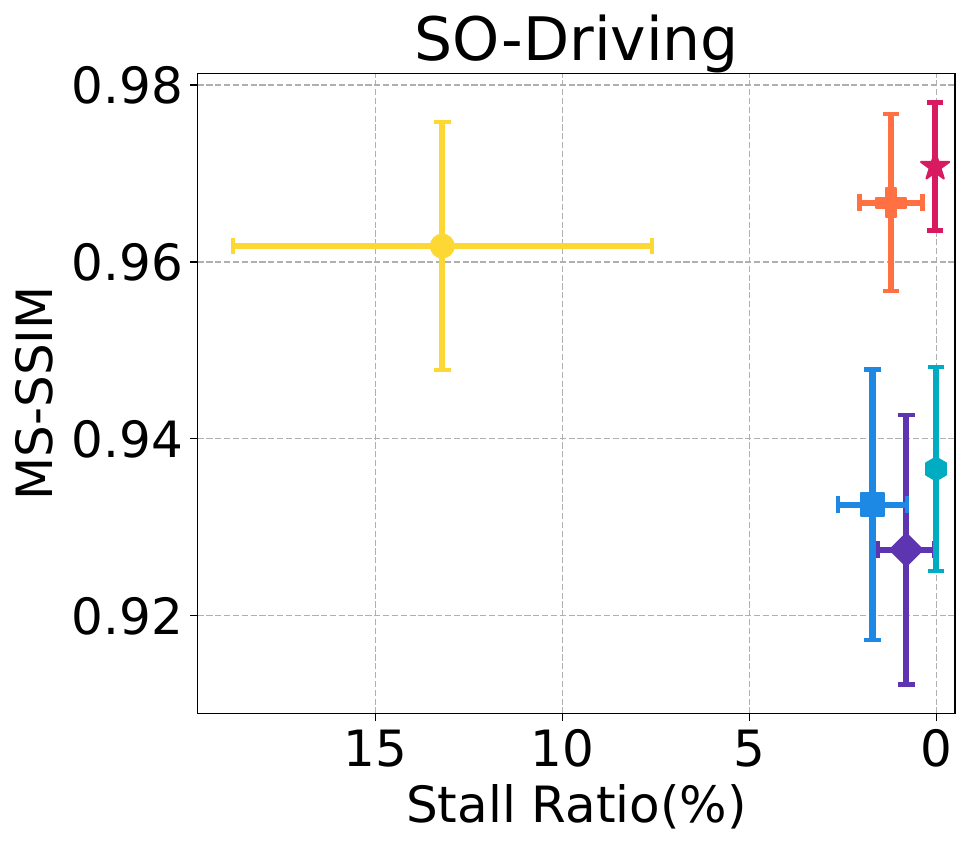}}
    \subfigure{\includegraphics[width=.24\textwidth]{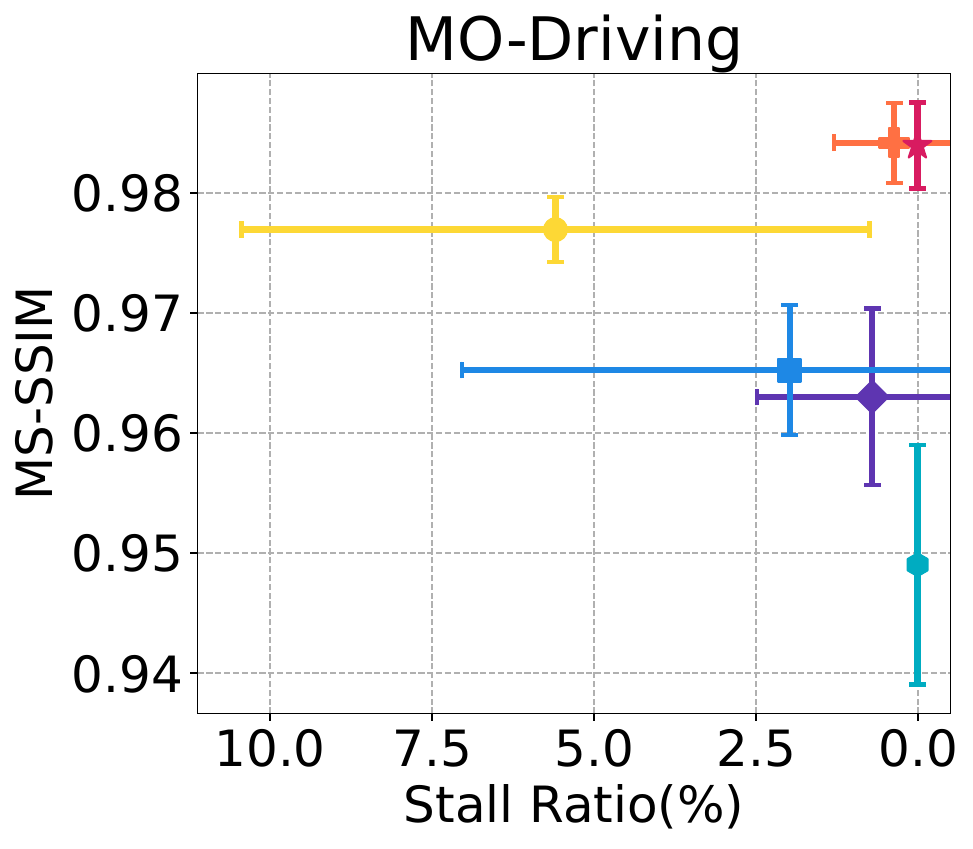}}
    \vspace{-4mm}
    \caption{Performance in the 4 real-world 5G network \diff{scenarios}. Error bars show 95\% confidence intervals.}
    \vspace{-8mm}
    \label{f:real_world_scenario}
\end{figure*}

\noindent
% \textbf{Performance Across Different Mobility Conditions} 
\textbf{Stationary vs. Walking vs. Driving.} 
% driving:
% MDC-Stripe are 0.016661, 0.958190
% MDC-Pin are 0.000720, 0.966863
% h265-Stripe are 0.216417, 0.944118
Fig~\ref{f:real_world_scenario} shows single-operator (SO) performance across mobility conditions.  As mobility increases, visual quality degrades, and stall ratios rise.  However, NeuralMDC with path-pinning consistently achieves the best performance, maintaining the highest visual quality and keeping  stall ratio below 0.32\%. 
H.265 stalls escalate sharply with mobility, as frequent throughput drops prevent timely full-chunk delivery. SHVC stalls also increase with mobility, and neither pinning nor striping dominates across all conditions, highlighting the challenge of designing effective path mapping for layered codecs under varying dynamics. GRACE maintains low stalls in static (0.01\%) but degrades under driving (0.76\%), as burst losses increasingly leave frames with no received packets.    

\noindent
\textbf{Single Operator vs. Multiple Operators } Fig.~\ref{f:real_world_scenario} \diff{shows} MO-Driving and SO-Driving \diff{performance}\shepherd{. We} find that 
%We compare the multipath streaming performance in driving mobility using a single operator and multiple operators (MO) in Fig.~\ref{f:real_world_scenario}. 
although using multiple independent operators improves aggregated network throughput and reliability, which enables higher-quality video streaming, the throughput dynamics of each operator still make bitrate adaptation challenging for H.265 and proper layer-to-path allocation difficult for SHVC, leading to video stalls. %GRACE achieves near-zero stalls though at lower quality. 
In contrast, NeuralMDC’s fine-grained streams and dependency-free decoding effectively leverage the benefits of multiple operators while adapting to throughput fluctuations, resulting in higher video quality and nearly zero stall time. % real-world streaming performance
\section{Codec Properties for Multipath Streaming} 
\label{sec:codec_performance}
%\todo{Eurosys reviewer: \\
%1. Lack of comparison between the codec and traditional system on quality. compared your quality vs H264 or H265.\\
%2. Does encoder-decoder framework lack generalization for different datasets? Or the solution is independent of the dataset it is trained? What kind of impact will it have if this solution is adopted in reality?
%}

This section evaluates the codec-level properties of NeuralMDC that underpin its streaming performance: independent decodability, mutual refinability, compression efficiency, and runtime overhead. We use the same test videos and baseline codecs as in \S\ref{s:setup}. In addition, we compare with VCT~\cite{vct}, a transformer-based neural \hl{monolithic} codec\footnote{\hl{VCT lacks public training code and pretrained checkpoints; we report only its published rate–distortion results.
%VCT’s public release does not include training code or pretrained model checkpoints, preventing partial stream evaluation and streaming evaluation. We report its rate-distortion performance using its released numbers.
}}, and H.264 via FFmpeg~\cite{ffmpeg}. \label{s:codec_setup} We use bits per pixel (\diff{bpp}) to measure video size after compression and Peak Signal to Noise Ratio (PSNR) and MS-SSIM~\cite{wang2003multiscale} for visual quality.

\begin{figure}[t]
    \centering
    \subfigure{\includegraphics[width=.45\columnwidth]{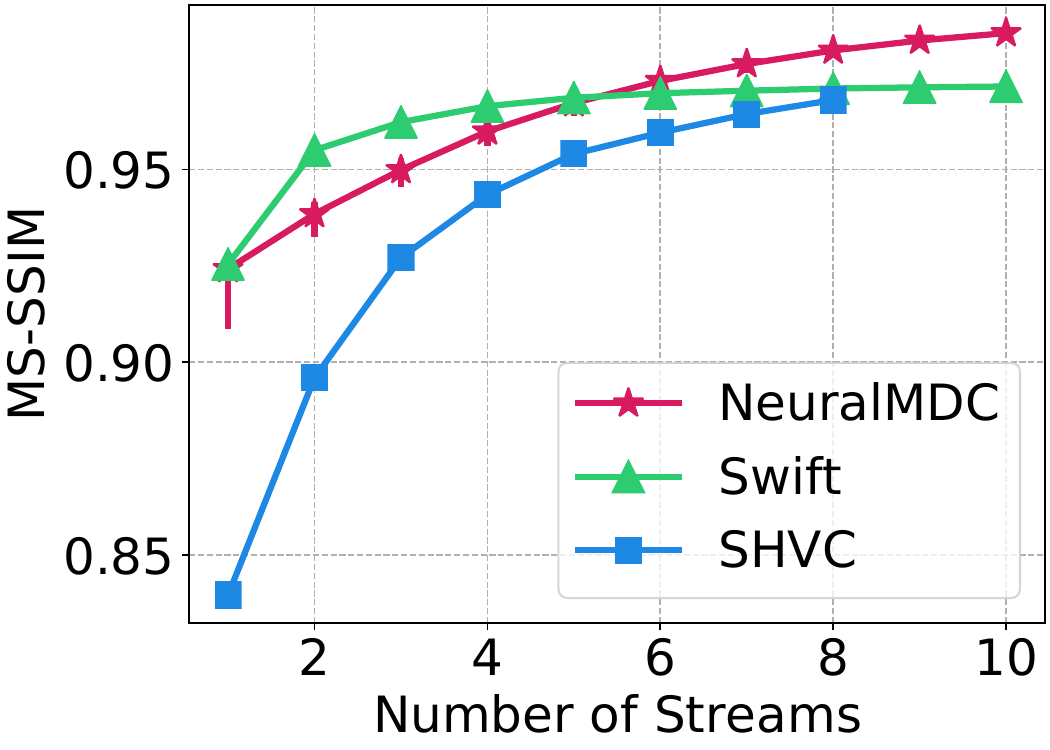}}
    \subfigure{\includegraphics[width=.45\columnwidth]{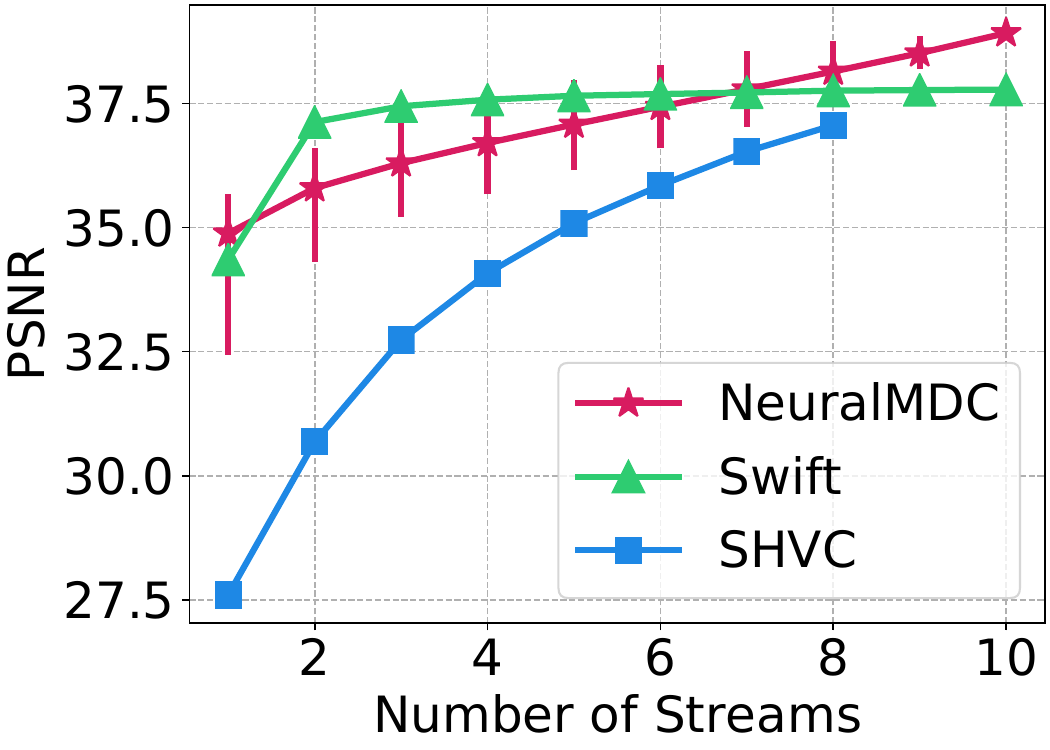}}
    \vspace{-4mm}
    \caption{Impact of stream number on video quality. }%For each codec, the total bpp of all streams is similar.}
    \vspace{-6mm}
    \label{reconquality}
\end{figure}

% \begin{figure}[htbp]
% \centering
% \begin{minipage}{0.49\linewidth}
%     \vspace{-2mm}
%     %\subfigure{\includegraphics[width=.48\linewidth]{figures/stream_msssim.pdf}}
%     \subfigure{\includegraphics[width=.49\linewidth]{figures/ssim_layer.pdf}}
%     % \subfigure{\includegraphics[width=.455\linewidth]{figures/stream_psnr.pdf}}
% \subfigure{\includegraphics[width=.49\linewidth]{figures/psnr_layer.pdf}}
%     \vspace{-4mm}
%     % \Description{}
%     \caption{ Impact of stream number on video quality. For each codec, the total bpp of all streams is similar.}
%     \vspace{-4mm}
%     \label{reconquality}
% \end{minipage}
% \hfill
% \begin{minipage}{0.49\linewidth}  
%     \vspace{-2mm}
%     \subfigure{\includegraphics[width=.49\linewidth]{figures/psnr_partial_stream.pdf}}
%     \subfigure{\includegraphics[width=.49\linewidth]{figures/MS-SSIM_partial_stream.pdf}}
%     \vspace{-4mm}
%     \caption{Impact of partial stream on video quality. We set both codecs to have similar bpp.}
%     \vspace{-4mm}
%     \label{f:stream_loss}
%     \end{minipage}
% \end{figure}
%%%%%%%%%%%%% Impact of stream number %%%%%%%%%%

\subsection{Impact of Stream Number on Video Quality }\label{sec:codec-evaluation}
%Goals: 1) show any combination of stream can work. 2) better video quality. 

Both NeuralMDC and layered codecs adapt bitrate on the fly by adjusting the number of streams or layers. 
We compress videos into 10 streams/layers\footnote{We set stream number to 10 to match the maximum layers Swift can generate.} at comparable total bpp. 
 The maximum number of layers SHVC can generate is 8.
%The 10 MDC streams are generated using the pyramid source information splitting method. 
%The 10 Swift layers are generated through a chain of AutoEncoders on frame residuals. %encoding and decoding operations on frame residuals. 
SHVC layers are configured with quantization parameters to achieve a pyramid structure similar to NeuralMDC.
%The 8 SHVC layers are created by adjusting layer quantization parameters to achieve a similar pyramid structure as NeuralMDC. 

Fig. \ref{reconquality} shows video quality across different numbers of streams. Since each MDC stream can be combined with any other, we plot the minimum, average, and maximum qualities of all combinations (see red error bars in Fig. \ref{reconquality}). %distribution (\ie minimum, average, and maximum) of all stream combinations. 
%For layered codecs, video quality is deterministic due to strict layer dependencies. 
The results show that video quality improves for all codecs as the number of streams/layers increases. However, Swift exhibits a different pattern where higher layers contribute less to quality improvement. This is because the entropy (\ie video information) is very high in Swift’s initial layers and decreases in later layers, resulting in progressively smaller layer sizes (see %\S 3.2 in~\cite{swift} and 
Fig. \ref{f:stream_size} in \S Appendix \ref{a:codec_benchmark} ). For NeuralMDC, video quality of a single stream depends on which stream is selected.  
%The smallest stream contains the least information, resulting in lower quality, while 
%The largest stream contains the most information, yielding highest quality. %\ali{Can we update Swift to have this feature as well?}
%As the number of streams increases, the variance in possible video quality narrows. Since the total bpp of all streams/layers is similar, NeuralMDC achieves higher MS-SSIM and PSNR than Swift and SHVC, indicating superior rate-distortion performance. 
As more streams are added, the quality variance narrows.  Overall, NeuralMDC achieves higher MS-SSIM and PSNR than Swift and SHVC at similar total bpp, indicating superior rate-distortion performance.                     

%Besides, unlike layered codecs that require full layer reception for decoding, NeuralMDC allows partial streams to be decoded and used in combination with other streams without dependency constraints (see Appendix \ref{a:partial_stream}). Reconstructed frame samples from NeuralMDC with 50\% token reception are provided in \fig \ref{f:reconstruct} in the Appendix \ref{a:reconstruction_sample}.

%Note that  a partial stream can be decoded and used in combi- nation with other streams.

% \begin{figure}[!ht]
%     \centering
%     \vspace{-6pt}
%     \subfloat{\includegraphics[width=.48\linewidth]{figures/UVG_MCL_psnr.pdf}}
%     \subfloat{\includegraphics[width=.48\linewidth]{figures/UVG_MCL_msssim.pdf}}
%     \vspace{-12pt}
%     \Description{}
%     \caption{Compression efficiency}
%     \vspace{-12pt}
%     \label{f:rate-distortion}
% \end{figure}

\begin{figure}[t]
\centering
    \subfigure{\includegraphics[width=.45\columnwidth]{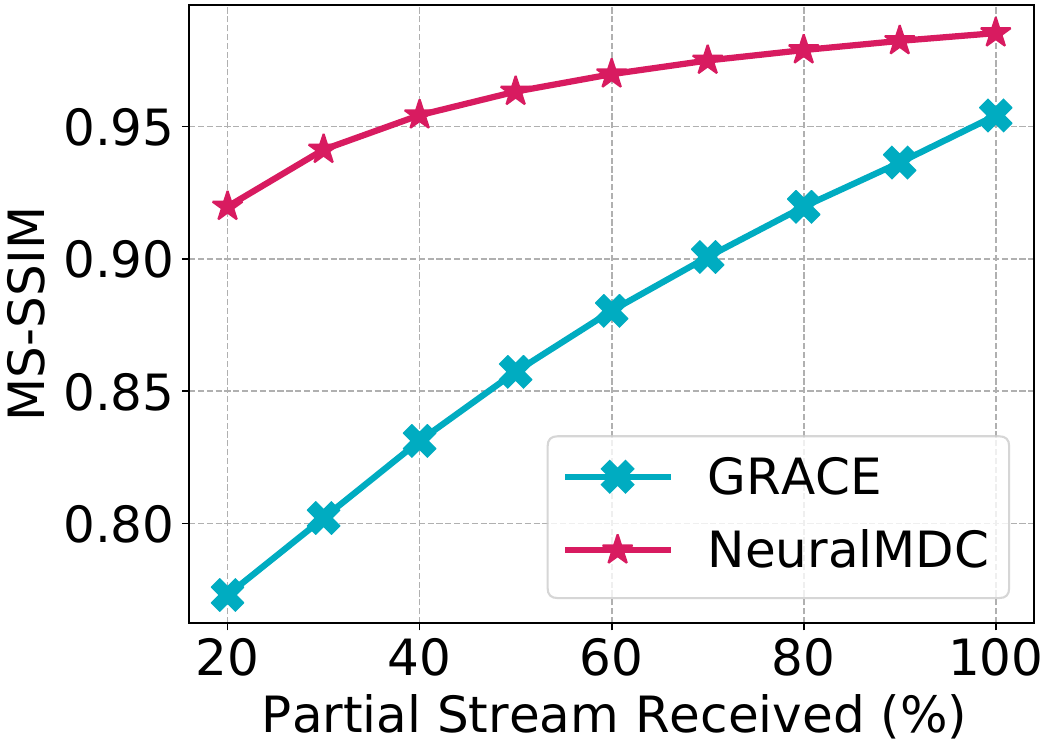}} 
    \subfigure{\includegraphics[width=.45\columnwidth]{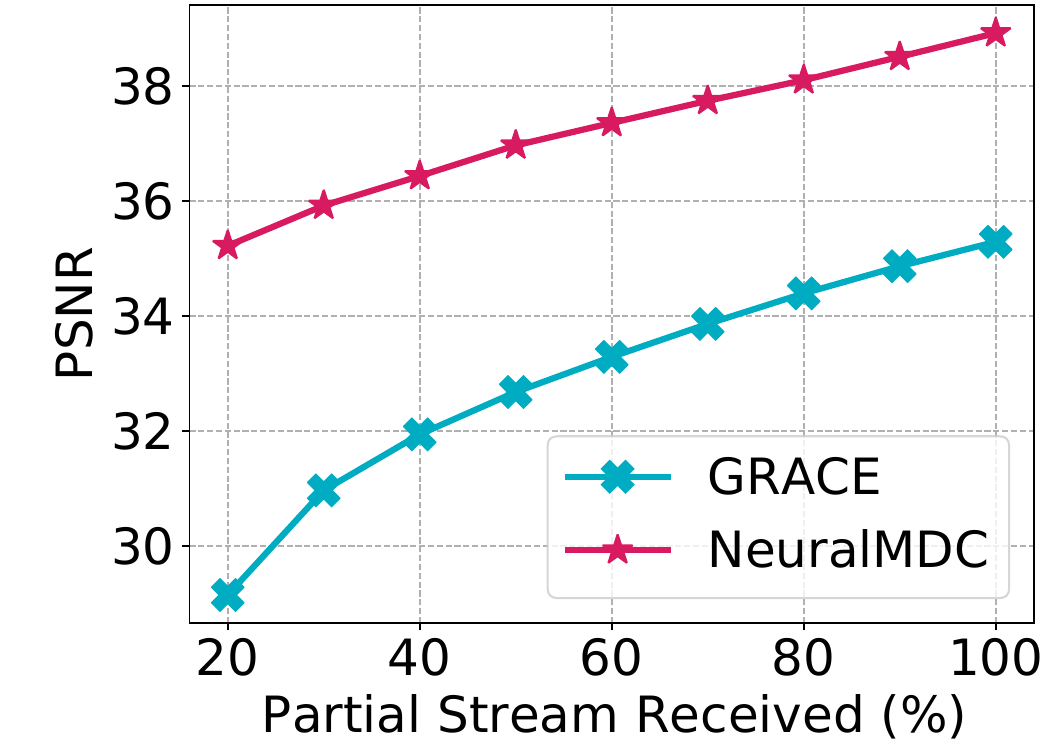}}
    \vspace{-4mm}
    \caption{Impact of partial stream on video quality.}% We set both codecs to have similar bpp.}
    \vspace{-6mm}
    \label{f:stream_loss}
\end{figure}

% Compression efficiency figure moved to combined figure* below with microbenchmarks

\begin{figure*}[t]
    \centering
    \begin{minipage}{0.4\textwidth}
         \subfigure{\includegraphics[width=.49\linewidth]{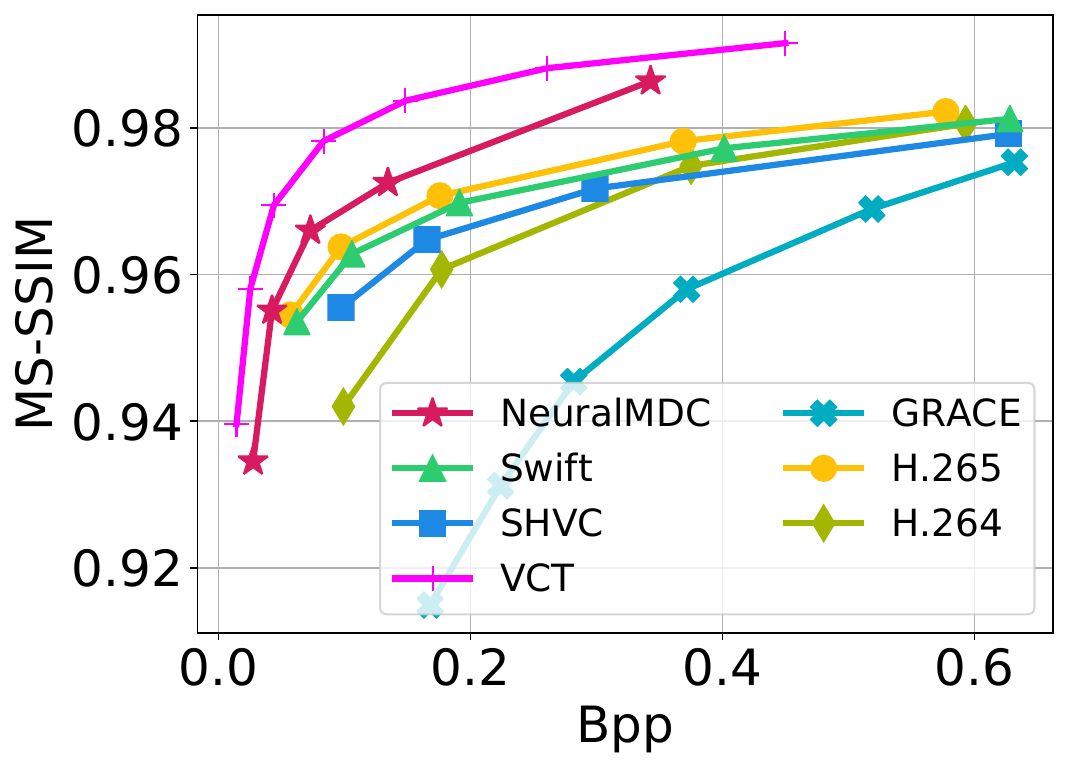}}%
        \subfigure{\includegraphics[width=.49\linewidth]{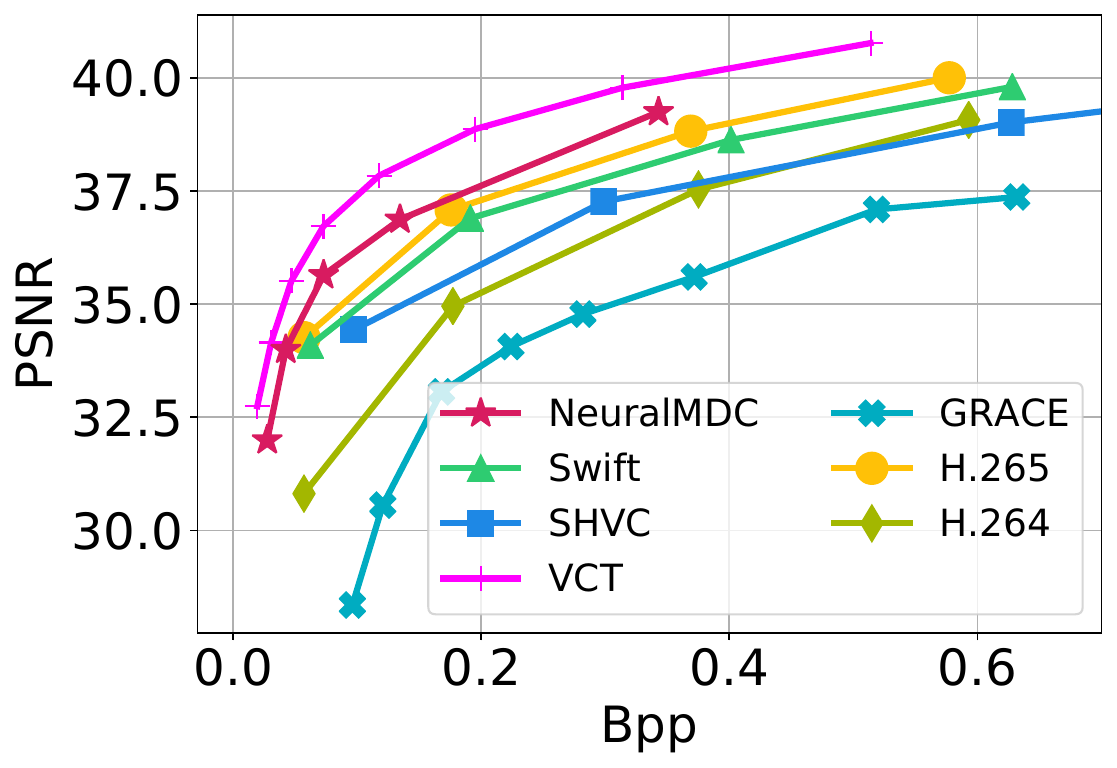}}
        \vspace{-4mm}
        \caption{Compression efficiency.}
        % \vspace{-4mm}
        \label{f:rate-distortion}
    \end{minipage}
    \begin{minipage}{0.22\textwidth}
        \centering
        \includegraphics[width=\linewidth]{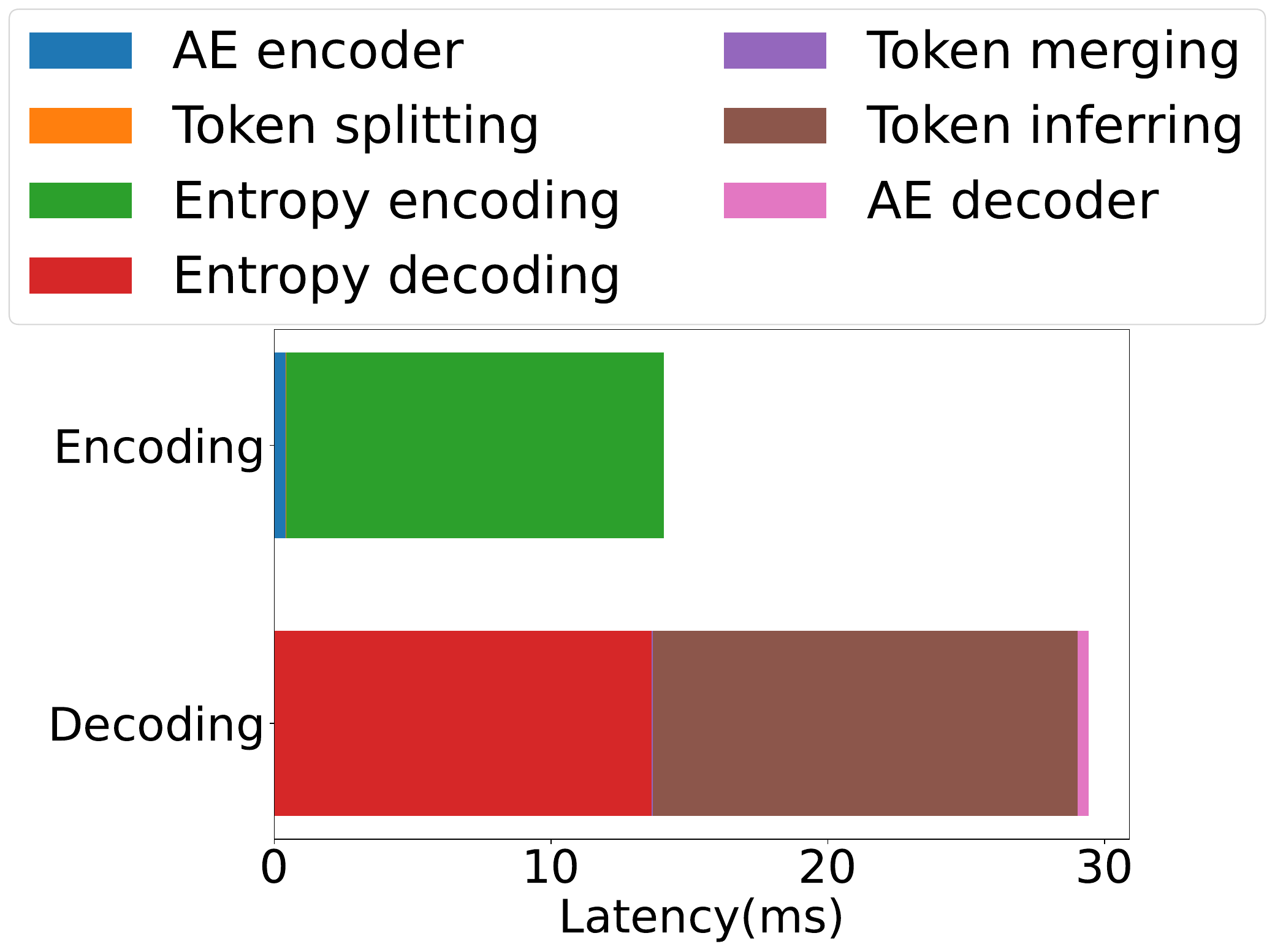}
        \vspace{-6mm}
        \caption{Runtime for a 1080p video frame.}
        \vspace{-4mm}
        \label{f:runtime_infer}
    \end{minipage}
    \begin{minipage}{0.36\textwidth}
    \centering
    \includegraphics[width=.48\columnwidth]{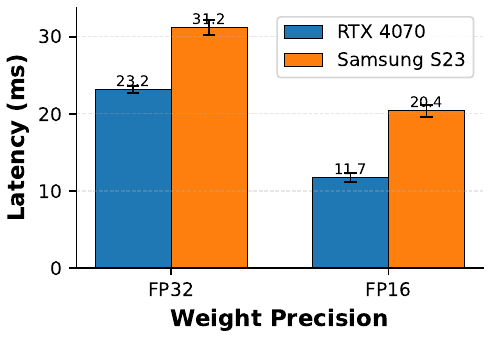}\label{fig:latency_benchmark}
    \includegraphics[width=.48\columnwidth]{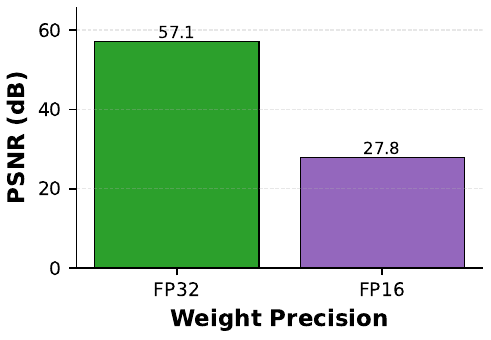}\label{fig:model_size_benchmark}
    \caption{Performance Optimization Comparison.}
    \label{fig:deployment_performance}
    \end{minipage}
    \vspace{-2mm}
\end{figure*}

\subsection{Impact of Partial Stream on Video Quality} \label{a:partial_stream}

%\todo{Eurosys reviewer: \\
%When you are comparing against Grace, did you set the arrival probability of each stream of MDC to be 20\% to 100\%? (I don't understand your description in that part.)}\\
%Goals: 1) portion/part of a stream is still decodable. 2) better loss resilient than Grace. 
%During video streaming over lossy 5G networks, part of a stream may be received before its playback time. However, each layer in layered codecs must be fully received to be decodable, as a partial layer would disrupt the dependency structure and result in unusable data. In contrast, since NeuralMDC compresses tokens within a stream independently and there is no dependency structure among tokens, a partial stream can be decoded and combined with other streams.

In layered codecs, each layer must be fully received to be decodable, as a partial layer disrupts the dependency structure and \diff{results} in unusable data. In contrast, NeuralMDC compresses tokens within a stream independently and there is no dependency structure among tokens, so a partial stream can be decoded and combined with other streams.

To evaluate the video quality of partial streams,  we compare NeuralMDC with GRACE~\cite{cheng2024grace}, a loss-resilient single-stream neural codec. We set both codecs to have similar bpp and configure streams to be partially received (from 20\% to 100\%).  
Fig. \ref{f:stream_loss} shows that NeuralMDC achieves better video quality than GRACE  at the same reception percentage. This superior performance stems from NeuralMDC's equally important and correlated tokens, whereas in GRACE, motion  and residual vectors carry different levels of information and are weakly correlated, making it more difficult to efficiently estimate the lost one. Reconstructed frames by NeuralMDC with partial reception are shown in \fig \ref{f:reconstruct} in Appendix \ref{a:reconstruction_sample}. %\todo{grace reconstruction samples}

% \begin{figure}[ht]
%     \centering
%     \vspace{-2mm}
%     %     \subfigure{\includegraphics[width=.49\linewidth]{figures/UVG_MCL_psnr-vct.pdf}}
%     % \subfigure{\includegraphics[width=.49\linewidth]{figures/UVG_MCL_msssim-vct.pdf}}
%      \subfigure{\includegraphics[width=.3\linewidth]{figures/UVG_MCL_psnr.pdf}}
%     \subfigure{\includegraphics[width=.31\linewidth]{figures/UVG_MCL_msssim.pdf}}
%     \Description{}
%     \vspace{-4mm}
%     \caption{ Compression efficiency}
%     \vspace{-6mm}
%     \label{f:rate-distortion}
% \end{figure}

% \vspace{-0.2in}
\subsection{Compression Efficiency}\label{sec:codec-compression}

Fig. \ref{f:rate-distortion}  shows the rate-distortion performance of all codecs. 
We compare the visual quality at different compression rates in terms of bits per pixel. 
%For NeuralMDC and Swift, 10 streams/layers are used; for SHVC, 8 layers. 
NeuralMDC improves compression efficiency by 13\% over H.265 and 27\% over Swift. 
The superior performance of NeuralMDC stems from its context-based conditional coding framework, which leverages transformers to capture richer inter-frame contexts than motion prediction and warping~\cite{vct, li2021deep}. 
%richer contexts between frames to improve compression efficiency. Unlike motion prediction and warping operations, which rely on simple additive and subtractive relationships and introduce architectural biases, transformers identify arbitrary dependencies between frames for more effective compression~\cite{vct, li2021deep}. 
However, compared to VCT,  NeuralMDC incurs reduced compression efficiency because splitting  source information into multiple descriptions lowers intra-description correlation. 
\hl{GRACE’s compression efficiency is content-dependent; it underperforms H.264 on videos with high spatial and temporal complexity~\cite{cheng2024grace}, and our test videos exhibit high SI and TI (see Table \ref{tab:video_si_ti} in Appendix \ref{a:si_ti}).} 

%For NeuralMDC and Swift, a total of 10 streams/layers are used to generate the plots, with each point representing the combined performance of all streams/layers. For SHVC, 8 layers are used. %For H.264 and H.265, each point corresponds to the independent compression performance at the respective compression rate. 
%arbitrary relationships between frames while avoiding the limitations of handcrafted architectural biases and priors, such as motion prediction and warping operations (refer to previous work~\cite{vct, li2021deep} on condition coding-based codecs. ).
%\ali{Is there a way to show higher contextual information leads to better compression rate?}

\subsection{Microbenchmarking}
\label{s:codec_benchmark}

\noindent\textbf{Latency Breakdown.} Fig. \ref{f:runtime_infer} %presents a detailed time cost breakdown  of NeuralMDC. It 
shows that splitting and merging descriptions incur negligible time costs. %, indicating that latency is independent of the number of streams.  
The time costs of AutoEncoder are amortized across a batch of frames. The main source of latency is the Masked Transformer. % for entropy coding and inferring missing streams. 
With the optimizations in \S \ref{s:runtime_opt}, 
NeuralMDC achieves 70 fps encoding and 34 fps decoding for 1080p on an Nvidia A6000 GPU.
          
%NeuralMDC achieves encoding of a 1080p frame in 14.1 ms (70 fps) and decoding in 29.4 ms (34 fps) on an Nvidia A6000 GPU. Additional runtime benchmarks on a desktop GPU (NVIDIA RTX 4070) and a smartphone (Snapdragon 8 Gen 2) are reported in  Appendix \ref{a:deployability}. %\todo{move Appendix D here?}

\noindent\textbf{Deployability.} We further benchmark NeuralMDC with two weight precisions (FP32, FP16) on a high-end desktop GPU (NVIDIA RTX 4070) and a resource-constrained mobile phone (Samsung S23, Snapdragon 8 Gen 2). \fig \ref{fig:deployment_performance} shows that the quantization optimizations are crucial for enabling real-time video processing across this hardware spectrum. For one $224\times224$ resolution video frame, on RTX 4070, FP16 reduces per-frame latency from 23.2ms to 11.7ms ($2.0\times$ speedup). On mobile NPU, latencies are 31.2ms (FP32) and 20.4ms (FP16). FP16 also halves model size (334.8MB $\rightarrow$ 167.5MB), \diff{suggesting feasibility on accelerator-equipped mobile platforms.}
\section{Related Work}\label{sec:related}

\noindent
\textbf{Multipath Video Streaming:}
Studies~\cite{ye2024dissecting,rochman2024comprehensive,fezeu2023mid,chen2025large} show that mobile networks are inherently heterogeneous, spanning various frequency bands and operator-specific infrastructures. 
%This drives research into resilient wireless connectivity solutions~\cite{CarrierAggregation_Wikipedia, DualConnectivity_EverythingRF, hassan2024case,cellfusion-sigcomm23}. 
Multi-path video streaming has emerged as a crucial technique for enhancing video transmission quality. Multipath techniques, such as MPTCP~\cite{MPTCP_Wikipedia}, MPQUIC~\cite{viernickel2018multipath}, and XLINK~\cite{zheng2021xlink}, dynamically distribute video segments across different paths, but were originally designed for wired networks with similar path characteristics, making them less effective in adapting to heterogeneous wireless links with highly dynamic conditions. 
The subsequent work, MuSher~\cite{saha2019musher}, optimizes MPTCP scheduling for heterogeneous channels. Chorus~\cite{lv2024chorus} coordinates multipath scheduling with adaptive streaming via cross-layer feedback control loops between the server and client. Habitus~\cite{zhang2024habitus} further enhances streaming by leveraging viewer pose  to predict mmWave throughput. More recently, \hl{PDStream~\cite{xiao2025pdstream} mitigates long-tail latency via pseudo-dual streaming with separate keyframe and non-keyframe delivery.} 
However, these methods require tight server–client synchronization, protocol modifications, or additional hardware sensing inputs. In contrast, our approach is a user-space design that does not depend on extra sensory devices and remains compatible with unmodified HTTP-based video delivery. \diff{COMPACT~\cite{chaudhary2025compact} and STORM~\cite{hu2025storm} improve mobile video delivery via tile-aware adaptation and transport-layer scheduling; NeuralMDC is complementary, focusing on codec-level stream independence for robust 5G multipath delivery.}

%However, these methods face deployment challenges due to their reliance on server-user synchronization or additional hardware requirements. 
%In contrast, our approach operates entirely in user space without requiring extra sensory devices, making it easier to deploy. 
%Additionally, we design a neural network-based MDC codec to more effectively adapt to channel variations.

\noindent\textbf{Video Compression:}
Video compression has long been studied to reduce spatial and temporal redundancy, using intra-frame coding, inter-frame prediction, transform and entropy coding. %Building on these principles, 
Standardized codecs such as AVC/HEVC~\cite{ffmpeg} have achieved high compression efficiency, while extensions including SVC/SHVC~\cite{schwarz2007overview,shvc} and MDC codecs~\cite{goyal2001multiple, kazemi2014review} aim to improve adaptability and robustness under network impairments. 
\diff{Classical MDC codecs split source information in the spatial~\cite{shirani2006content}, temporal~\cite{radulovic2009multiple}, or transform domains~\cite{conci2007real}, but these designs often incur high redundancy, multi-decoder complexity, limited scalability, and inflexible rate allocation.} %has studied quantization-based rate--distortion trade-offs~\cite{fleming1999generalized} and later 
Neural video codecs~~\cite{lu2019dvc,hu2021fvc,ladune2021conditional,li2023neural,li2022hybrid,li2021deep,rippel2021elf, vct} replace handcrafted modules with neural networks for improved rate-distortion performance. \diff{Recent neural MDC work mainly targets image coding~\cite{le2023inr}, while neural MDC video coding remains largely unexplored, with prior work~\cite{hu2021multiple} primarily improving reconstruction quality for traditional MDC video codecs.} 
\hl{More recently, neural codec streaming systems integrate learned codecs into streaming pipelines: GRACE~\cite{cheng2024grace} enhances single-stream delivery with loss recovery via denoising and packet-level redundancy, while SWIFT~\cite{swift} improves the compression efficiency of layered codecs by chaining multiple autoencoders. 
In contrast, NeuralMDC  targets multipath delivery by generating multiple \emph{independently decodable and mutually refinable} streams, enabling graceful quality refinement from any subset of received streams. Unlike SWIFT, which preserves inter-layer dependencies, NeuralMDC produces fully independent streams with flexible rate allocation, making it better suited for heterogeneous and dynamic wireless channels.}

\section{Conclusion}
\label{sec:conclusion}
%\vspace{-1em}
\diff{NeuralMDC improves robustness by shifting multipath delivery from dependent packets/layers to independently decodable descriptions, but this benefit comes with computational overhead. Our current prototype targets accelerator-equipped clients or edge-assisted deployment rather than CPU-only execution on commodity smartphones. Nevertheless, compute, memory, and energy consumption remain important deployment constraints, especially for sustained high-resolution mobile streaming. Further model compression, hardware-aware compilation, mobile NPU optimization, and adaptive edge offloading are important directions for making NeuralMDC practical on more resource-constrained devices.}
% As shown in our microbenchmarks, the Masked Transformer dominates runtime, while FP16 quantization and batching substantially reduce latency and memory footprint.

In conclusion, our novel neural network-based MDC video codec, combined with a simple yet effective multipath mapping algorithm, significantly improves the resilience and performance of video streaming over real-world dynamic 5G environments. By encoding video frames into multiple independent and mutually refinable streams, our system enables flexible, fine-grained mapping of streams to network paths, thereby overcoming the limitations of traditional \hl{monolithic coding} and \hl{layered coding} streaming methods. Our approach reduces video stalls and improves video quality, demonstrating superior performance across key QoE metrics. This enhanced robustness and adaptability to fluctuating 5G network conditions make our multi-stream video streaming system a strong candidate for NextG applications.

% [Mention that while our streaming system is designed for a ``explicit multi-path'' network environment, namely, explicitly using multiple 5G channels for video streaming, state that our algorithm also works well for  an ``implicit multi-path'' environment, and hints at the potential that our multi-stream video streaming can enable more advanced  in-network ``cross-layer''  mechanisms to better support video streaming.] 

\bibliographystyle{IEEEtran}
\bibliography{reference}

@inproceedings{zhang2024habitus,
author = {Zhang, Anlan and Wang, Chendong and Hu, Yuming and Hassan, Ahmad and Zhang, Zejun and Han, Bo and Qian, Feng and Xu, Shichang},
title = {Habitus: boosting mobile immersive content delivery through full-body pose tracking and multipath networking},
year = {2024},
isbn = {978-1-939133-39-7},
publisher = {USENIX Association},
address = {USA},
booktitle = {Proceedings of the 21st USENIX Symposium on Networked Systems Design and Implementation},
articleno = {92},
numpages = {19},
location = {Santa Clara, CA, USA},
series = {NSDI'24}
}

@INPROCEEDINGS{wong20205g,
  author={Wong, Kin-Lu},
  booktitle={2020 IEEE Asia-Pacific Microwave Conference (APMC)}, 
  title={5G/B5G Multi-Gbps Antennas for User Terminals and Their Throughput Verification}, 
  year={2020},
  volume={},
  number={},
  pages={366-368},
  publisher={IEEE},
  address={Hong Kong},
  doi={10.1109/APMC47863.2020.9331539}
}

@inproceedings{sigcomm-ross,
author = {K. Fezeu, Rostand A. and Fiandrino, Claudio and Ramadan, Eman and Carpenter, Jason and de Freitas, Lilian Coelho and Bilal, Faaiq and Ye, Wei and Widmer, Joerg and Qian, Feng and Zhang, Zhi-Li},
title = {Unveiling the 5G Mid-Band Landscape: From Network Deployment to Performance and Application QoE},
year = {2024},
isbn = {9798400706141},
publisher = {Association for Computing Machinery},
address = {New York, NY, USA},
url = {https://doi.org/10.1145/3651890.3672269},
doi = {10.1145/3651890.3672269},
booktitle = {Proceedings of the ACM SIGCOMM 2024 Conference},
pages = {358–372},
location = {Sydney, NSW, Australia},
series = {ACM SIGCOMM '24}
}

@inproceedings{Ahmad-Hotmobile23,
  author = {Hassan, Ahmad and Ye, Wei and Zhang, Anlan and Carpenter, Jason and Zhu, Ruiyang and Jin, Shuowei and Qian, Feng and Mao, Z. Morley and Zhang, Zhi-Li},
  title = {The Case for Boosting Mobile Application QoE via Smart Band Switching in 5G/xG Networks},
  year = {2024},
  isbn = {9798400704970},
  publisher = {Association for Computing Machinery},
  address = {New York, NY, USA},
  url = {https://doi.org/10.1145/3638550.3641132},
  doi = {10.1145/3638550.3641132},
  booktitle = {Proceedings of the 25th International Workshop on Mobile Computing Systems and Applications},
  pages = {127–132},
  location = {San Diego, CA, USA},
  series = {HOTMOBILE '24},
  abbr = {HotMobile},
  selected = {true},
  bibtex_show = {true}
}

@article{schwarz2007overview,
  title={Overview of the scalable video coding extension of the H. 264/AVC standard},
  author={Schwarz, Heiko and Marpe, Detlev and Wiegand, Thomas},
  journal={IEEE Transactions on circuits and systems for video technology},
  volume={17},
  number={9},
  pages={1103--1120},
  year={2007},
  publisher={IEEE}
}

@ARTICLE{hu2021fvc,
  author={Hu, Zhihao and Xu, Dong and Lu, Guo and Jiang, Wei and Wang, Wei and Liu, Shan},
  journal={IEEE Transactions on Pattern Analysis and Machine Intelligence}, 
  title={FVC: An End-to-End Framework Towards Deep Video Compression in Feature Space}, 
  year={2023},
  volume={45},
  number={4},
  pages={4569-4585}, 
  doi={10.1109/TPAMI.2022.3210652}}

@article{adelson1984pyramid,
  title={Pyramid methods in image processing},
  author={Adelson, Edward H and Anderson, Charles H and Bergen, James R and Burt, Peter J and Ogden, Joan M},
  journal={RCA engineer},
  volume={29},
  number={6},
  pages={33--41},
  year={1984}
}

@article{witten1987arithmetic,
  title={Arithmetic coding for data compression},
  author={Witten, Ian H and Neal, Radford M and Cleary, John G},
  journal={Communications of the ACM},
  volume={30},
  number={6},
  pages={520--540},
  year={1987},
  publisher={ACM New York, NY, USA}
}

@inproceedings {chen2024lifter,
author = {Bo Chen and Zhisheng Yan and Yinjie Zhang and Zhe Yang and Klara Nahrstedt},
title = {{LiFteR}: Unleash Learned Codecs in Video Streaming with Loose Frame Referencing},
booktitle = {21st USENIX Symposium on Networked Systems Design and Implementation (NSDI 24)},
year = {2024},
isbn = {978-1-939133-39-7},
address = {Santa Clara, CA},
pages = {533--548},
url = {https://www.usenix.org/conference/nsdi24/presentation/chen-bo},
publisher = {USENIX Association},
month = apr
}

@inproceedings{wang2016mcl,
  title={MCL-JCV: a JND-based H. 264/AVC video quality assessment dataset},
  author={Wang, Haiqiang and Gan, Weihao and Hu, Sudeng and Lin, Joe Yuchieh and Jin, Lina and Song, Longguang and Wang, Ping and Katsavounidis, Ioannis and Aaron, Anne and Kuo, C-C Jay},
  booktitle={2016 IEEE international conference on image processing (ICIP)},
  pages={1509--1513},
  year={2016},
  address={Phoenix, AZ, USA},
  publisher={IEEE}
}

@inproceedings{mercat2020uvg,
  title={UVG dataset: 50/120fps 4K sequences for video codec analysis and development},
  publisher = {Association for Computing Machinery},
  address = {New York, NY, USA},
  author={Mercat, Alexandre and Viitanen, Marko and Vanne, Jarno},
  booktitle={Proceedings of the 11th ACM Multimedia Systems Conference},
  pages={297--302},
  year={2020}
}

@article{xue2019video,
  title={Video enhancement with task-oriented flow},
  author={Xue, Tianfan and Chen, Baian and Wu, Jiajun and Wei, Donglai and Freeman, William T},
  journal={International Journal of Computer Vision},
  volume={127},
  pages={1106--1125},
  year={2019},
  publisher={Springer}
}

@inproceedings{minnen2018joint,
 author = {Minnen, David and Ball\'{e}, Johannes and Toderici, George D},
 booktitle = {Advances in Neural Information Processing Systems},
 editor = {S. Bengio and H. Wallach and H. Larochelle and K. Grauman and N. Cesa-Bianchi and R. Garnett},
 pages = {},
 publisher = {Curran Associates, Inc.},
 title = {Joint Autoregressive and Hierarchical Priors for Learned Image Compression},
 address={MONTREAL},
 url = {https://proceedings.neurips.cc/paper_files/paper/2018/file/53edebc543333dfbf7c5933af792c9c4-Paper.pdf},
 volume = {31},
 year = {2018}
}

@inproceedings{minnen2020channel,
  title={Channel-wise autoregressive entropy models for learned image compression},
  publisher={IEEE},
  address={Abu Dhabi, United Arab Emirates},
  author={Minnen, David and Singh, Saurabh},
  booktitle={2020 IEEE International Conference on Image Processing (ICIP)},
  pages={3339--3343},
  year={2020},
  organization={IEEE}
}

@inproceedings{wang2003multiscale,
  title={Multiscale structural similarity for image quality assessment},
  author={Wang, Zhou and Simoncelli, Eero P and Bovik, Alan C},
  booktitle={The thrity-seventh asilomar conference on signals, systems \& computers, 2003},
  volume={2},
  pages={1398--1402},
  year={2003},
  organization={Ieee}
}

@article{ladune2021conditional,
  title={Conditional coding and variable bitrate for practical learned video coding},
  author={Ladune, Th{\'e}o and Philippe, Pierrick and Hamidouche, Wassim and Zhang, Lu and D{\'e}forges, Olivier},
  journal={arXiv preprint arXiv:2104.09103},
  year={2021}
}

@inproceedings{li2023neural,
  title={Neural video compression with diverse contexts},
  author={Li, Jiahao and Li, Bin and Lu, Yan},
  booktitle={Proceedings of the IEEE/CVF conference on computer vision and pattern recognition},
  pages={22616--22626},
  publisher={IEEE},
  address={Vancouver, BC, Canada},
  year={2023}
}

@inproceedings{li2022hybrid,
  title={Hybrid spatial-temporal entropy modelling for neural video compression},
  author={Li, Jiahao and Li, Bin and Lu, Yan},
  booktitle={Proceedings of the 30th ACM International Conference on Multimedia},
  pages={1503--1511},
  year={2022}
}

@article{li2021deep,
  title={Deep contextual video compression},
  author={Li, Jiahao and Li, Bin and Lu, Yan},
  journal={Advances in Neural Information Processing Systems},
  volume={34},
  pages={18114--18125},
  year={2021}
}

@article{hu2021multiple,
  title={Multiple description coding for best-effort delivery of light field video using GNN-based compression},
  author={Hu, Xinjue and Pan, Yuxuan and Wang, Yumei and Zhang, Lin and Shirmohammadi, Shervin},
  journal={IEEE Transactions on Multimedia},
  volume={25},
  pages={690--705},
  year={2021},
  publisher={IEEE}
}

@article{ffmpeg,
  title={Converting video formats with FFmpeg},
  author={Tomar, Suramya},
  journal={Linux journal},
  volume={2006},
  number={146},
  pages={10},
  year={2006},
  publisher={Belltown Media Houston, TX}
}

@article{shvc,
  title={Overview of SHVC: Scalable extensions of the high efficiency video coding standard},
  author={Boyce, Jill M and Ye, Yan and Chen, Jianle and Ramasubramonian, Adarsh K},
  journal={IEEE Transactions on Circuits and Systems for Video Technology},
  volume={26},
  number={1},
  pages={20--34},
  year={2015},
  publisher={IEEE}
}

@inproceedings{lu2019dvc,
  title={Dvc: An end-to-end deep video compression framework},
  author={Lu, Guo and Ouyang, Wanli and Xu, Dong and Zhang, Xiaoyun and Cai, Chunlei and Gao, Zhiyong},
  booktitle={Proceedings of the IEEE/CVF conference on computer vision and pattern recognition},
  pages={11006--11015},
  publisher={IEEE},
  address={Long Beach, CA, USA},
  year={2019}
}

@inproceedings{rippel2021elf,
  title={Elf-vc: Efficient learned flexible-rate video coding},
  author={Rippel, Oren and Anderson, Alexander G and Tatwawadi, Kedar and Nair, Sanjay and Lytle, Craig and Bourdev, Lubomir},
  booktitle={Proceedings of the IEEE/CVF International Conference on Computer Vision},
  pages={14479--14488},
  publisher={IEEE},
  address={Montreal, QC, Canada},
  year={2021}
}

@inproceedings{pensieve,
  title={Neural adaptive video streaming with pensieve},
  author={Mao, Hongzi and Netravali, Ravi and Alizadeh, Mohammad},
  booktitle={Proceedings of the conference of the ACM special interest group on data communication},
  publisher = {Association for Computing Machinery},
  address = {New York, NY, USA},
  pages={197--210},
  year={2017}
}

@inproceedings{cheng2024grace,
  title={$\{$GRACE$\}$:$\{$Loss-Resilient$\}$$\{$Real-Time$\}$ Video through Neural Codecs},
  address = {Santa Clara, CA},
  publisher = {USENIX Association},
  author={Cheng, Yihua and Zhang, Ziyi and Li, Hanchen and Arapin, Anton and Zhang, Yue and Zhang, Qizheng and Liu, Yuhan and Du, Kuntai and Zhang, Xu and Yan, Francis Y and others},
  booktitle={21st USENIX Symposium on Networked Systems Design and Implementation (NSDI 24)},
  pages={509--531},
  year={2024}
}

@inproceedings{he2022elic,
  title={Elic: Efficient learned image compression with unevenly grouped space-channel contextual adaptive coding},
  author={He, Dailan and Yang, Ziming and Peng, Weikun and Ma, Rui and Qin, Hongwei and Wang, Yan},
  booktitle={Proceedings of the IEEE/CVF Conference on Computer Vision and Pattern Recognition},
  pages={5718--5727},
  publisher={IEEE},
  address={New Orleans, LA, USA},
  year={2022}
}

@inproceedings{vct,
author = {Mentzer, Fabian and Toderici, George and Minnen, David and Hwang, Sung Jin and Caelles, Sergi and Lucic, Mario and Agustsson, Eirikur},
title = {VCT: a video compression transformer},
year = {2022},
isbn = {9781713871088},
publisher = {Curran Associates Inc.},
address = {Red Hook, NY, USA},
booktitle = {Proceedings of the 36th International Conference on Neural Information Processing Systems},
articleno = {951},
numpages = {13},
location = {New Orleans, LA, USA},
series = {NIPS '22}
}

@inproceedings{li2023mage,
  title={Mage: Masked generative encoder to unify representation learning and image synthesis},
  author={Li, Tianhong and Chang, Huiwen and Mishra, Shlok and Zhang, Han and Katabi, Dina and Krishnan, Dilip},
  booktitle={Proceedings of the IEEE/CVF Conference on Computer Vision and Pattern Recognition},
  pages={2142--2152},
  year={2023}
}

@INPROCEEDINGS{yu2023magvit,
  author={Yu, Lijun and Cheng, Yong and Sohn, Kihyuk and Lezama, José and Zhang, Han and Chang, Huiwen and Hauptmann, Alexander G. and Yang, Ming-Hsuan and Hao, Yuan and Essa, Irfan and Jiang, Lu},
  booktitle={2023 IEEE/CVF Conference on Computer Vision and Pattern Recognition (CVPR)}, 
  title={MAGVIT: Masked Generative Video Transformer}, 
  publisher={IEEE},
  address={Vancouver, BC, Canada},
  year={2023},
  volume={},
  number={},
  pages={10459-10469},
  doi={10.1109/CVPR52729.2023.01008}
}

@inproceedings{xiang2022mimt,
  title={Mimt: Masked image modeling transformer for video compression},
  author={Xiang, Jinxi and Tian, Kuan and Zhang, Jun},
  booktitle={The Eleventh International Conference on Learning Representations},
  year={2022}
}

@inproceedings{m2t,
  title={M2t: Masking transformers twice for faster decoding},
  author={Mentzer, Fabian and Agustson, Eirikur and Tschannen, Michael},
  booktitle={Proceedings of the IEEE/CVF International Conference on Computer Vision},
  pages={5340--5349},
  publisher={IEEE},
  address={Paris, France},
  year={2023}
}

@inproceedings{swift,
  title={Swift: Adaptive video streaming with layered neural codecs},
  author={Dasari, Mallesham and Kahatapitiya, Kumara and Das, Samir R and Balasubramanian, Aruna and Samaras, Dimitris},
  booktitle={19th USENIX Symposium on Networked Systems Design and Implementation (NSDI 22)},
  address = {Renton, WA},
  publisher = {USENIX Association},
  pages={103--118},
  year={2022}
}

@inproceedings{chaudhary2025compact,
  title={Compact: Content-aware multipath live video streaming for online classes using video tiles},
  author={Chaudhary, Shubham and Mishra, Navneet and Gambhir, Keshav and Rajore, Tanmay and Bhattacharya, Arani and Maity, Mukulika},
  booktitle={Proceedings of the 16th ACM Multimedia Systems Conference},
  pages={201--213},
  year={2025}
}

@inproceedings{hu2025storm,
  title={$\{$STORM$\}$: a Multipath $\{$QUIC$\}$ Scheduler for Quick Streaming Media Transport under Unstable Mobile Networks},
  author={Hu, Liekun and Li, Changlong},
  booktitle={2025 USENIX Annual Technical Conference (USENIX ATC 25)},
  pages={851--866},
  year={2025}
}

@inproceedings{le2023inr,
  title={Inr-mdsqc: Implicit neural representation multiple description scalar quantization for robust image coding},
  author={Le, Trung Hieu and Pic, Xavier and Antonini, Marc},
  booktitle={2023 IEEE 25th International Workshop on Multimedia Signal Processing (MMSP)},
  pages={1--6},
  year={2023},
  organization={IEEE}
}

@article{conci2007real,
  title={Real-time multiple description intra-coding by sorting and interpolation of coefficients},
  author={Conci, Nicola and De Natale, Francesco GB},
  journal={Signal, Image and Video Processing},
  volume={1},
  number={1},
  pages={1--10},
  year={2007},
  publisher={Springer}
}

@article{radulovic2009multiple,
  title={Multiple description video coding with H. 264/AVC redundant pictures},
  author={Radulovic, Ivana and Frossard, Pascal and Wang, Ye-Kui and Hannuksela, Miska M and Hallapuro, Antti},
  journal={IEEE Transactions on Circuits and Systems for Video Technology},
  volume={20},
  number={1},
  pages={144--148},
  year={2009},
  publisher={IEEE}
}

@article{shirani2006content,
  title={Content-based multiple description image coding},
  author={Shirani, Shahram},
  journal={IEEE transactions on multimedia},
  volume={8},
  number={2},
  pages={411--419},
  year={2006},
  publisher={IEEE}
}

@inproceedings{le2023multiple,
  title={Multiple description video coding for real-time applications using HEVC},
  author={Le, Trung Hieu and Antonini, Marc and Lambert, Marc and Alioua, Karima},
  booktitle={2023 IEEE International Conference on Image Processing (ICIP)},
  pages={2580--2584},
  year={2023},
  organization={IEEE}
}

@article{franchi2005multiple,
  title={Multiple description video coding for scalable and robust transmission over IP},
  author={Franchi, Nicola and Fumagalli, Marco and Lancini, Rosa and Tubaro, Stefano},
  journal={IEEE Transactions on circuits and systems for video technology},
  volume={15},
  number={3},
  pages={321--334},
  year={2005},
  publisher={IEEE}
}

@ARTICLE{5g-mobility,
  author={Zhang, Haijun and Liu, Na and Chu, Xiaoli and Long, Keping and Aghvami, Abdol-Hamid and Leung, Victor C. M.},
  journal={IEEE Communications Magazine}, 
  title={Network Slicing Based 5G and Future Mobile Networks: Mobility, Resource Management, and Challenges}, 
  year={2017},
  volume={55},
  number={8},
  pages={138-145},
  doi={10.1109/MCOM.2017.1600940}
}

@inproceedings{robust-mpc,
author = {Yin, Xiaoqi and Jindal, Abhishek and Sekar, Vyas and Sinopoli, Bruno},
title = {A Control-Theoretic Approach for Dynamic Adaptive Video Streaming over HTTP},
year = {2015},
isbn = {9781450335423},
publisher = {Association for Computing Machinery},
address = {New York, NY, USA},
url = {https://doi.org/10.1145/2785956.2787486},
doi = {10.1145/2785956.2787486}, 
booktitle = {Proceedings of the 2015 ACM Conference on Special Interest Group on Data Communication},
pages = {325–338},
location = {London, United Kingdom},
series = {SIGCOMM '15}
}

@article{goyal2001multiple,
  title={Multiple description coding: Compression meets the network},
  author={Goyal, Vivek K},
  journal={IEEE Signal processing magazine},
  volume={18},
  number={5},
  pages={74--93},
  year={2001},
  publisher={IEEE}
}

@article{kazemi2014review,
  title={A review of multiple description coding techniques for error-resilient video delivery},
  author={Kazemi, Mohammad and Shirmohammadi, Shervin and Sadeghi, Khosrow Haj},
  journal={Multimedia Systems},
  volume={20},
  pages={283--309},
  year={2014},
  publisher={Springer}
}

@inproceedings {yan2020learning,
author = {Francis Y. Yan and Hudson Ayers and Chenzhi Zhu and Sadjad Fouladi and James Hong and Keyi Zhang and Philip Levis and Keith Winstein},
title = {Learning in situ: a randomized experiment in video streaming },
booktitle = {17th USENIX Symposium on Networked Systems Design and Implementation (NSDI 20)},
year = {2020},
isbn = {978-1-939133-13-7},
address = {Santa Clara, CA},
pages = {495--511},
url = {https://www.usenix.org/conference/nsdi20/presentation/yan},
publisher = {USENIX Association},
}

@inproceedings{netravali2015mahimahi,
  title={Mahimahi: accurate $\{$Record-and-Replay$\}$ for $\{$HTTP$\}$},
  author={Netravali, Ravi and Sivaraman, Anirudh and Das, Somak and Goyal, Ameesh and Winstein, Keith and Mickens, James and Balakrishnan, Hari},
  booktitle={2015 USENIX Annual Technical Conference (USENIX ATC 15)},
  pages={417--429},
  publisher = {USENIX Association},
  address = {USA},
  year={2015}
}

@misc{onnx,
  title = {Accelerated Mobile Machine Learning.},
  author = {{ONNX Runtime}},
  howpublished = {\url{https://onnxruntime.ai/}},
  year = {2025},
  note = {Accessed: 2025-03-11}
}

@misc{onnx-transformer,
  title = {Transformer Model Optimization Tool Overview.},
  author = {{ONNX Runtime}},
  howpublished = {\url{https://onnxruntime.ai/docs/performance/transformers-optimization.html}},
  year = {2024},
  note = {Accessed: 2026-08-21}
}

@inproceedings{xu2020understanding,
author = {Xu, Dongzhu and Zhou, Anfu and Zhang, Xinyu and Wang, Guixian and Liu, Xi and An, Congkai and Shi, Yiming and Liu, Liang and Ma, Huadong},
title = {Understanding Operational 5G: A First Measurement Study on Its Coverage, Performance and Energy Consumption},
year = {2020},
isbn = {9781450379557},
publisher = {Association for Computing Machinery},
address = {New York, NY, USA},
url = {https://doi.org/10.1145/3387514.3405882},
doi = {10.1145/3387514.3405882},
booktitle = {Proceedings of the Annual Conference of the ACM Special Interest Group on Data Communication on the Applications, Technologies, Architectures, and Protocols for Computer Communication},
pages = {479–494},
location = {Virtual Event, USA},
series = {SIGCOMM '20}
}

@misc{xcal,
	title = {{XCAL}},
	howpublished =  {\url{https://www.accuver.com/sub/products/view.php?idx=6&ckattempt=2}},
      author={Accuver},
	note = {Accessed: 2025-08-18}
}

@article{ahokangas2019business,
  title={Business models for local 5G micro operators},
  author={Ahokangas, Petri and Matinmikko-Blue, Marja and Yrj{\"o}l{\"a}, Seppo and Sepp{\"a}nen, Veikko and H{\"a}mm{\"a}inen, Heikki and Jurva, Risto and Latva-Aho, Matti},
  journal={IEEE Transactions on Cognitive Communications and Networking},
  volume={5},
  number={3},
  pages={730--740},
  year={2019},
  publisher={IEEE}
}

@article{liu20205g,
  title={5G deployment: Standalone vs. non-standalone from the operator perspective},
  author={Liu, Guangyi and Huang, Yuhong and Chen, Zhuo and Liu, Liang and Wang, Qixing and Li, Na},
  journal={IEEE Communications Magazine},
  volume={58},
  number={11},
  pages={83--89},
  year={2020},
  publisher={IEEE}
}

@misc{shvc_implement,
  author = {{Fraunhofer HHI}},
  title = {A Reference Implementation of SHVC (Scalable Extension to HEVC)},
  year = 2025,
  howpublished = {\url{https://hevc.hhi.fraunhofer.de/shvc}},
  note = {Accessed: 2025-03-30}
}

@inproceedings{lv2024chorus,
  title={Chorus: Coordinating Mobile Multipath Scheduling and Adaptive Video Streaming},
  author={Lv, Gerui and Wu, Qinghua and Liu, Yanmei and Li, Zhenyu and Tan, Qingyue and Yang, Furong and Chen, Wentao and Ma, Yunfei and Guo, Hongyu and Chen, Ying and others},
  booktitle={Proceedings of the 30th Annual International Conference on Mobile Computing and Networking},
  publisher = {Association for Computing Machinery},
  address = {New York, NY, USA},
  pages={246--262},
  year={2024}
}

@inproceedings{liu2020grad,
  title={Grad: Learning for overhead-aware adaptive video streaming with scalable video coding},
  publisher = {Association for Computing Machinery},
  address = {New York, NY, USA},
  author={Liu, Yunzhuo and Jiang, Bo and Guo, Tian and Sitaraman, Ramesh K and Towsley, Don and Wang, Xinbing},
  booktitle={Proceedings of the 28th ACM International Conference on Multimedia},
  pages={349--357},
  year={2020}
}

@ARTICLE{PREDICT,
  author={Minovski, Dimitar and Ögren, Niclas and Mitra, Karan and Åhlund, Christer},
  journal={IEEE Transactions on Mobile Computing}, 
  title={Throughput Prediction Using Machine Learning in LTE and 5G Networks}, 
  year={2023},
  volume={22},
  number={3},
  pages={1825-1840},
  doi={10.1109/TMC.2021.3099397}}

@inproceedings{ye2024dissecting,
author = {Ye, Wei and Hu, Xinyue and Sleder, Steven and Zhang, Anlan and Dayalan, Udhaya Kumar and Hassan, Ahmad and Fezeu, Rostand A. K. and Jajoo, Akshay and Lee, Myungjin and Ramadan, Eman and Qian, Feng and Zhang, Zhi-Li},
title = {Dissecting Carrier Aggregation in 5G Networks: Measurement, QoE Implications and Prediction},
year = {2024},
isbn = {9798400706141},
publisher = {Association for Computing Machinery},
address = {New York, NY, USA},
url = {https://doi.org/10.1145/3651890.3672250},
doi = {10.1145/3651890.3672250},
booktitle = {Proceedings of the ACM SIGCOMM 2024 Conference},
pages = {340–357},
location = {Sydney, NSW, Australia},
series = {ACM SIGCOMM '24}
}

@inproceedings{narayanan2020first,
  title={A first look at commercial 5G performance on smartphones},
  publisher = {Association for Computing Machinery},
  address = {New York, NY, USA},
  author={Narayanan, Arvind and Ramadan, Eman and Carpenter, Jason and Liu, Qingxu and Liu, Yu and Qian, Feng and Zhang, Zhi-Li},
  booktitle={Proceedings of The Web Conference 2020},
  pages={894--905},
  year={2020}
}

@inproceedings{narayanan2020lumos5g,
  title={Lumos5G: Mapping and predicting commercial mmWave 5G throughput},
  publisher = {Association for Computing Machinery},
  address = {New York, NY, USA},
  author={Narayanan, Arvind and Ramadan, Eman and Mehta, Rishabh and Hu, Xinyue and Liu, Qingxu and Fezeu, Rostand AK and Dayalan, Udhaya Kumar and Verma, Saurabh and Ji, Peiqi and Li, Tao and others},
  booktitle={Proceedings of the ACM internet measurement conference},
  pages={176--193},
  year={2020}
}

@inproceedings{de2019multipathtester,
  title={Multipathtester: Comparing mptcp and mpquic in mobile environments},
  author={De Coninck, Quentin and Bonaventure, Olivier},
  booktitle={2019 Network Traffic Measurement and Analysis Conference (TMA)},
  pages={221--226},
  year={2019},
  publisher={IEEE},
  address={Paris, France},
  organization={IEEE}
}

@inproceedings{viernickel2018multipath,
  title={Multipath QUIC: A deployable multipath transport protocol},
  author={Viernickel, Tobias and Froemmgen, Alexander and Rizk, Amr and Koldehofe, Boris and Steinmetz, Ralf},
  booktitle={2018 IEEE International Conference on Communications (ICC)},
  pages={1--7},
  year={2018},
  address={Kansas City, MO, USA},
  publisher={IEEE}
}

@inproceedings{zheng2021xlink,
author = {Zheng, Zhilong and Ma, Yunfei and Liu, Yanmei and Yang, Furong and Li, Zhenyu and Zhang, Yuanbo and Zhang, Jiuhai and Shi, Wei and Chen, Wentao and Li, Ding and An, Qing and Hong, Hai and Liu, Hongqiang Harry and Zhang, Ming},
title = {XLINK: QoE-driven multi-path QUIC transport in large-scale video services},
year = {2021},
isbn = {9781450383837},
publisher = {Association for Computing Machinery},
address = {New York, NY, USA},
url = {https://doi.org/10.1145/3452296.3472893},
doi = {10.1145/3452296.3472893},
booktitle = {Proceedings of the 2021 ACM SIGCOMM 2021 Conference},
pages = {418–432},
location = {Virtual Event, USA},
series = {SIGCOMM '21}
}

@ARTICLE{saha2019musher,
  author={Aggarwal, Shivang and Saha, Swetank Kumar and Khan, Imran and Pathak, Rohan and Koutsonikolas, Dimitrios and Widmer, Joerg},
  journal={IEEE/ACM Transactions on Networking}, 
  title={MuSher: An Agile Multipath-TCP Scheduler for Dual-Band 802.11ad/ac Wireless LANs}, 
  year={2022},
  volume={30},
  number={4},
  pages={1879-1894},
  doi={10.1109/TNET.2022.3158678}}

@inproceedings{Dimitrios-Uplink-IMC23,
author = {Ghoshal, Moinak and Khan, Imran and Kong, Z. Jonny and Dinh, Phuc and Meng, Jiayi and Hu, Y. Charlie and Koutsonikolas, Dimitrios},
title = {Performance of Cellular Networks on the Wheels},
year = {2023},
isbn = {9798400703829},
publisher = {Association for Computing Machinery},
address = {New York, NY, USA},
url = {https://doi.org/10.1145/3618257.3624814},
doi = {10.1145/3618257.3624814},
booktitle = {Proceedings of the 2023 ACM on Internet Measurement Conference},
pages = {678–695},
location = {Montreal QC, Canada},
series = {IMC '23}
}

@misc{MPTCP_Wikipedia,
  author       = "{Wikipedia contributors}",
  title        = "{Multipath TCP -- Wikipedia, The Free Encyclopedia}",
  year         = "2025",
  url          = "https://en.wikipedia.org/wiki/Multipath_TCP",
  note         = "Accessed: 2025-02-08"
}

@techreport{mptcp,
  title={TCP extensions for multipath operation with multiple addresses},
  author={Ford, Alan and Raiciu, Costin and Handley, Mark and Bonaventure, Olivier},
  institution={Internet Engineering Task Force},
  year={2013}
}

@inproceedings{mpquic:1,
  title={Multipath quic: Design and evaluation},
  publisher = {Association for Computing Machinery},
  address = {New York, NY, USA},
  author={De Coninck, Quentin and Bonaventure, Olivier},
  booktitle={Proceedings of the 13th international conference on emerging networking experiments and technologies},
  pages={160--166},
  year={2017}
}

@inproceedings{Eman-IMC21,
author = {Ramadan, Eman and Narayanan, Arvind and Dayalan, Udhaya Kumar and Fezeu, Rostand A. K. and Qian, Feng and Zhang, Zhi-Li},
title = {Case for 5G-aware video streaming applications},
year = {2021},
isbn = {9781450386364},
publisher = {Association for Computing Machinery},
address = {New York, NY, USA},
url = {https://doi.org/10.1145/3472771.3474036},
doi = {10.1145/3472771.3474036},
booktitle = {Proceedings of the 1st Workshop on 5G Measurements, Modeling, and Use Cases},
pages = {27–34},
location = {Virtual Event},
series = {5G-MeMU '21}
}

@inproceedings{rochman2024comprehensive,
  title={A comprehensive real-world evaluation of 5g improvements over 4g in low-and mid-bands},
  author={Rochman, Muhammad Iqbal and Ye, Wei and Zhang, Zhi-Li and Ghosh, Monisha},
  booktitle={2024 IEEE International Symposium on Dynamic Spectrum Access Networks (DySPAN)},
  pages={257--266},
  year={2024},
  address={Washington, DC, USA},
  publisher={IEEE}
}

@misc{fezeu2023mid,
      title={Mid-Band 5G: A Measurement Study in Europe and US}, 
      author={Rostand A. K. Fezeu and Jason Carpenter and Claudio Fiandrino and Eman Ramadan and Wei Ye and Joerg Widmer and Feng Qian and Zhi-Li Zhang},
      year={2023},
      eprint={2310.11000},
      archivePrefix={arXiv},
      primaryClass={cs.NI},
      url={https://arxiv.org/abs/2310.11000}, 
}

@inproceedings{chen2025large,
  title={A Large-Scale Study of the Potential of Multi-carrier Access in the 5G Era},
  author={Chen, Fukun and Ghoshal, Moinak and Nan, Enfu and Dinh, Phuc and Khan, Imran and Jonny Kong, Z and Charlie Hu, Y and Koutsonikolas, Dimitrios},
  booktitle={International Conference on Passive and Active Network Measurement},
  address={virtual},
  publisher={Springer Nature Switzerland},
  pages={469--484},
  year={2025},
  organization={Springer}
}

@inproceedings{cellfusion-sigcomm23,
  title={Cellfusion: Multipath vehicle-to-cloud video streaming with network coding in the wild},
  author={Ni, Yunzhe and Zheng, Zhilong and Lin, Xianshang and Gao, Fengyu and Zeng, Xuan and Liu, Yirui and Xu, Tao and Wang, Hua and Zhang, Zhidong and Du, Senlang and others},
  booktitle={Proceedings of the ACM SIGCOMM 2023 Conference},
  pages={668--683},
  publisher = {Association for Computing Machinery},
  address = {New York, NY, USA},
  year={2023}
}

@misc{qualcomm_DSDA,
  title = {Two birds, one stone: Unleashing the full potential for simultaneous 5G cellular connections, thanks to our new Qualcomm DSDA Gen 2 with dual data},
  howpublished = {\url{https://www.qualcomm.com/news/onq/2023/05/unleashing-full-potential-for-simultaneous-5g-cellular-connections-qualcomm-dsda-gen-2-with-dual-data}},
  note = {Accessed: 2025-07-18}
}

@misc{car_DSDA,
  title = {Seamless car connectivity via Dual-SIM Dual-Active},
  howpublished = {\url{https://www.gi-de.com/en/digital-security/connectivity-iot/automotive/dual-sim-dual-active#our-dsda-solution}},
  note = {Accessed: 2025-07-18}
}

@misc{Qualcomm_NPU,
  title = {Unlocking on-device generative AI
with an NPU and heterogeneous computing},
  howpublished = {\url{https://www.qualcomm.com/content/dam/qcomm-martech/dm-assets/documents/Unlocking-on-device-generative-AI-with-an-NPU-and-heterogeneous-computing.pdf}},
  note = {Accessed: 2025-07-18}
}

@misc{qai_mae,
  title = {Video-MAE},
  howpublished = {\url{https://aihub.qualcomm.com/models/video_mae}},
  note = {Accessed: 2025-07-18}
}

@inproceedings{xiao2025pdstream,
  title={PDStream: Slashing Long-Tail Delay in Interactive Video Streaming via Pseudo-Dual Streaming},
  author={Xiao, Xuedou and Zuo, Yingying and Yan, Mingxuan and Liu, Kezhong and Wang, Wei},
  booktitle={IEEE INFOCOM 2025-IEEE Conference on Computer Communications},
  pages={1--10},
  year={2025},
  organization={IEEE}
}

% The camera-ready version ends after the references.  The arXiv entry point
% defines \ARXIVVERSION so the supplementary appendix is included there.
\ifdefined\ARXIVVERSION
  \clearpage

%\appendix
\appendices
\section*{\centering Appendix}
% Override IEEEtran's two-line appendix section format to match subsection style
\makeatletter
\def\@seccntformat#1{\csname the#1\endcsname.\hskip 0.5em}
\def\section{\@startsection{section}{1}{\z@}{1.5ex plus 1.5ex minus 0.5ex}%
{0.7ex plus 1ex minus 0ex}{\normalfont\itshape}}
\makeatother

\section{More Emulated 5G Video Streaming Results}
\begin{figure}[t]
        \centering
        \includegraphics[width=.48\columnwidth]{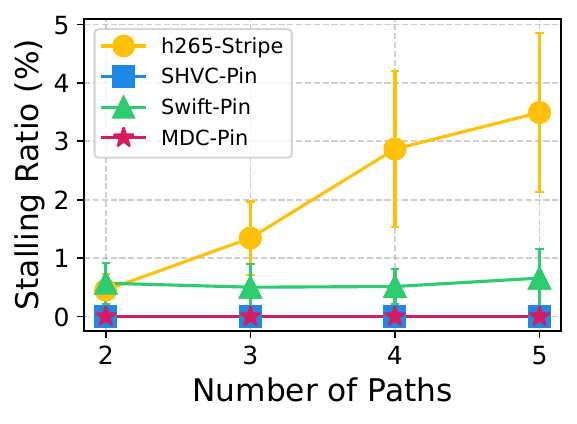}\label{fig:increase_number_of_path_stalling}
        \includegraphics[width=.48\columnwidth]{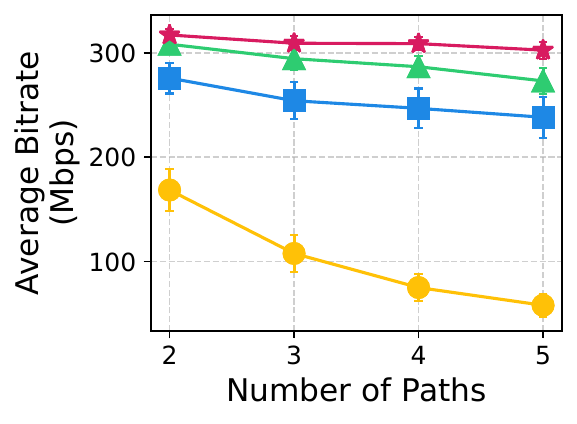}\label{fig:increased_number_of_path_bitrate}
        % \vspace{-4mm}
    \caption{Performance comparison over path numbers}
    \label{fig:increased_number_of_path}
\end{figure}
\noindent\textbf{Impact of Path Number:} 
We simulate the impact of increasing the number of available paths. Additional paths are introduced by randomly sampling from existing traces, with per-path throughputs proportionally scaled to hold aggregate bandwidth constant. 
As shown in Fig.~\ref{fig:increased_number_of_path}, NeuralMDC and SHVC both sustain near-zero stall ratios across all path counts as the number of paths grow, while Swift remains stable at around 0.5--0.7\%. H.265 degrades markedly with additional paths: stall ratios increase from $\sim$0.5\% to $\sim$3.5\%, and average bitrate falls from $\sim$165~Mbps to $\sim$50~Mbps. This degradation arises because striping distributes each chunk across all paths, so the completion time is dominated by the slowest path; with more paths, the probability of encountering a degraded path at any given time grows accordingly. Pinning schemes are insulated from this effect since each stream is confined to a single path. A slight decline in bitrate is nonetheless observed for NeualMDC and layered codecs as the number of paths grows, reflecting increasing mismatch between discrete stream/layer sizes and individual path bandwidths. NeualMDC consistently achieves the highest bitrate ($\sim$300--325~Mbps) across all configurations, and its margin over SHVC widens with more paths.
% Finally, we simulate the impact of increasing the number of available paths. By introducing additional paths—selected randomly from existing traces—and proportionally scaling their throughputs to keep the aggregate bandwidth constant, we study how stalling ratios and average bitrate evolve. Fig.~\ref{fig:increased_number_of_path} demonstrates that NeuralMDC and SHVC both sustain near-zero stalling ratios as the number of paths grows, while Swift and H.265 experience a marked increase in stalls. Moreover, the advantage of NeuralMDC over Swift and SHVC becomes increasingly pronounced with more available paths, whereas H.265 quickly becomes impractical under such conditions.
%
\begin{figure}
    \centering
    \includegraphics[width=1\linewidth]{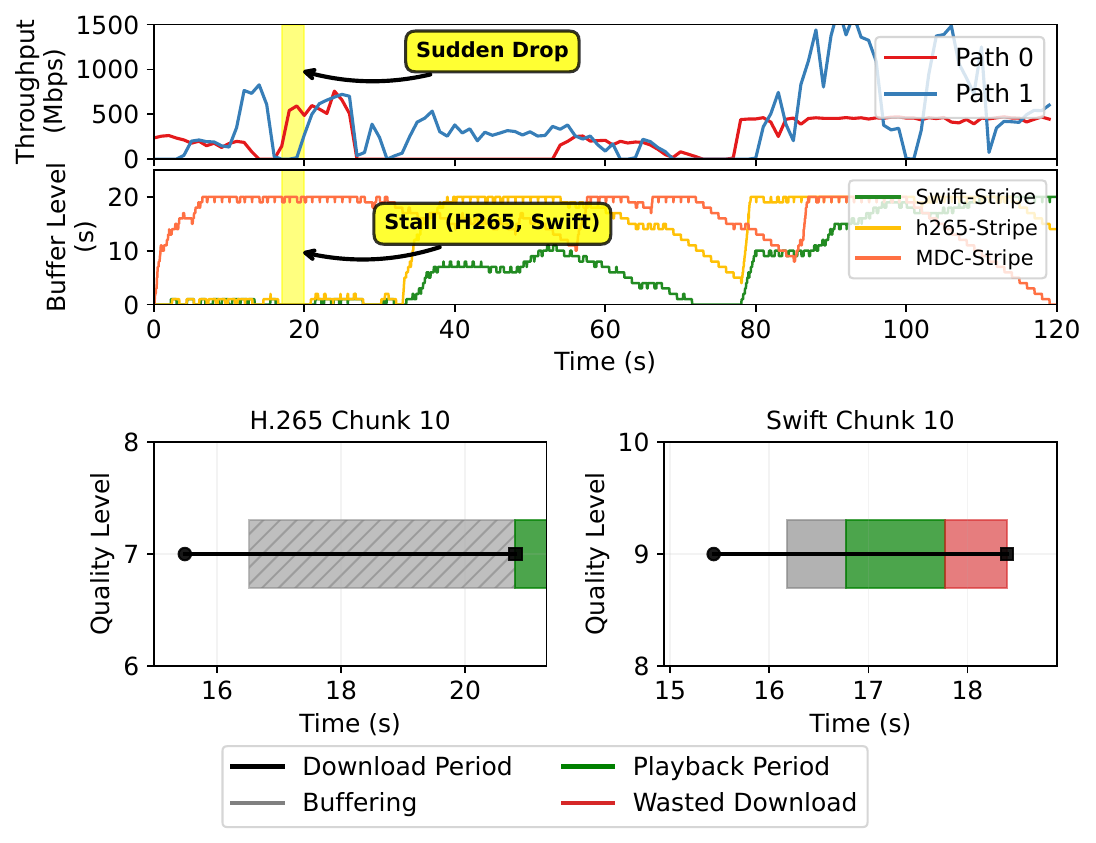}
    \caption{ H.265 and Swift struggle to maintain buffer size during network fluctuations when chunk/base layer bitrate exceeds  throughput.}\label{fig:without_dominant_path}
\end{figure}
\begin{figure}
    \centering
    \includegraphics[width=1\linewidth]{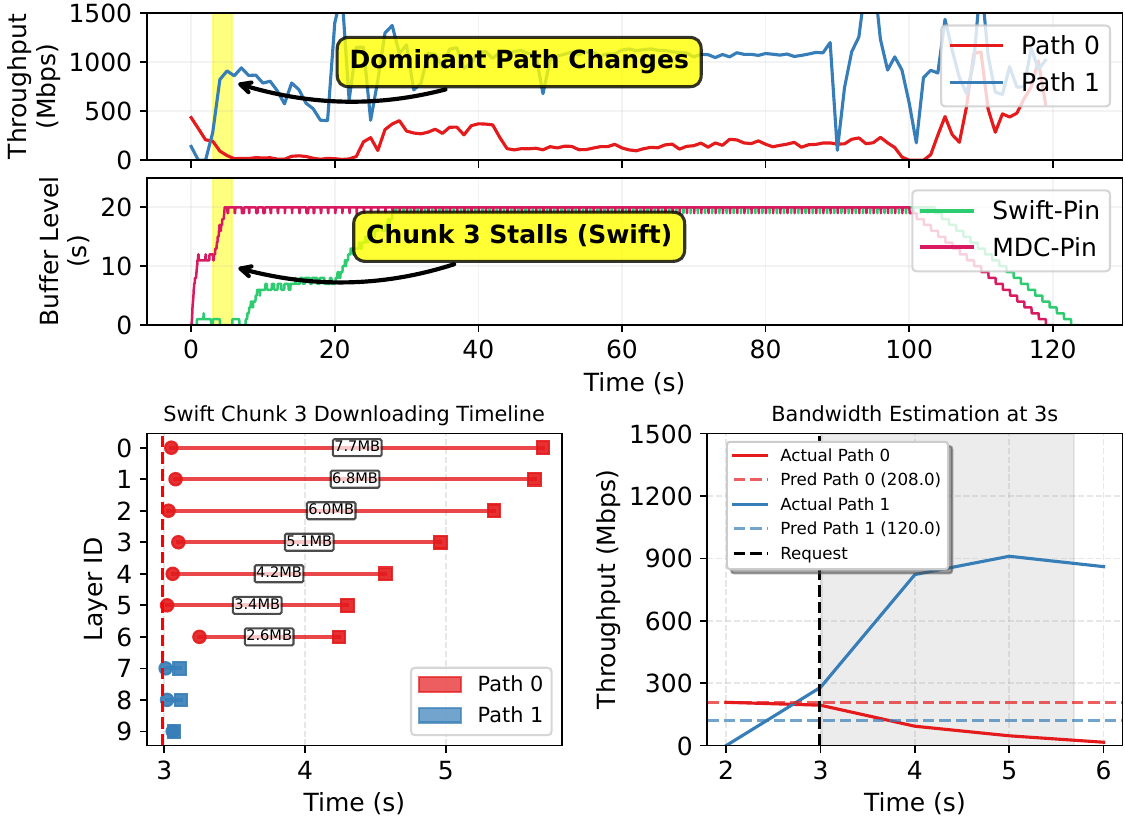}
   \caption{ Incorrect path mapping for Swift causes the lower layer to arrive last, resulting in rebuffering.}
    \label{fig:with_dominant_path}
\end{figure}
% \begin{figure*}
%     \vspace{-0.2cm}
%     \begin{minipage}{.45\textwidth}
%     \includegraphics[width=\textwidth]{plots/hol_swift_H265.pdf}
%     \vspace{-8mm} % Adjust the value to reduce the space
%     %\caption{ H.265 and Swift have challenges building up the buffer during fluctuations due to the selection of high-quality level/large base layer size.}\label{fig:without_dominant_path}
%      \caption{ H.265 and Swift struggle to maintain buffer size during network fluctuations when chunk/base layer bitrate exceeds  throughput.}\label{fig:without_dominant_path}
%     \end{minipage}\hfill
%     \begin{minipage}{.45\textwidth}
%     {\includegraphics[width=\textwidth]{plots/impact_path_prediction.pdf}}
%     \vspace{-8mm} % Adjust the value to reduce the space
%     \caption{ Incorrect path mapping for Swift causes the lower layer to arrive last, resulting in rebuffering.
%     }\label{fig:with_dominant_path}
%     \end{minipage}
%     \vspace{-6mm}
% \end{figure*}

\begin{figure*}[t]
    \centering
    \subfigure[Bitrate loss of SHVC compared to NeuralMDC]{
        \includegraphics[width=0.45\textwidth]{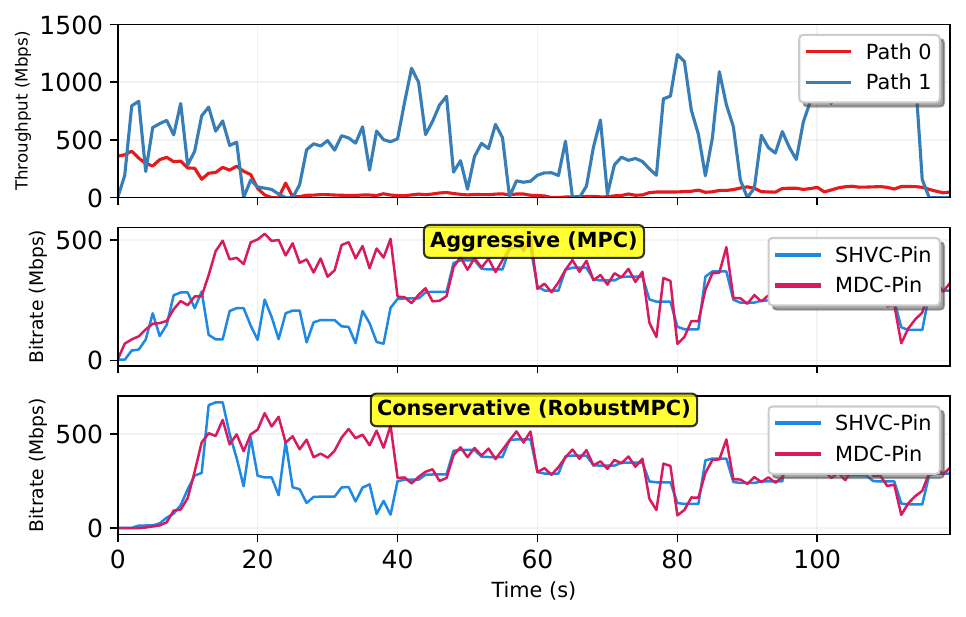}
        \label{subfig:rate_adaptation}
    }
    \subfigure[Transmission Timeline Analysis for Chunk 23]{
        \includegraphics[width=0.45\textwidth]{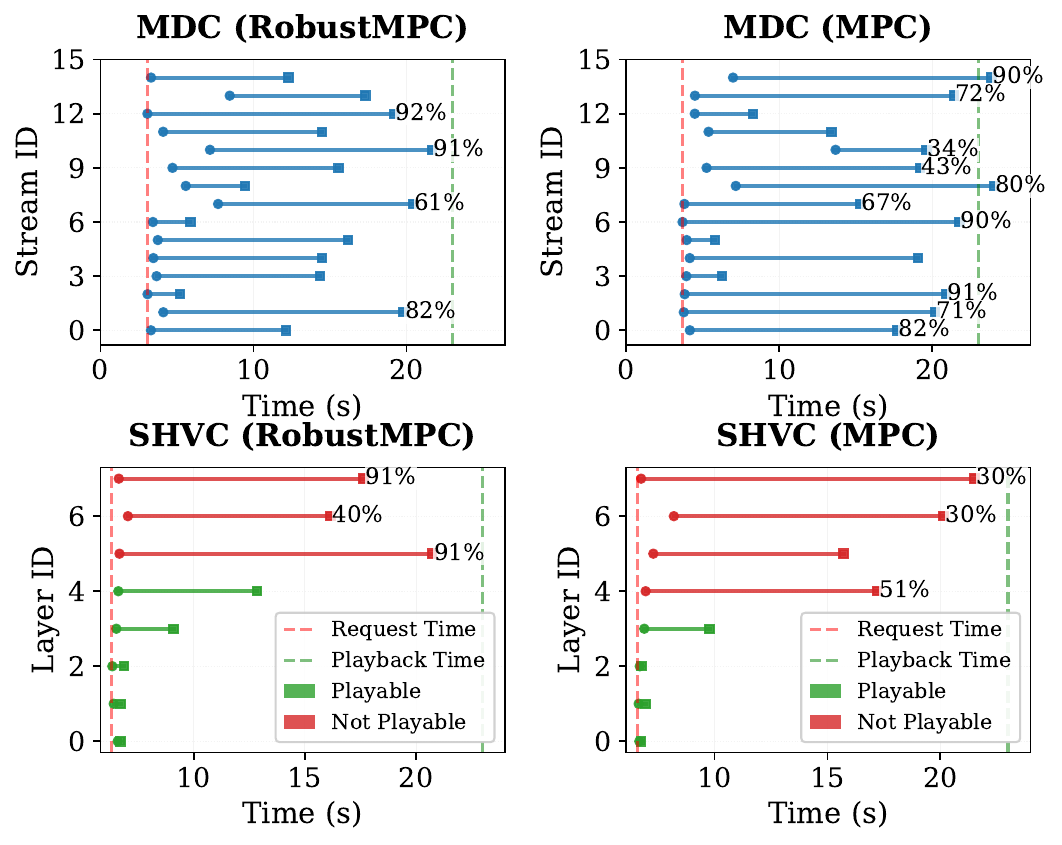}
        \label{subfig:wasted_transmission}
    }
    \Description{}
    \caption{Simplifying QoE Decision Making: SHVC has to abort layer data, causing wasteful transmission.}
    \label{fig:case_study_qoe_decision}
\end{figure*}

In the following, we provides detailed case studies that explain the rationale behind NeuralMDC performance gains and demonstrate how it simplifies multi-path video streaming over 5G.
% The purpose of section is to use concrete cases to illustrate why NeuralMDC can simplify the design of video streaming system because of its robustness brought by indepent decodability. We show that H.265 and Swift heavily depend on the path scheduling mechanisms for smooth playback, and SHVC risks of ineffective transmission due to its dependency hierarchy for decoding.

\noindent\textbf{Case Study1: Simplifying Dynamic Scheduling.} During significant traffic fluctuations, striping-based path mapping requires frequent rescheduling for H.265 and layered codecs, as both require full chunk delivery (H.265) or a complete base layer (layered codecs) before playback can continue. 
Fig.~\ref{fig:without_dominant_path} illustrates stalls in streaming with Swift and H.265 caused by a sudden drop in a particular 5G channel. Around 18–19 seconds, a sudden network drop disrupts H.265 and Swift, hindering downloading and depleting playback buffer. 
In contrast, NeuralMDC recovers more easily due to its finer-grained stream size and partial decodability. 
For simplicity, we omit SHVC, as its base layer, though small, can still be affected by network fluctuations due to the decoding dependency on the base layer. 

\noindent\textbf{Case Study2: Simplifying Video-Stream-to-Path Mapping.} Focusing on path pinning, we observe that Swift can still be vulnerable to suboptimal path selection. Fig.~\ref{fig:with_dominant_path} shows that during the transmission of chunk 3, historical performance data incorrectly favored path 0 over path 1. As a result, the critical lower-layer data for Swift is transmitted through the weaker path, precipitating a stall. In contrast, NeuralMDC does not depend on exact throughput predictions because every arriving stream is useful for the codec. This independence from precise path quality estimation makes NeuralMDC more robust to mapping missteps.

\noindent\textbf{Case Study3: Simplifying QoE Decision Making.} The independent decodability of NeuralMDC streams also simplifies the task of QoE decision making compared with the rigid layer dependencies of SHVC. An analysis shown in Figure \ref{fig:case_study_qoe_decision} using two ABR strategies—MPC and RobustMPC—reveals that under the aggressive MPC strategy, SHVC inefficiently selects enhancement layers that cannot be fully sustained (for example, Layer 4 achieves only 51\% transfer by playback, rendering subsequent layers ineffective). In contrast, NeuralMDC effectively leverages all available data, substantially reducing wasted bandwidth. Even with the more conservative RobustMPC, SHVC still encounters issues related to overestimation and under-utilization during fluctuations, resulting in an inferior bitrate compared to NeuralMDC.

% \noindent\textbf{(Case Study) Simplifying QoE Decision Making:} We further show how the advantages of flexible layer independence of NeuralMDC over rigid layer depenency of SHVC simplify QoE decision making. To show the contrast of "aggressive" and "conservative" decision making, the analysis examines the bitrate under two ABR strategies, MPC and RobustMPC. Under MPC, which aggressively utilizes available throughput, SHVC suffers from inefficient bandwidth use due to its rigid layer dependencies, often selecting enhancement layers that cannot be fully sustained (Layer 4 is only 51\% transferred upon the playback, rendering higher layers completely unusable even they have arrived). In contrast, NeuralMDC can utilize all available data, significantly reducing wasted bandwidth. With RobustMPC, which adopts a more conservative approach by prioritizing buffer occupancy, SHVC mitigates waste but cannot completely avoid the issue because the overestimation can still happen during fluctuation, and risks under-utilization because of the conservativeness, as illustrated by its inferior bitrate compared to NeuralMDC during 20-40s. MDC, benefiting from its independent streams, ensures efficient bandwidth utilization across varying network conditions, regardless of the adaptation strategies used.

\section{Real-world 5G Video Streaming Case Study}
\begin{figure}[t]
    \centering
    \subfigure{\includegraphics[width=.48\columnwidth]{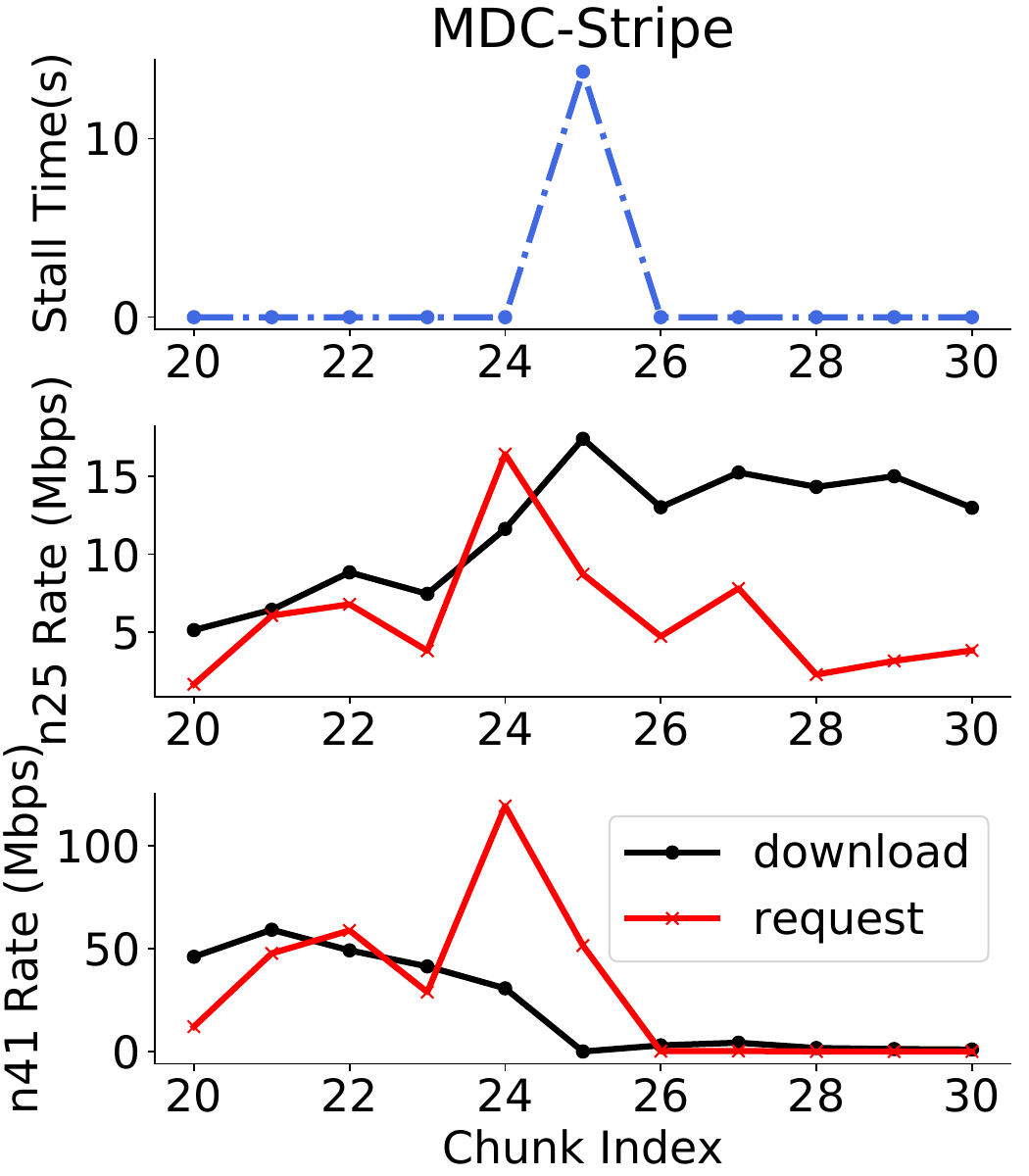}}
    \subfigure{\includegraphics[width=.48\columnwidth]{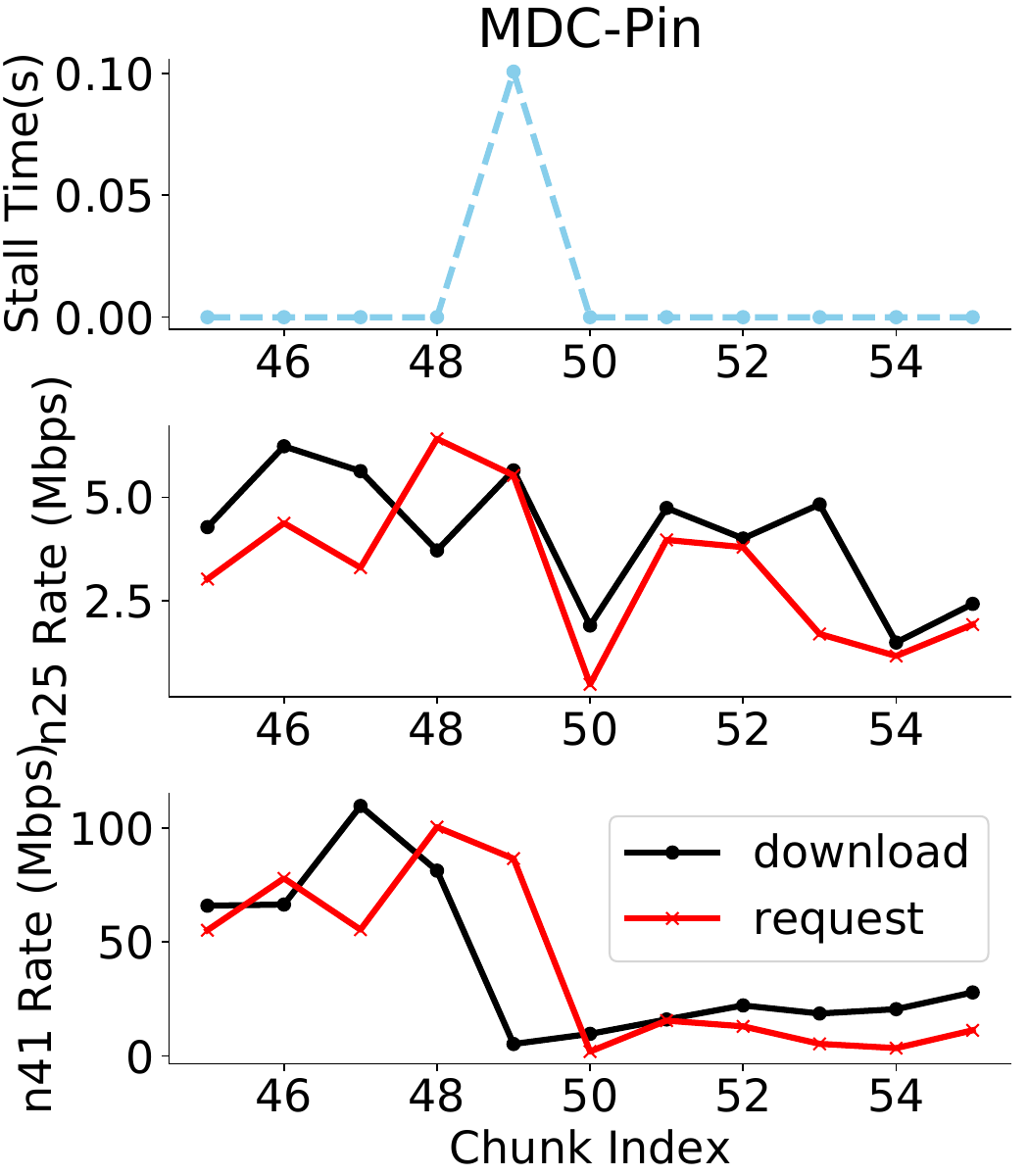}}
    \vspace{-4mm}
    \Description{}
    \caption{A snapshot of streaming under similar 5G network conditions. Path-striping encounters a large stall due to throughput temporarily dropping to zero in the n41 band, while Path-pinning experiences a slight stall despite the n41 throughput dropping to zero.}
    \vspace{-4mm}
    \label{f:real_world_stripe_pin}
\end{figure}
\noindent
\textbf{Path-Pinning vs Path-Striping}
When comparing the two NeuralMDC variants, the path-pinning scheduling algorithm decouples multiple paths to independently deliver streams and hence mitigates the impact of sudden throughput drops in a 5G band. In contrast, the path-striping scheduling algorithm may suffer from longer stream completion times due to slow paths and inappropriate load distribution.  

Fig. \ref{f:real_world_stripe_pin} illustrates a detailed comparison of streaming behaviors between the two NeuralMDC variants in a walking scenario, where both schemes experience similar 5G network conditions (with a chunk throughput cosine similarity of 0.76). In the path-striping streaming session, the n41 bandwidth drops to nearly zero at the 25th chunk, while in the path-pinning session, this occurs at the 49th chunk. Consequently, path-striping incurs a stall time of 13.7 seconds when downloading the 25th chunk due to inappropriately distributing a portion of each stream to n41. In contrast, path-pinning encounters only a 0.1-second stall at the 49th chunk, as the chunk can be decoded as long as at least one stream, distributed to the n25 band, is fully received. The temporarily inactive n41 band does not impact stream completion time in this case.

% \begin{figure}[!htp]
%     \centering
%     \vspace{-4mm}
%     \subfloat{\includegraphics[width=0.5\textwidth]{plots/hol_swift_H265.pdf}}
%     % \subfigure[Todo Buffering Comparison]{\includegraphics[width=0.24\textwidth]{figures/ExplicitMultiPath/explicit_overall_boxplot_stalling.pdf}}
%     % \subfigure[Todo QoE Comparison]{\includegraphics[width=0.24\textwidth]{figures/ExplicitMultiPath/explicit_overall_boxplot_stalling.pdf}} 
%     \vspace{-10mm} % Adjust the value to reduce the space
%     \Description{}
%     \caption{\small Simplifying Dynamic Scheduling: H.265 and Swift has challenges building up the buffer during fluctuations, due to selection of high quality level/large base layer size.
%     }
%     \vspace{-4mm} 
%     \label{fig:without_dominant_path}
% \end{figure}

%%%%%%%%%%%%%%%%%%%% BACK UP %%%%%%%%%%%%%%%
\iffalse
\begin{figure}[!htp]
    \centering
    \captionsetup[subfigure]{skip=1pt} % Minimize space between subfigure and caption
    \setlength{\abovecaptionskip}{1pt} % Reduce space above caption
    \setlength{\belowcaptionskip}{1pt} % Reduce space below caption

    \subfigure{
        \includegraphics[width=0.22\textwidth]{plots/high_variance_stall_qoe.pdf}
        \label{fig:high_variance}
    }
    \hspace{2pt} % Minimize horizontal space
    \subfigure{
        \includegraphics[width=0.22\textwidth]{plots/low_variance_stall_qoe.pdf}
        \label{fig:low_variance}
    }

    \caption{ \small Comparison under different throughput variance.}
    \label{fig:different_variance_comparison}
\end{figure}
\fi

\section{Spatio-temporal Complexity of Test Videos}
\label{a:si_ti}

\begin{table}[h]
\caption{Spatial Information (SI) and Temporal Information (TI) of test videos.}
\label{tab:video_si_ti}
\begin{tabular}{lcc}
\toprule
Video & SI & TI \\
\midrule
UVG\_Beauty & 22.36 & 10.75 \\
UVG\_Bosphorus & 31.69 & 5.84 \\
UVG\_HoneyBee & 25.23 & 2.79 \\
UVG\_Jockey & 35.13 & 29.65 \\
UVG\_ReadySteadyGo & 79.66 & 37.77 \\
UVG\_ShakeNDry & 29.77 & 8.42 \\
UVG\_YachtRide & 54.88 & 14.42 \\
\midrule
MCL-JCV\_videoSRC01 & 21.49 & 5.89 \\
MCL-JCV\_videoSRC02 & 38.21 & 29.75 \\
MCL-JCV\_videoSRC03 & 41.59 & 12.37 \\
MCL-JCV\_videoSRC04 & 97.57 & 33.05 \\
MCL-JCV\_videoSRC05 & 79.70 & 30.86 \\
MCL-JCV\_videoSRC06 & 27.20 & 10.13 \\
MCL-JCV\_videoSRC07 & 33.07 & 17.16 \\
MCL-JCV\_videoSRC08 & 33.93 & 17.23 \\
MCL-JCV\_videoSRC09 & 82.12 & 18.13 \\
MCL-JCV\_videoSRC10 & 96.30 & 58.58 \\
MCL-JCV\_videoSRC11 & 52.79 & 31.43 \\
MCL-JCV\_videoSRC12 & 66.60 & 5.07 \\
MCL-JCV\_videoSRC13 & 114.06 & 8.59 \\
MCL-JCV\_videoSRC14 & 47.72 & 30.05 \\
MCL-JCV\_videoSRC15 & 56.07 & 12.34 \\
MCL-JCV\_videoSRC16 & 25.60 & 11.08 \\
MCL-JCV\_videoSRC17 & 27.75 & 17.42 \\
MCL-JCV\_videoSRC18 & 66.70 & 26.93 \\
MCL-JCV\_videoSRC19 & 57.54 & 31.08 \\
MCL-JCV\_videoSRC20 & 53.31 & 57.07 \\
MCL-JCV\_videoSRC21 & 37.46 & 38.64 \\
MCL-JCV\_videoSRC22 & 54.09 & 56.15 \\
MCL-JCV\_videoSRC23 & 42.59 & 16.60 \\
MCL-JCV\_videoSRC24 & 47.82 & 18.87 \\
MCL-JCV\_videoSRC25 & 85.44 & 61.38 \\
MCL-JCV\_videoSRC26 & 47.17 & 66.25 \\
MCL-JCV\_videoSRC27 & 56.66 & 64.81 \\
MCL-JCV\_videoSRC28 & 53.67 & 6.00 \\
MCL-JCV\_videoSRC29 & 8.80 & 15.22 \\
MCL-JCV\_videoSRC30 & 23.25 & 15.67 \\
\bottomrule
\end{tabular}
\end{table}

Table~\ref{tab:video_si_ti} shows that the test videos span a broad range of spatiotemporal complexity, with SI values from 8.80 to 114.06 and TI values from 2.79 to 66.25, covering low-motion close-up sequences and high-motion complex scenes. As reported in~\cite{cheng2024grace}, GRACE achieves favorable compression efficiency relative to H.264 only on content with low spatial complexity; for videos with SI exceeding 22 and TI approaching 10, GRACE's rate-distortion performance falls below H.264. Since the majority of our test videos exceed these thresholds, GRACE underperforms H.264 on our test set, which is consistent with the codec evaluation results in \S\ref{sec:codec-compression}.

\section{NeuralMDC Codec Microbenchmarking}
\begin{figure}[t]
    \centering
    \begin{minipage}{0.25\textwidth}
        \centering
        \includegraphics[width=\linewidth]{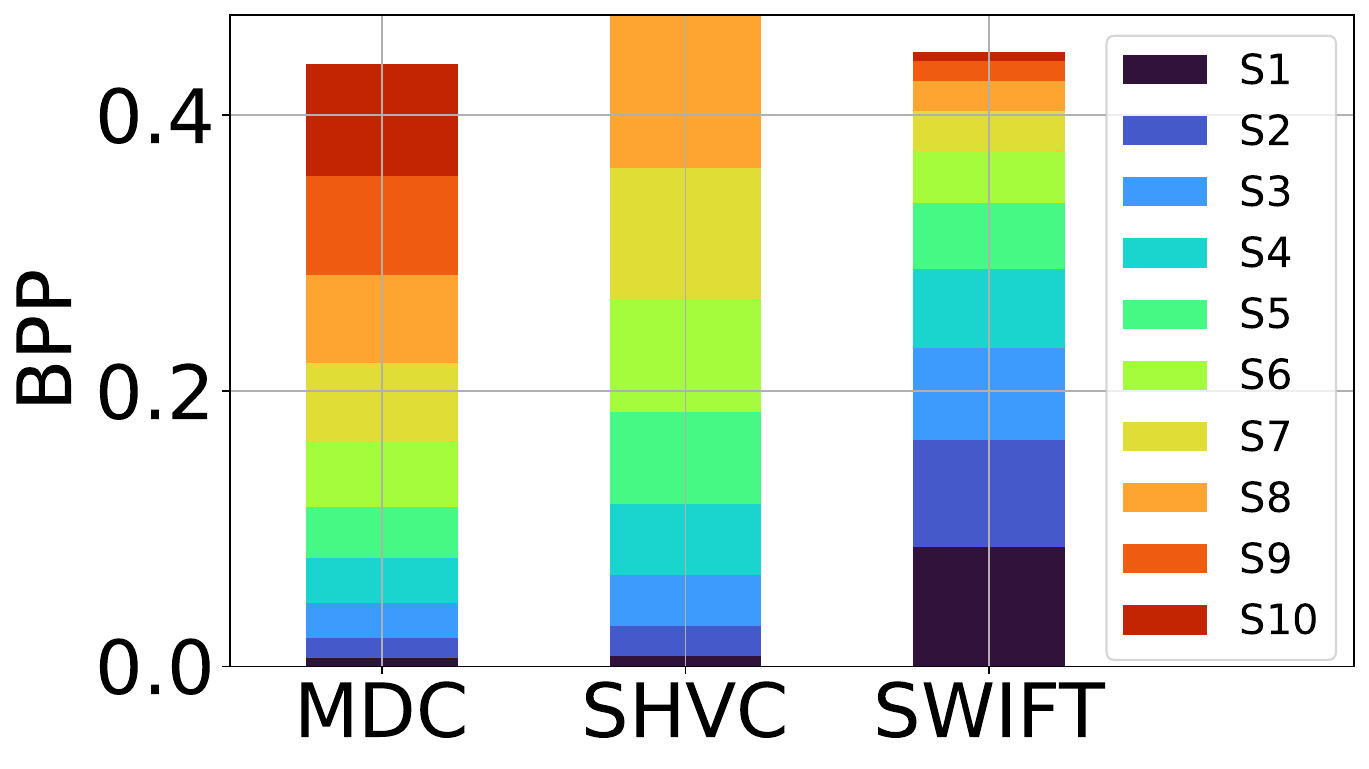}
        \vspace{-6mm}
        \caption{Stream/Layer size.}
        \vspace{-4mm}
        \label{f:stream_size}
    \end{minipage}
    \hfill
    \begin{minipage}{0.2\textwidth}
        \centering
        \includegraphics[width=\linewidth]{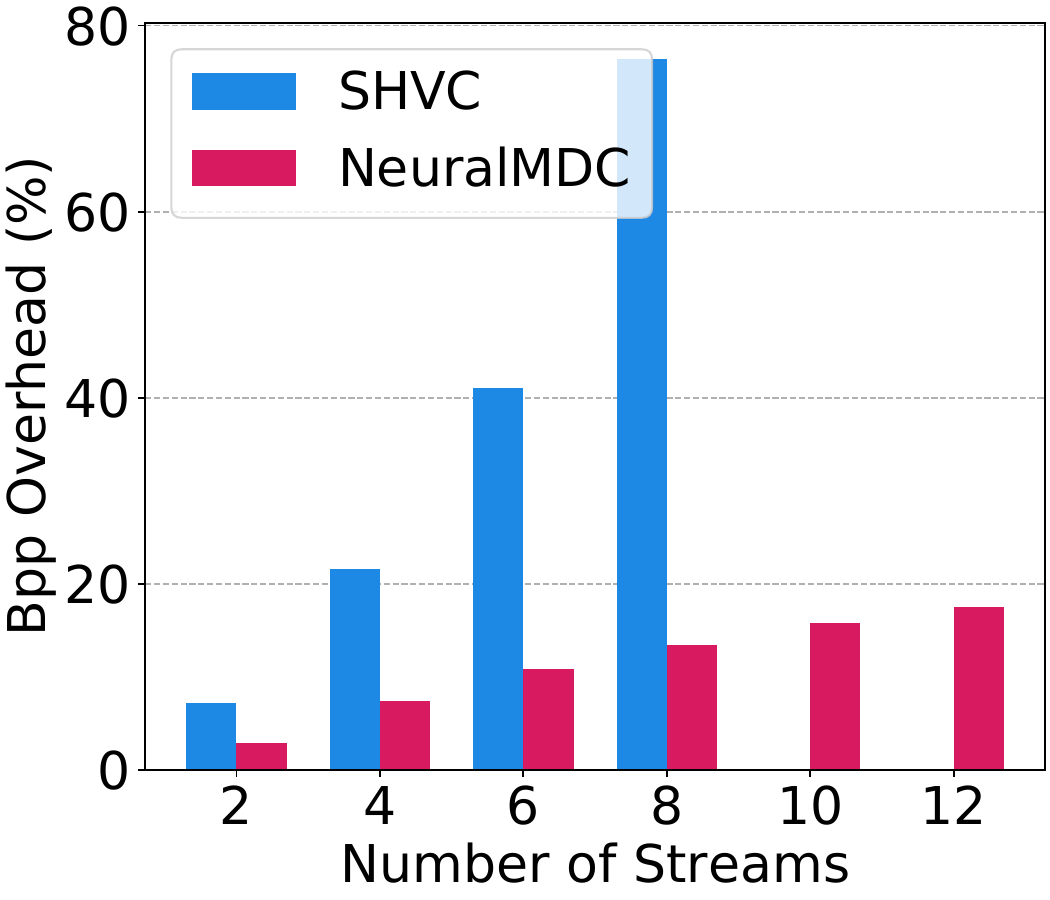}
        \vspace{-6mm}
        \caption{\shepherd{bpp} overhead vs. one stream.}
        \vspace{-4mm}
        \label{f:bpp_overhead}
    \end{minipage}
    \vspace{-4mm}
\end{figure}
\label{a:codec_benchmark}
\noindent
% \textbf{Stream Size Distribution} Fig. \ref{f:stream_size} shows the bits per pixel(BPP) of each stream/layer. We observed that Swift's base layer has the largest size, with each subsequent layer progressively smaller due to the decreasing entropy of subsequent residuals. %as the decoded frame approaches the input frame. 
% In contrast, NeuralMDC and SHVC offer more configurable stream/layer sizes, enabling finer-grained control. Considering the importance of the base layer, the flexibility of stream size configuration, and the independence among streams, NeuralMDC can offer significantly more flexible bitrate adaptation. %compared to Swift and SHVC. 
\textbf{Stream Size Distribution} Fig.~\ref{f:stream_size} shows the \shepherd{bpp} of each stream/layer for MDC, SHVC, and Swift, all with 10 streams/layers at comparable total \shepherd{bpp}. NeuralMDC's pyramid splitting (\S\ref{s:spliting}) produces streams of progressively increasing size from S1 to S10, providing fine-grained control over the bitrate contribution of each stream. SHVC also offers configurable layer sizes by adjusting the layer quantization parameters. In contrast, Swift's distribution is heavily skewed: S1 (the base layer) alone accounts for roughly half the total \shepherd{bpp}, with subsequent layers shrinking rapidly as residual entropy decreases. This concentration of information in the base layer underpins Swift's vulnerability to path failures and bandwidth fluctuations, as the base layer must be fully received before any decoding can proceed.

\noindent\textbf{Bitrate Overhead.} Fig. \ref{f:bpp_overhead} shows the relative bitrate increase of NeuralMDC and SHVC as the number of streams increases, measured against the single-stream baseline at equal visual quality\footnote{Swift encodes only the residuals from previous layers, without incurring any cross-layer compression overhead~\cite{swift}.}. SHVC is known for significant cross-layer compression overheads. 
NeuralMDC also incurs bitrate overhead as adding more streams reduces latent tokens per stream, decreasing intra-stream correlation and compression efficiency. 
However, this overhead reaches an upper limit as more streams are added, since the previous frame provides the primary temporal-spatial context for compressing latent tokens. Compared to SHVC, Neural MDC incurs a more acceptable bitrate overhead.

\section{Inference Example with Partial Reception}
\label{a:reconstruction_sample}

%We present some reconstruction examples of NeuralMDC  when 50\% tokens are received in Figure \ref{f:reconstruct} together with the original frame. Also, the metric PSNR and MS-SSIM are attached at the bottom of each example. Clearly, the examples show the capacity of NeuralMDC's superior inference and reconstruction performance.
Fig.~\ref{f:reconstruct} presents reconstruction examples from NeuralMDC under 50\% token reception alongside the corresponding original frames. Across content types with varying spatiotemporal complexity, NeuralMDC produces perceptually coherent reconstructions: a close-up scene (PSNR: 33.68~dB, MS-SSIM: 0.96), an outdoor scene (PSNR: 32.85~dB, MS-SSIM: 0.96), and a fast-motion scene (PSNR: 29.92~dB, MS-SSIM: 0.88). The relatively lower quality on the fast-motion content is consistent with the inherent difficulty of inferring missing tokens under high temporal variation, where the conditional distribution estimated from the previous frame is a less accurate prior.
\begin{figure*}[ht]
    \centering
    % % original
    \subfigure[Original]{\includegraphics[width=.30\linewidth]{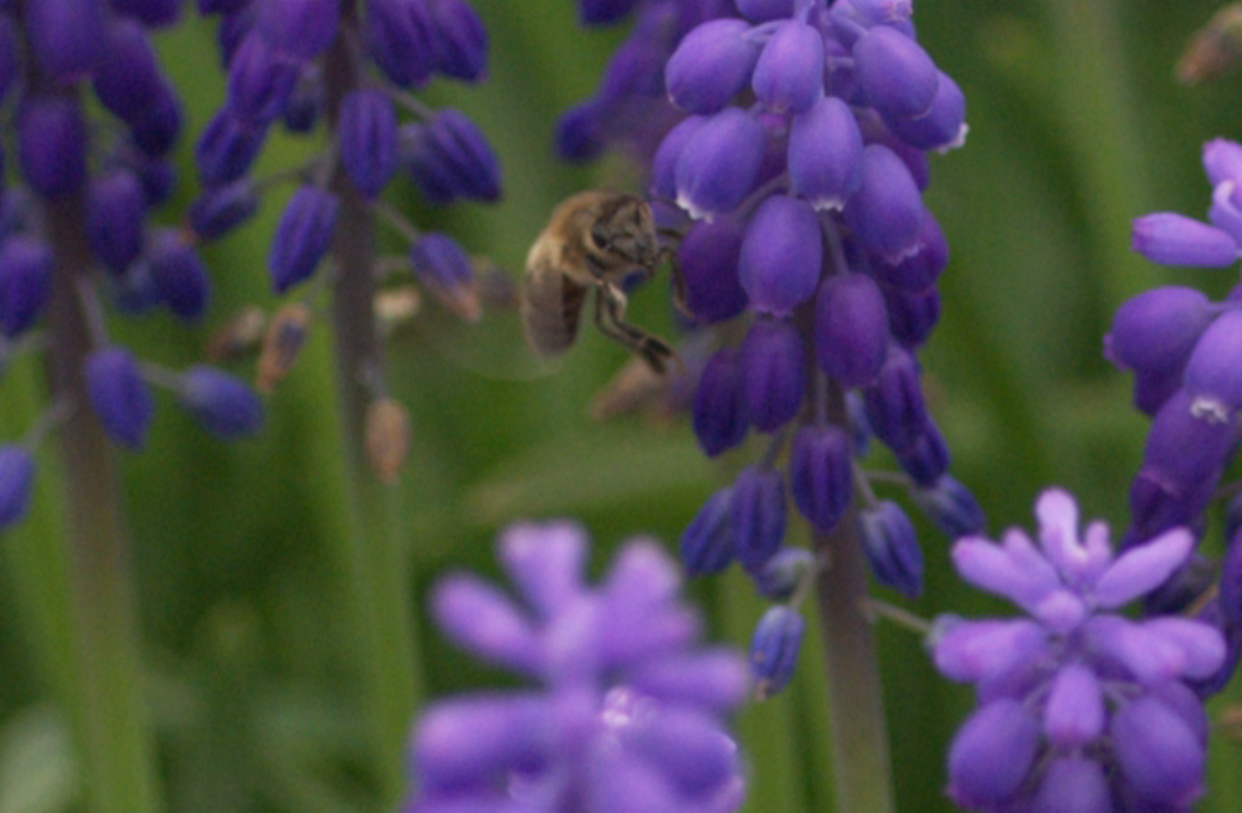}}
    \subfigure[Original]{\includegraphics[width=.28\linewidth]{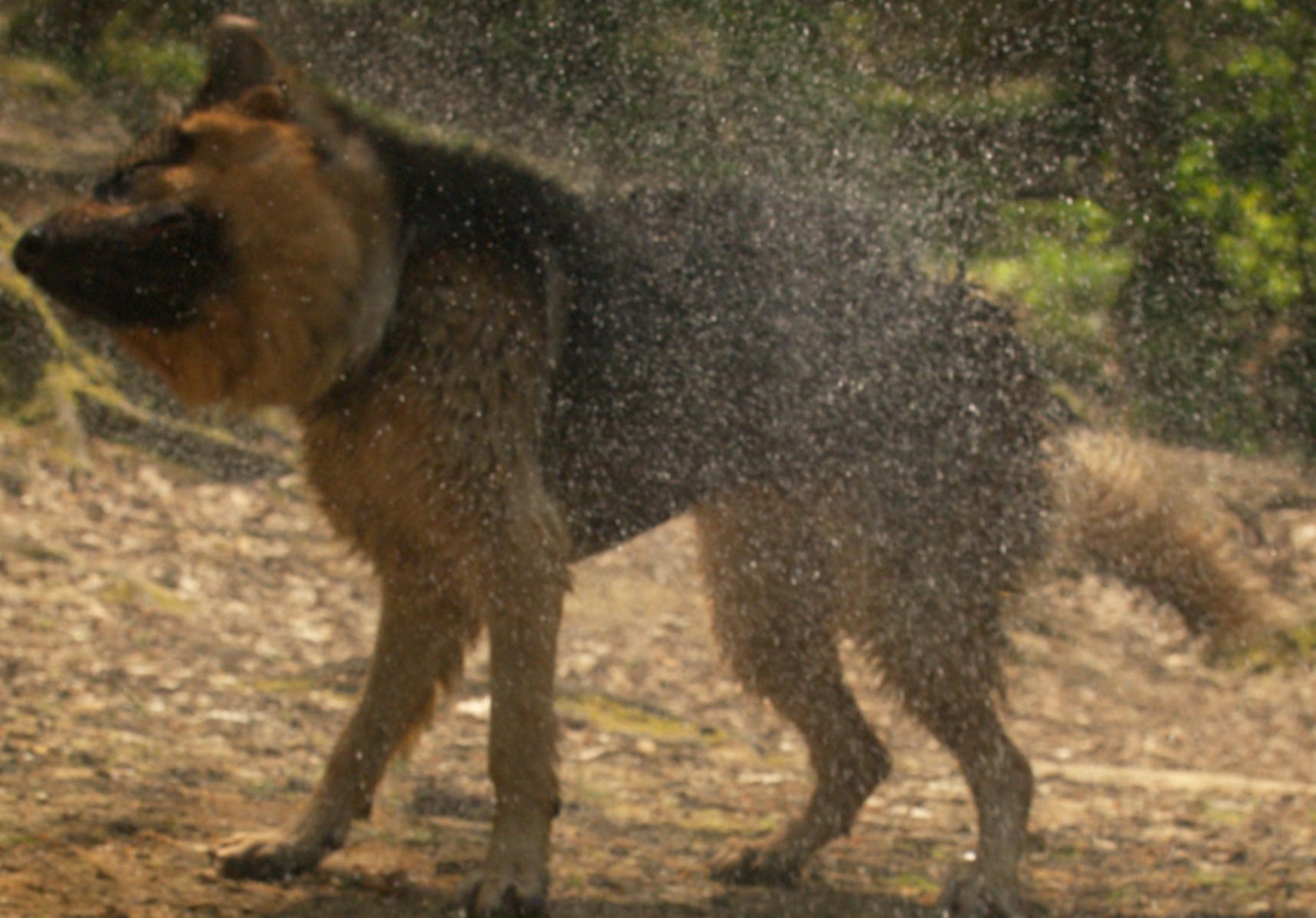}}
    \subfigure[Original]{\includegraphics[width=.37\linewidth]{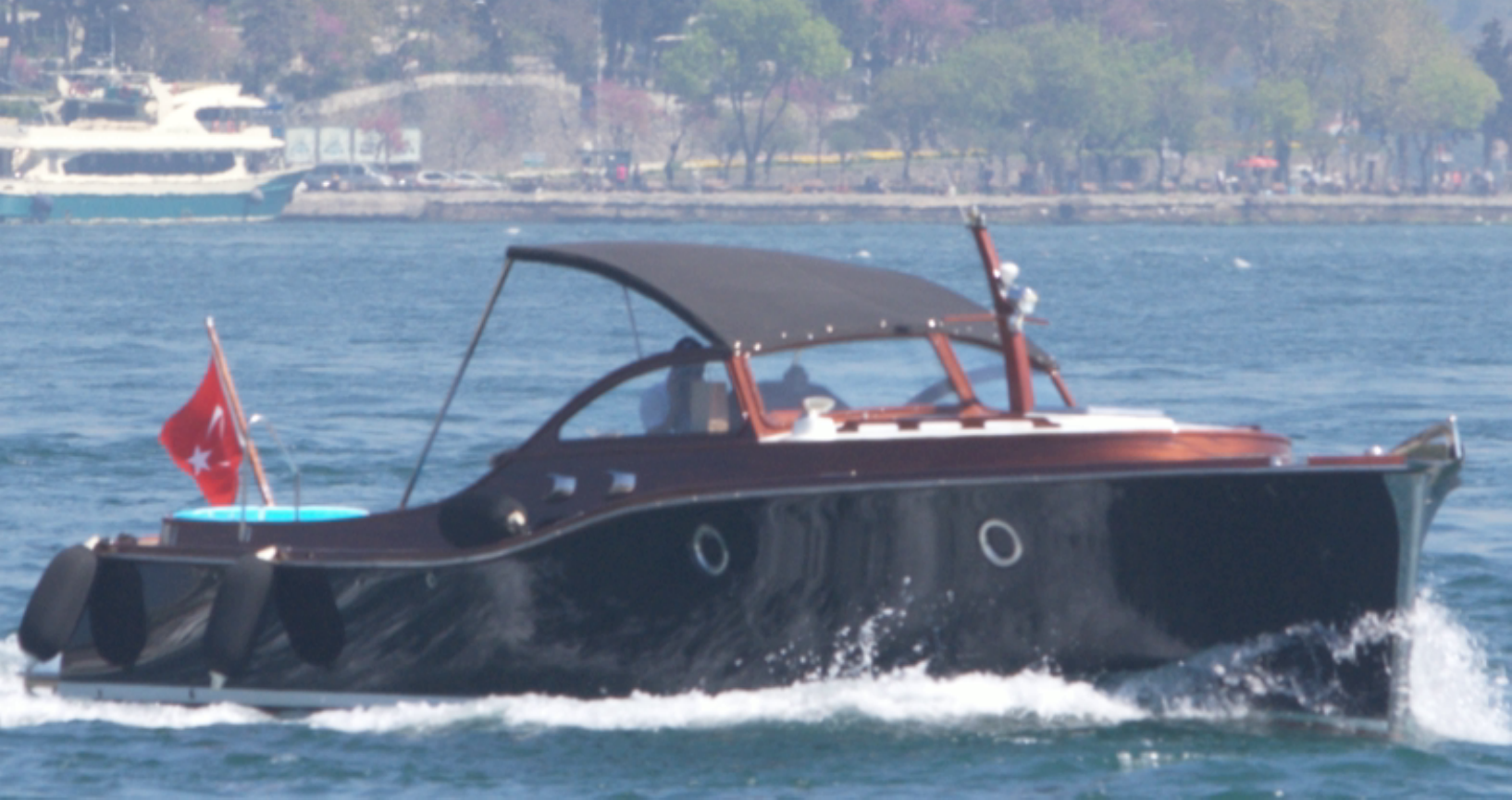}}
    \\
    % % reconstruction
    \subfigure[{PSNR: 33.68  MS-SSIM: 0.96, NeuralMDC}]{\includegraphics[width=.30\linewidth]{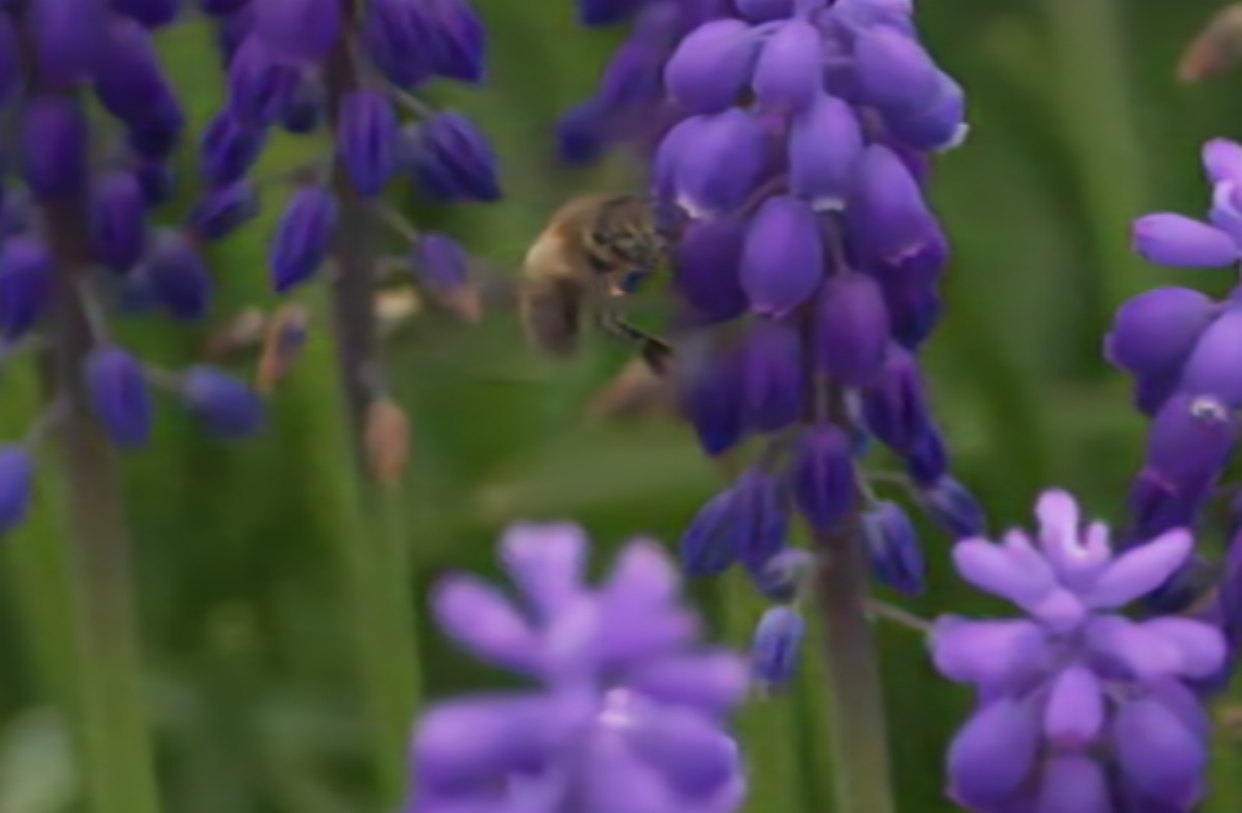}}
    \subfigure[{PSNR: 29.92 MS-SSIM: 0.88, NeuralMDC}]{\includegraphics[width=.28\linewidth]{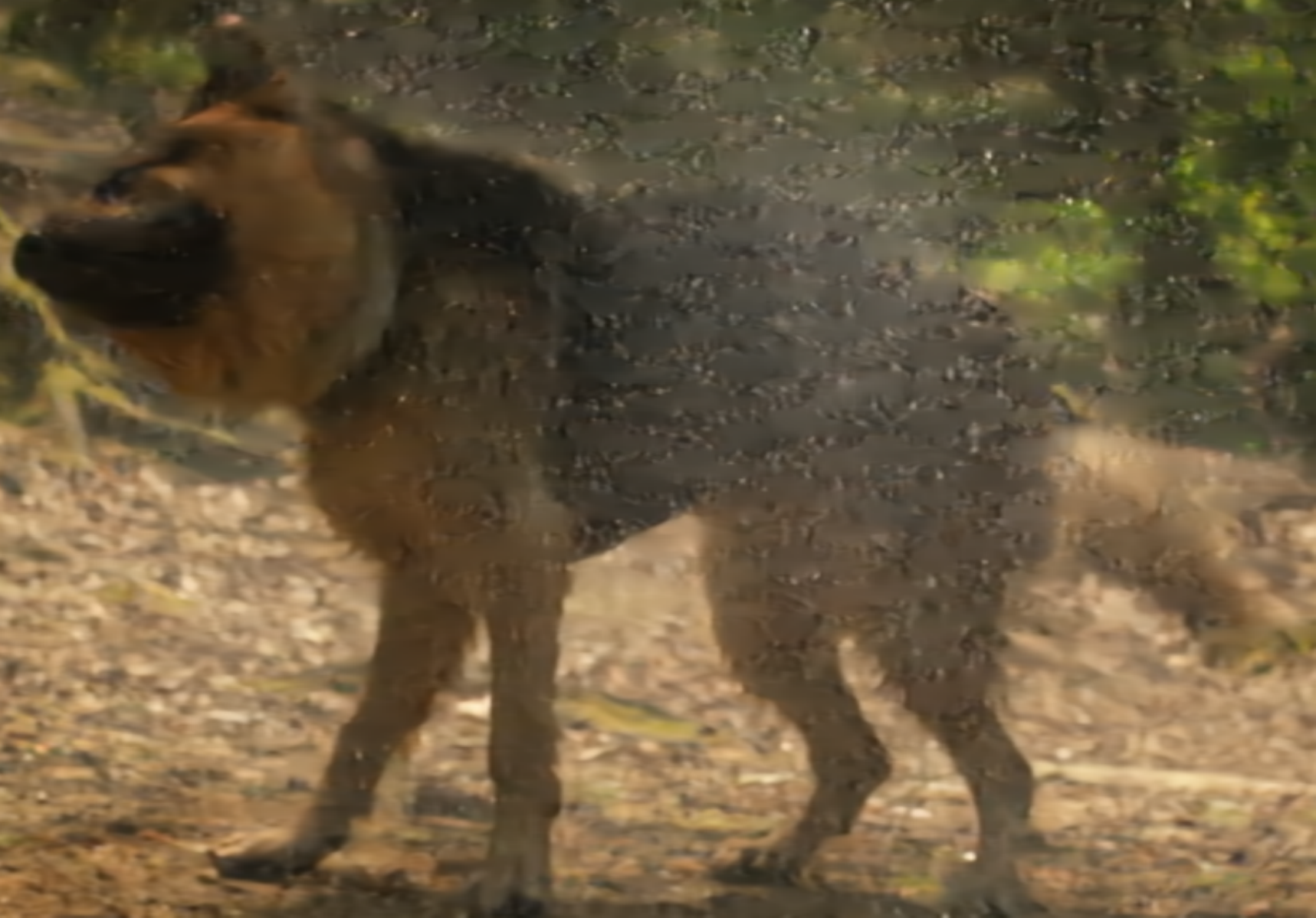}}
    \subfigure[{PSNR: 32.85 MS-SSIM: 0.96, NeuralMDC}]{\includegraphics[width=.37\linewidth]{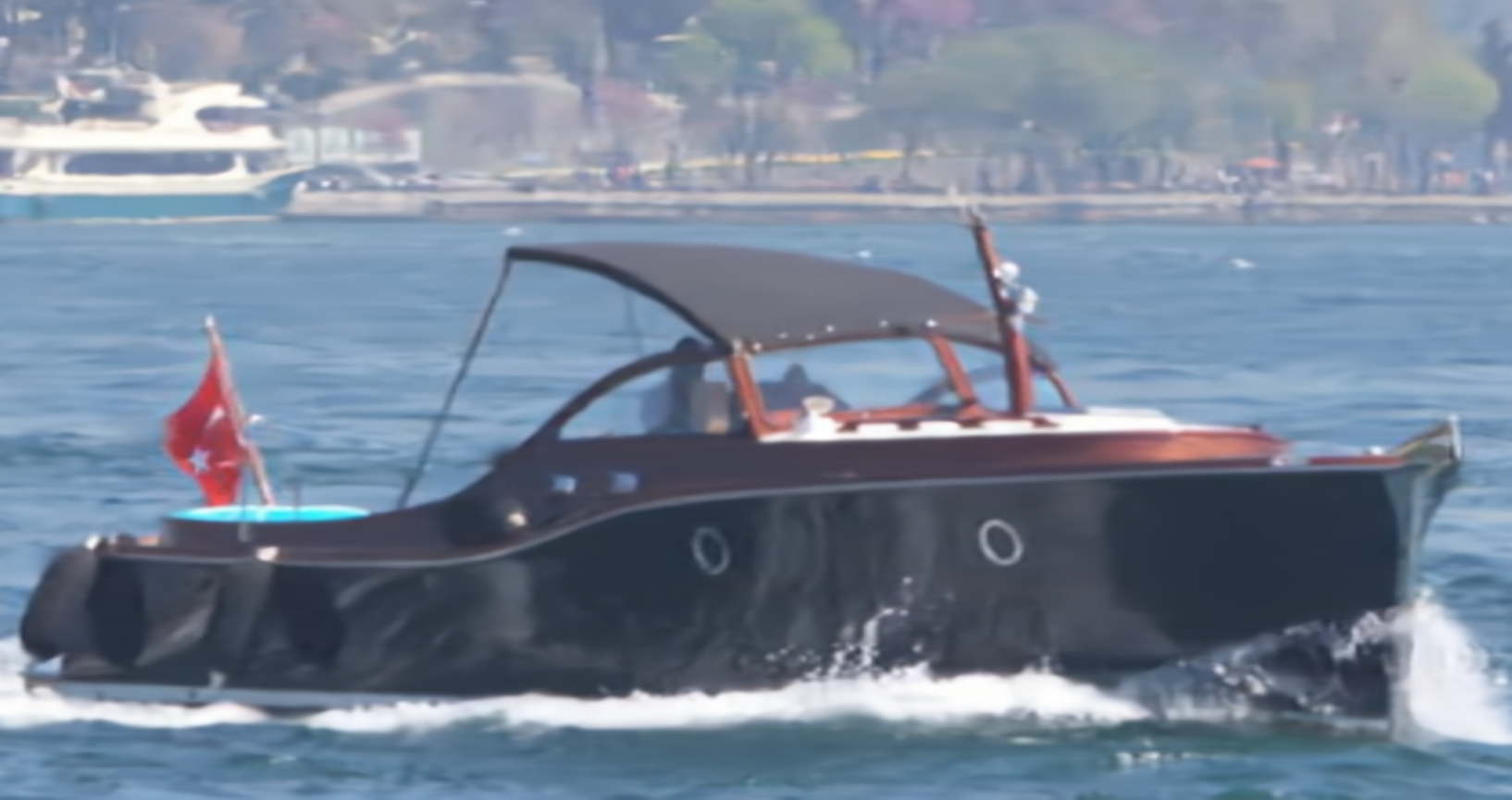}}\\
    \caption{Inference examples from NeuralMDC with 50\% tokens reception.}
    \label{f:reconstruct}
\end{figure*}

\fi

\end{document}